\documentclass[12pt]{article}

\usepackage{amsthm,color}
\usepackage{amsfonts}
\usepackage{graphicx}
\usepackage{mathrsfs}
\usepackage{amsmath}
\usepackage{amssymb}
\usepackage{float}
\usepackage{natbib}
\usepackage{booktabs}
\usepackage{url}
\usepackage{multirow}
\usepackage{arydshln}
\usepackage{threeparttable}
\usepackage{courier}
\usepackage{authblk} 

\usepackage{multicol}
\usepackage{latexsym}
\usepackage{psfrag}
\usepackage[usenames,dvipsnames]{xcolor}

\usepackage{breakcites}
\usepackage{bbm}
\usepackage{epsfig,epstopdf}
\usepackage[colorlinks, linkcolor=black, citecolor=black]{hyperref}

\usepackage[utf8]{inputenc}
\usepackage[english]{babel}
\usepackage{caption}
\usepackage{setspace} 

\usepackage{amssymb,amsmath,epsfig,epsf,psfrag,graphicx,rotating,times,color}
\usepackage{multirow}
\usepackage{amssymb,amsmath,epsfig,epsf,psfrag,graphicx,rotating,times}
\usepackage{color,latexsym,amsfonts,amsthm,amscd}
\usepackage{multirow}
\usepackage{booktabs}
\usepackage{mathabx}

\usepackage{graphicx}

\usepackage{sectsty}
\sectionfont{\fontsize{12}{15}\selectfont}
\subsectionfont{\fontsize{12}{15}\selectfont}

\usepackage{booktabs}
\newtheorem{theorem}{Theorem}
\newtheorem{lemma}{Lemma}
\newtheorem{definition}{Definition}
\newtheorem{corollary}{Corollary}

\newtheorem{example}{Example}

\newcommand{\bm}[1]{\mbox{\boldmath$#1$}}

\def\mathF{\mathcal{F}}
\def\mathG{\mathcal{G}}
\def\mathX{\mathcal{X}}
\def\mathB{\mathcal{B}}
\def\mathY{\mathcal{Y}}
\def\mathD{\mathcal{D}}

\def\mathJ{\mathcal{J}}
\def\mathW{\mathcal{W}}

\def\mathO{\mathcal{O}}

\def\bbR{\mathbb{R}}
\def\bbF{\mathbb{F}}

\def\bbU{\mathbb{U}}

\def\endpf{\quad \blacksquare}
\def\epf{\quad $\blacksquare$}
\newtheorem{remark}{Remark}

\def\pf{\it Proof:}
\allowdisplaybreaks[1]

\begin{document}

{\centering {\Large {\bf Maximum pairwise-rank-likelihood-based inference for the semiparametric transformation model}}}

\begin{center}
{By Tao Yu\\
Department of Statistics \& Applied Probability,
National University of Singapore, Singapore, 117546\\
Email: stayt@nus.edu.sg\\
}
\vspace{0.1in}

{Pengfei Li\\
Department of Statistics and Actuarial Science, University of Waterloo, Waterloo, Ontario, Canada N2L 3G1\\
Email: pengfei.li@uwaterloo.ca\\
}
\vspace{0.1in}

{Baojiang Chen\\
Department of Biostatistics and Data Science,  
University of Texas Health Science Center at Houston,
School of Public Health in Austin,
Austin, Texas 78701, USA\\}
\vspace{0.1in}

{Ao Yuan\\
Department of Biostatistics, Bioinformatics and Biomathematics,
Georgetown University, Washington DC, 20057,  USA\\
Email: ay312@georgetown.edu\\}
 
\vspace{0.1in}

{Jing Qin\\
National Institute of Allergy and Infectious Diseases, National Institutes of Health, MD 20892, USA\\
Email: jingqin@niaid.nih.gov
}

\end{center}

\vspace{0.1in}

\hrule

{\small
\begin{abstract}
\noindent
In this paper, we study the linear transformation model in the most general setup. This model includes many important and popular models in
statistics and econometrics as special cases. Although it has been studied for many years, the methods in the literature are based on kernel-smoothing techniques or make use of only the ranks of the responses in the estimation of the parametric components. The former approach needs a tuning parameter, which is not easily optimally specified in practice; and the latter is computationally expensive and may not make full use of the information in the data. 
In this paper, we propose two methods: a pairwise rank likelihood method and a score-function-based method based on this pairwise rank likelihood. We also explore the theoretical properties of the proposed estimators. Via extensive numerical studies, we demonstrate that our methods are appealing in that the estimators are not only robust to the distribution of the random errors but also lead to mean square errors that are in many cases comparable to or smaller than those of existing methods.

\vspace{0.3cm}

\end{abstract}
}

\hrule

\section{Introduction}
In this paper, we consider the linear transformation model in the most general setup. Let $\{X_i, Y_i\}$, $i=1,\ldots,n$ be independent and identically distributed (i.i.d.) copies of $\{X, Y\}$, where $Y$ denotes the response and $X$ is a $p\times 1$ random vector {\color{black} of covariates}. The transformation model assumes that
\begin{equation}\label{Transformation model}
H(Y_i)=X_i^T\bm{\beta}+\epsilon_i ,
\end{equation}
where  $\epsilon_i$ is the random error with unknown cumulative distribution function (c.d.f.) $F_\epsilon(\cdot)$, {\color{black} and we define $F(\cdot)$ to be the c.d.f. of $\epsilon_i - \epsilon_j$, $i\neq j$}. We assume that $H(\cdot)$, {\color{black} $F(\cdot)$}, and $\bm{\beta}$ are all unknown parameters, where $H(\cdot)$ is a strictly increasing function, {and $E(\epsilon_i)=0$}. {\color{black} Furthermore, we need to impose some conditions so that this model is identifiable. One such condition is that $\|\bm{\beta}\|_2 = 1$, $F_0(\cdot)$ is strictly increasing, and the support of $X_i$, denoted by $\mathX$, contains at least one interior point in $\bbR^p$. Here, $F_0(\cdot)$ denotes the true value of $F(\cdot)$; see Remark \ref{remark-identifiablity} in the Appendix for more discussion of the identifiability of the model. 
}


The class of linear transformation models is of great practical value; it includes many important and popular models in statistics and econometrics as special cases. A comprehensive review of these special cases is as follows. The popular Box--Cox model (Box and Cox 1964) can be viewed as a special case of the linear transformation model with both $H(\cdot)$ and the distribution of the random errors modeled parametrically; in particular, the distribution of the random errors is typically assumed to follow a normal distribution (see Carroll and Ruppert 1988).
The log-linear regression and accelerated failure time models model the distribution of the random errors nonparametrically but assume that the transformation function $H(\cdot)$ satisfies a specific parametric form (see Bickel and Doksum 1981 and the references therein). The parametric and semiparametric proportional hazard models (Cox 1972) can also be formulated  as special cases of linear transformation models.
Specifically, they can be formulated as the linear transformation model with the transformation function modeled nonparametrically but the distribution of the random errors modeled parametrically (Zeng and Lin 2007); see Horowitz (1996) for a more detailed discussion.

Developing appropriate methodology and fast algorithms that accurately estimate the parameters in transformation models is a challenging task because of the nonparametric unknown components. To overcome this difficulty, there are two general strategies.  {\color{black} The first is kernel-smoothing techniques (see for example Song~et~al.~2007; Lin~and~Peng~2013; and Zhang~et~al.~2018).
These methods rely on a tuning parameter, which may not be specified optimally.} The second strategy is rank-based methods: the main idea is to use $\mbox{Rank}(Y_i)$ to replace $H(Y_i)$ in the loss function in the estimation of the parametric parameters. This is motivated by the fact that $\mbox{Rank}(Y_i) = \mbox{Rank}(H(Y_i))$ because of the monotonicity of $H(\cdot)$. Examples include the maximum rank correlation (MRC) method (Han 1987), the monotone rank (MR) method (Cavanagh and Sherman 1998), and the pairwise-difference rank (PDR) method (Abrevaya 1999a,b). These methods are fast and do not need a tuning parameter. However, they use the rank information (rather than the full information) of the responses and may   {\color{black} be less efficient. Furthermore, estimating the transformation function $H(\cdot)$ and the c.d.f. $F(\cdot)$ may be of particular interest in some applications. For example, the estimate of $H(\cdot)$, together with that of $\bm{\beta}$, can be used to predict the response $Y$ for every given $X$. The estimates of $H(\cdot)$ and $F(\cdot)$ can be used to check the model assumptions of some parametric and semiparametric models, e.g., the multiple linear regression model and the Cox PH model reviewed in Section \ref{section-coxPH}. Estimation methods for  $H(\cdot)$ are available in  the literature (e.g., Horowitz~1996; Chen~2002; Zhang~2013), but those for $F(\cdot)$ are not. 
}


We consider the transformation model (\ref{Transformation model}) by treating $H(\cdot), F(\cdot)$,  and $\bm{\beta}$ as unknown parameters. We propose two methods for estimating {\color{black} $\bm{\beta}$ and $F(\cdot)$}. First,
{in the spirit of the pseudo-likelihood method (Besag 1975), we propose a pairwise rank likelihood method}; then, using a strategy similar to that of Groeneboom and Hendrickx (2018), we propose a score-function-based method based on this pairwise rank likelihood.  Numerical studies show that both methods {\color{black} lead to desirable $\bm{\beta}$ and $F(\cdot)$ estimates}. Furthermore, we explore the asymptotic properties of the proposed estimators.
We expect that the methodology, theoretical results, and technical tools of this paper will benefit the study of similar models (e.g., the Box--Cox model). Via extensive simulation studies, we show that our methods are robust to the distribution function $F_\epsilon(\cdot)$ and {\color{black} lead to mean square errors (MSEs) that are in many cases comparable to or smaller than those of} existing methods.

The rest of the paper is organized as follows. Section \ref{section-exist} reviews two popular methods; they are the main competitors of our methods in the numerical studies.  Section \ref{section-estimaton} presents our pairwise likelihood and score-function-based methods for the estimation of the unknown parameters in the linear transformation model (\ref{Transformation model}).  {Section \ref{section-3} investigates the asymptotic properties of these estimates, and Section \ref{section-4} discusses the simulation studies.} Section \ref{section-5} applies our methods to a real-data example, and Section \ref{section-6} ends the paper with a discussion. The technical conditions are relegated to the Appendix, and the  technical details are given in the supplementary material.

\section{Existing Methods} \label{section-exist}

We now briefly review several existing methods that are relevant to our work. Our numerical studies will use them for comparison purposes.

\subsection{Pairwise-difference rank estimators} \label{section-PDR}
Abrevaya (2003) gave a thorough review of rank estimators for the linear transformation model (\ref{Transformation model}). These include the MRC estimator (Han 1987), the MR estimator (Cavanagh and Sherman 1998), and the PDR estimators (Abrevaya 1999a, 1999b). There are several versions of the PDR estimators. PDR3 is based on observation triples, and PDR4 is based on observation quadruples. As demonstrated by Abrevaya (2003), PDR4 has the best performance of all the above estimators.

PDR4 considers the following objective function:
\begin{eqnarray*}
S_{4}(\bm{\beta})=\sum_{(i,j,k,l) \in \mathcal{C}_4}I\{(X_i-X_j)^T\bm{\beta}>(X_k-X_l)^T\bm{\beta}\}\cdot \{I(Y_i>Y_j)-I(Y_k>Y_l)\},
\end{eqnarray*}
with
$$\mathcal{C}_4=\{(i,j,k,l): i,j,k,l\in \{1,\ldots,n\}, i\neq j,i\neq k,i\neq l, j\neq k,j\neq l, k\neq l\}$$
being the quadruples of the index set $\{1,\ldots,n\}$. The estimator of the unknown parameter $\bm{\beta}$ is defined to be
\begin{eqnarray*}
\widehat{\bm{\beta}}_{PDR4} = \mbox{argmax}_{\bm{\beta}} S_4(\bm{\beta}).
\end{eqnarray*}

Although PDR4 performs well, it does not make full use of the data information, and it does not result in an estimator for the nonparametric function {\color{black} $F(\cdot)$}. We propose a {pairwise rank likelihood} method for estimating both {\color{black} $F(\cdot)$} and the parametric parameter $\bm{\beta}$. 

\subsection{Proportional hazard model} \label{section-coxPH}

The proportional hazard model (Cox 1972, 1975), called the Cox PH model hereafter, has been frequently studied and widely applied to analyze censored survival data from the medical sciences. It is a special case of the linear transformation model (Doksum 1987). Let $Y_i$ be a survival time with c.d.f. $F_i(y)$ and probability density function (p.d.f.) $f_i(y)$, and let $X_i$ be the $p\times 1$ clinical covariate. Let $r_i(y) = f_i(y)/\{1-F_i(y)\}$ be the hazard rate. The Cox PH model assumes
\begin{eqnarray*}
r_i(y) = r(y) \exp(X_i^T \bm{\beta}),
\end{eqnarray*}
which has an equivalent form (Lehmann 1953):
\begin{eqnarray*}
F_i(y) = 1-\{1-h(y)\}^{\exp(X_i^T \bm{\beta})},
\end{eqnarray*}
where $h(y) = 1-\exp\left\{-\int_0^y r(t) dt \right\}$. This immediately leads to the model
\begin{eqnarray*}
H(Y_i)   = - X_i^T \bm{\beta} + \epsilon_i,
\end{eqnarray*}
where $H(y) = \log\left[ - \log\{ 1-h(y) \} \right]$
and the $\epsilon_i$'s are i.i.d.~with the extreme value distribution.
 The partial likelihood method (Cox 1972, 1975) is commonly used to
estimate $\bm{\beta}$.
Since the error is specified as the extreme value distribution, the Cox PH model may not be robust if this distribution is misspecified.  For example, another popular model in survival analysis is the proportional odds ratio model where the error distribution is the logistic distribution (Bennett, 1983a, 1983b).
If this is the underlying model, then a method based on the Cox model may not be robust and {\color{black} may lead to larger errors in the $\bm{\beta}$ estimation}.

\section{Proposed Estimation Methods} \label{section-estimaton}

Existing methods for semiparametric regression models
include the profile likelihood method (Breslow, 1972), the partial likelihood method (Cox 1972, 1975), and the rank likelihood method (Kalbfleisch and Prentice, 1973). Because of the existence of sets of infinitely many nuisance parameters $F_\epsilon(\cdot)$ and $H(\cdot)$, none of these methods can be applied directly.
In this section, we propose two methods to estimate {\color{black} both $F(\cdot)$ and $\bm{\beta}$} in the transformation model (\ref{Transformation model}) effectively. First,
{in the spirit of the pseudo-likelihood method (Besag 1975), we propose a pairwise rank likelihood method}. Second, using this pairwise rank likelihood and a strategy similar to that of Groeneboom and Hendrickx (2018), we propose a score-function-based method.

\subsection{Pairwise rank likelihood estimation} \label{section-2}

We propose the {pairwise rank likelihood} and establish a fast algorithm for the estimation of $\bm{\beta}$ and $F(\cdot)$, where we recall that $F(\cdot)$ is the c.d.f. of $\epsilon_i-\epsilon_j$, $i\neq j$. 

{Given the} monotonicity of $H(\cdot)$, we immediately have
\begin{eqnarray}
P(Y_i>Y_j|X_i,X_j)&=&P\left(H(Y_i)>H(Y_j)|X_i,X_j\right) \nonumber\\
&=&P(X_i^T\bm{\beta}+\epsilon_i>X_j^T\bm{\beta}+\epsilon_j|X_i,X_j) \nonumber\\
&=&P\left(\epsilon_j-\epsilon_i<(X_i-X_j)^T\bm{\beta}|X_i,X_j\right) \nonumber\\
&=&F\left((X_i-X_j)^T\bm{\beta}\right). \label{eq-pseudo-l-added}
\end{eqnarray}
This motivates the {pairwise rank log-likelihood}, given by
\begin{eqnarray}
\ell(\bm{\beta},F)&=& \sum_{i\neq j}\Big[I(Y_i>Y_j)\log\{ P(Y_i>Y_j|X_i,X_j) \} \nonumber \\ && \hspace{0.3in} +I(Y_i\leq Y_j)\log \{P(Y_i\leq Y_j|X_i,X_j)\}\Big]\nonumber\\
&=&\sum_{i\neq j}\Big[I(Y_i>Y_j)\log\{ F((X_i-X_j)^T\bm{\beta})\} \nonumber \\&& \hspace{0.3in}+I(Y_i\leq Y_j)\log \{1-F((X_i-X_j)^T\bm{\beta})\}\Big].  \label{pseudo-lik}
\end{eqnarray}
{Note that $\ell(\bm{\beta}, F)$ is essentially the log-likelihood function of the generalized linear model with
$I(Y_i>Y_j), i, j=1,\ldots,n$ and $i\neq j$ being the responses and $F(\cdot)$ being the link function. Consequently}
\begin{eqnarray}
(\widehat{\bm{\beta}}, \widehat F) = \mbox{argmax}_{\bm{\beta}\in \mathB, F \in \mathF} \ell(\bm{\beta}, F), \label{pseudo-like-estimate}
\end{eqnarray}
where $\mathB$ is a compact subspace of $\bbR^p$ and 
$$
\mathF = \{F(\cdot): F(x)\in [0,1] \mbox{ and is monotonically increasing}\}. 
$$
We propose the following two-stage algorithm that solves the optimization problem (\ref{pseudo-like-estimate}) numerically:

\begin{enumerate}
\item[] Stage 1. For a given $\bm{\beta}$,  profile $F$ to obtain
the profile likelihood of $\bm{\beta}$ through the following steps:
\begin{enumerate}
\item Let $(v_{1}(\bm{\beta}),\ldots, v_{K}(\bm{\beta}))$ be a vector composed of $\{ (X_i-X_j)^T\bm{\beta}: i\neq j\}$, and $(I_1,\ldots, I_K)$ be the corresponding vector composed of $\{I(Y_i>Y_j): i\neq j\}$. Sort $v_k(\bm{\beta})$ in increasing order:
$$v_{(1)}(\bm{\beta})\leq \ldots \leq v_{(K)}(\bm{\beta}).$$
The corresponding $I_j$'s are denoted by $I_1(\bm{\beta}),\ldots, I_K(\bm{\beta})$.
Substituting $v_{(j)}(\bm{\beta})$ and the corresponding $I_j(\bm{\beta})$ into (\ref{pseudo-lik}), we have
\begin{eqnarray*}
\ell(\bm{\beta},F)=\sum_{j=1}^K[I_j(\bm{\beta})\log\{ F(v_{(j)}(\bm{\beta}))\}+\{1-I_j(\bm{\beta})\}\log \{1-F(v_{(j)}(\bm{\beta}))\}].
\end{eqnarray*}

\item For any $\bm{\beta}$ and  the $\ell(\bm{\beta},F)$ given in (a),  let
\begin{eqnarray}
\widehat F_{\bm{\beta}} = \mbox{argmax}_{F\in \mathF}\ell(\bm{\beta}, F). \label{eq-F-profile}
\end{eqnarray}
\item The profile {pairwise rank log-likelihood} is given by
\begin{eqnarray}
\ell(\bm{\beta},\widehat{F}_{\bm{\beta}})&=&\sum_{i\neq j}\Big[I(Y_i>Y_j)\log\{ \widehat{F}_{\bm{\beta}}((X_i-X_j)^T\bm{\beta})\} \nonumber \\ && \hspace{0.3in} +I(Y_i\leq Y_j)\log \{1-\widehat{F}_{\bm{\beta}}((X_i-X_j)^T\bm{\beta})\}\Big]. \label{profile log-lik}
\end{eqnarray}
\end{enumerate}

\item[] Stage 2. Maximize (\ref{profile log-lik}) with respect to $\bm{\beta}$ to obtain $\widehat{\bm{\beta}}$. 
{\color{black} More details of the numerical implementations for $\widehat{\bm{\beta}}$ are given in Section 9 of the supplementary material.}
\end{enumerate}

Note that the optimization problem (\ref{eq-F-profile}) in Stage 1(b) can be readily solved by the classical pool-adjacent-violation-algorithm 
(PAVA; Ayer et al.~1955) and active
set methods (de Leeuw et~al.~2009). This is because, based on Ayer et al.~(1955), we immediately conclude that
\begin{eqnarray*}
\widehat F_{\bm{\beta}}(v_{(j)}(\bm{\beta}))=\max_{s\leq j}\min_{t \geq j}\frac{\sum_{h=s}^t I_h(\bm{\beta})}{t-s+1}, j=1,\ldots, K,
\end{eqnarray*}
which is equivalent to the solution from the classical isotonic regression problem (Robertson et~al.~1988),
\begin{eqnarray*}
\widehat F_{\bm{\beta}} &=& \mbox{argmin}_{F\in \mathF}\sum_{j=1}^K\left\{I_j(\bm{\beta})- F(v_{(j)}(\bm{\beta}))\right\}^2 \\ && \mbox{subject to }F\mbox{ being a nondecreasing function}.
\end{eqnarray*}


\begin{remark}
The linear transformation model (\ref{Transformation model}) is essentially a semiparametric model; it contains two nonparametric parameters, $F_{\epsilon}(\cdot)$ and $H(\cdot)$, and one $(p\times 1)$-dimensional parametric parameter $\bm{\beta}$. It may be difficult to construct and directly maximize a single objective function that includes all the unknown parameters (e.g., the full semiparametric likelihood). Our method solves this problem by first omitting the nuisance nonparametric component $H(\cdot)$ using the pseudo-likelihood technique; a profile-type method is then applied to further overcome the potential difficulty that may be caused by  $F(\cdot)$. Thus, the objective function carries only the parametric parameter $\bm{\beta}$, and it can be readily estimated by existing {\color{black} algorithms}.
\end{remark}

\begin{remark}
Our {pairwise rank likelihood} approach takes all the orders of $(X_i-X_j)^T\bm{\beta}$ into account, while PDR4 considers only the paired order of $(X_i-X_j)^T\bm{\beta}$ and $(X_k-X_l)^T\bm{\beta}$. {\color{black} Therefore, our method has used the information from the data more effectively}. Moreover, our method can be considered a pairwise-rank-based approach in which the transformation function $H(\cdot)$ is eliminated.
\end{remark}

{\begin{remark}
%

{
For presentational convenience, in all our analyses we have assumed that the responses $Y_i$ have no ties. However, in practice, ties may occasionally occur in the outcome data, e.g., if the collected responses are subject to some rounding mechanism. In Section 10 of the supplementary material, we suggest a solution for this problem in the framework of our pairwise rank log-likelihood. }

\end{remark}

}

\subsection{Score-function-based method based on the pairwise rank likelihood} \label{section-alt-beta}

As observed in Section \ref{section-3} and discussed in Remark \ref{remark-1}, it is difficult to establish the root $n$ consistency and asymptotic distribution for the maximum pairwise rank log-likelihood estimator $\widehat{\bm{\beta}}$. In this section, using a strategy similar to that of Groeneboom~and~Hendrickx~(2018), we consider an alternative estimator $\widetilde{\bm{\beta}}$ whose asymptotic distribution can be established; see Section \ref{section-asym-score}.

We need the following notation. For any $\bm{\beta}\in \mathB$, define
\begin{eqnarray*}
F_{\bm{\beta}}(t) &=& E\left\{I(Y_1>Y_2)\Big|(X_1-X_2)^T\bm{\beta} = t\right\}\\
&=& \int F_0(t + \bm{v}^T (\bm{\beta}_0 - \bm{\beta})) f_{(X_1-X_2)|(X_1-X_2)^T\bm{\beta} = t} (\bm{v}) d\bm{v}
\end{eqnarray*}
and
\begin{equation}
\varphi_{\bm{\beta}}(t) = E(X_1 - X_2|(X_1-X_2)^T \bm{\beta} = t), \quad \varphi_0(t) =  \varphi_{\bm{\beta}_0}(t). \label{def-varphi}
\end{equation}
We observe that for any $\bm{\beta}$, $\widehat F_{\bm{\beta}}$ defined by (\ref{eq-F-profile}) is essentially an estimator for $F_{\bm{\beta}}$, and
\begin{eqnarray*}
\widehat{\bm{\beta}} &=& \arg\max_{\bm{\beta}\in \mathB}\ell\left(\bm{\beta}, \widehat F_{\bm{\beta}} \right),
\end{eqnarray*}
where the form of $\ell\left(\bm{\beta}, \widehat F_{\bm{\beta}} \right)$ is given in (\ref{profile log-lik}).

The score function for $\ell\left(\bm{\beta}, \widehat F_{\bm{\beta}} \right)$ (if it exists) has the form
\begin{eqnarray*}
\overline{\psi}_n(\bm{\beta}) &=& \sum_{i\neq j} \xi(X_i,X_j;  \widehat F_{\bm{\beta}}, \bm{\beta})\left\{I(Y_i>Y_j)- \widehat F_{\bm{\beta}}\left((X_i-X_j)^T \bm{\beta} \right)\right\},
\end{eqnarray*}
 with $\xi(\cdot)$ being a function depending on the data and the unknown parameter $\bm{\beta}$.
 Then $\widehat{\bm{\beta}}$ can be viewed as the root of this score function. However, there are two aspects that complicate the development of the optimal convergence rate and asymptotic distribution for $\widehat{\bm{\beta}}$: (1) this score function may not exist, since $\widehat F_{\bm{\beta}}\left((X_i-X_j)^T \bm{\beta} \right)$ may not be differentiable with respect to $\bm{\beta}$; (2) the function $\xi(\cdot)$ has a complicated form and may depend on $\bm{\beta}$. To bypass these two difficulties, we first replace $\xi(X_i,X_j;  \widehat F_{\bm{\beta}}, \bm{\beta})$ by $X_i-X_j$ in this score function. Second, we would like to define $\widetilde{\bm{\beta}}$ to be the root of $\psi_n(\bm{\beta}) = 0$, where
\begin{equation}
\psi_n(\bm{\beta}) =\frac{1}{n^2} \sum_{i\neq j}(X_i-X_j)\left\{I(Y_i>Y_j)- \widehat F_{\bm{\beta}}\left((X_i-X_j)^T \bm{\beta} \right)\right\}. \label{def-psi-n}
\end{equation}
However, since $\psi_n(\bm{\beta})$ may not be continuous, this root may not exist, so, similarly to Groeneboom~and~Hendrickx~(2018), we define:
\begin{eqnarray*}
 \mbox{$\widetilde{\bm{\beta}}$ is the {\it zero-crossing} of $\psi_n(\bm{\beta})$},
\end{eqnarray*}
 where the zero-crossing of a function (or a mapping) is defined below.

\begin{definition}
For a function $\psi$: $\mathX \to \bbR$, $x$ is called the zero-crossing of $\psi$ if every open neighborhood of $x$ contains $x_1, x_2$ such that $\psi(x_1)\psi(x_2) \leq 0$.
For a mapping $\widetilde \psi$: $\mathX \to \bbR^d$, $x$ is called the zero-crossing of $\widetilde \psi$ if $x$ is the zero-crossing of each component of $\widetilde \psi$.
\end{definition}

{\color{black} Details of the numerical implementations of $\widetilde{\bm{\beta}}$ are given in Section 9 of the supplementary material. 
With $\widetilde{\bm{\beta}}$ and the corresponding techniques given in Section \ref{section-2},  we estimate $F(\cdot)$ by 
$\widetilde F(\cdot)=\widehat F_{\widetilde{\bm{\beta}}}(\cdot)$.
}

{\color{black} 
\begin{remark} \label{remark-H-est}
With the estimators of $\bm{\beta}$ and $F(\cdot)$, estimation methods for $H(\cdot)$ are available in the literature (see, e.g., Horowitz~1996; Chen~2002; Zhang~2013). Chen~(2002) proposed a rank-based estimator for $H(\cdot)$, assuming that the $\bm{\beta}$ estimate is available, and Zhang (2013) proposed a self-induced smoothing method based on Chen (2002)'s estimator.
If the estimate for $\bm{\beta}$ is root $n$ consistent, both estimates can achieve root $n$ consistency and converge weakly to Gaussian processes.  In Section~\ref{section-asym-score}, we show that $\widetilde{\bm{\beta}}$ is root $n$ consistent; therefore, with $\widetilde{\bm{\beta}}$, the corresponding $H(\cdot)$ estimates from both papers  have desirable asymptotic properties.

\end{remark}
}

{\color{black} 
\begin{remark}

In this paper, we have focused on the case where the $Y_i$'s are complete data. Our methods, however, can be extended to analyze the transformation model when some $Y_i$'s are right censored. See the supplementary material for this extension.

\end{remark}
}

\section{Asymptotic Properties}  \label{section-3}

In this section, we explore the asymptotic properties of our estimators. Similarly to Section~\ref{section-estimaton}, we organize this section into two subsections, respectively establishing the asymptotic properties of the estimators given in Sections~\ref{section-2} and \ref{section-alt-beta}. For space limitations and presentational continuity, we have relegated the technical conditions to the Appendix, and the technical details are in the supplementary material.

\subsection{Asymptotic properties for the pairwise rank likelihood estimators} \label{section-asym-prle}

We first investigate the asymptotic properties of the pairwise rank likelihood estimators proposed in Section \ref{section-2}.
Let
\begin{eqnarray}
D\Big(\bm{\beta}_2, F_2; \bm{\beta}_1, F_1\Big) &=& \bigg[\int \int \Big\{F_1((\bm{x}_1-\bm{x}_2)^T \bm{\beta}_1) \nonumber \\ && \hspace{0.5in}- F_2((\bm{x}_1-\bm{x}_2)^T \bm{\beta}_2) \Big\}^2 d F_{X}(\bm{x}_1)d F_{X}(\bm{x}_2)\bigg]^{1/2}, \label{def-D}
\end{eqnarray}
where $F_X(\cdot)$ is the c.d.f. of $X_i$. Theorem \ref{theorem-1} below establishes the convergence rates of $\widehat{\bm{\beta}}$ and $\widehat F(\cdot)$.
\begin{theorem}
\label{theorem-1}
Assume Conditions 0--6 in the Appendix. We have
\begin{itemize}
\item[(a)]
$
D\Big( \widehat{\bm{\beta}}, \widehat F; \bm{\beta}_0, F_0\Big) = O_p(n^{-1/3}),
$
\item[(b)] $\widehat{\bm{\beta}}-\bm{\beta}_0 = O_p(n^{-1/3})$,
\end{itemize}
where $\bm{\beta}_0$ and $F_0$ are the true values of $\bm{\beta}$ and $F$.
\end{theorem}

\begin{remark} \label{remark-1}
The development of the asymptotic properties given in the above theorem is a challenging task for two reasons. First, the structure of the {pairwise rank likelihood} function (\ref{pseudo-lik}) is complicated. It is not a sum of i.i.d.~components, and the M-estimator theory from empirical processes is not directly applicable. Second, the monotonic nonparametric components in the transformation model complicate the development of the convergence rate for the estimator of the parametric component. Such problems are long-standing. For example,
Huang and Wellner (1993) encountered a similar challenge.  They studied current status data under the accelerated failure time model assumption, and they proved only the $o_p(1)$ convergence of their estimators. In other words, they proved consistency but did not prove a convergence rate. In the theorem above, we show that the convergence rate for $\widehat{\bm{\beta}}$ is at least $O_p(n^{-1/3})$. However, we conjecture that this rate is not optimal; the best rate may be $O_p(n^{-1/2})$. We leave this important and interesting problem for future research.
%
\end{remark}


%

\subsection{Asymptotic distribution for the score-function-based estimators} \label{section-asym-score}

In this section, {\color{black} we establish the asymptotic distributions for $ \widetilde{\bm{\beta}}$ and $\widetilde F(\cdot)$. }

Recall the definition of $\psi_n(\cdot)$ in (\ref{def-psi-n}). Its population version $\psi_0(\bm{\beta})$ can be defined to be:
\begin{eqnarray}
 \psi_0(\bm{\beta}) = \int \int (\bm{x}_1 - \bm{x}_2)\left\{I(y_1>y_2) -  F_{\bm{\beta}}((\bm{x}_1 - \bm{x}_2)^T \bm{\beta})\right\} \nonumber \\ \times d F_{X,Y}(\bm{x}_1, y_1) d F_{X,Y}(\bm{x}_2, y_2),
 \label{def-psi-0}
\end{eqnarray}
where $F_{X,Y}(\bm{x}, y)$ is the joint c.d.f. of $(X,Y)$.

We have the following lemma for $\psi_0(\cdot)$.

\begin{lemma}\label{lemma-psi-0}
Assume Conditions 6, A1, and A2 in the Appendix. We have the following:

\begin{itemize}
\item[(1)] $\psi_0(\bm{\beta}_0) = 0$;

{\color{black} 
\item[(2)] $\psi_0'(\bm{\beta}_0) = \frac{\partial \psi_0(\bm{\beta})}{\partial \bm{\beta}}\Big|_{\bm{\beta} = \bm{\beta}_0}$ exists with rank $p-1$.

\item[(3)] Since $\|\bm{\beta}_0\|_2 = 1$, there exists an $i$ such that the $i$th component of $\bm{\beta}_0$ is nonzero. Define $A = \psi_0'(\bm{\beta}_0) + \bar A$, where $\bar A$ is a $p\times p$ matrix with $i$th row $\bm{\beta}_0^T$ and all other entries $0$. Then $A$ is of full rank, and
\begin{eqnarray*}
\psi_0(\bm{\beta}) = A\cdot (\bm{\beta}-\bm{\beta}_0) + o(\bm{\beta} - \bm{\beta}_0).
\end{eqnarray*}

}

\end{itemize}

\end{lemma}

We have the following theorem, which establishes the asymptotic properties of $\widetilde{\bm{\beta}}$.

\begin{theorem} \label{thm-normality}
Assume Conditions 0--2, 6, and A1--A3 in the Appendix. Denote by $\widetilde{\bm{\beta}}$ the zero-crossing of $\psi_n(\bm{\beta})$ (if it exists). We have the following:
\begin{itemize}

\item[(1)] When $n\to \infty$, a zero-crossing $\widetilde{\bm{\beta}}$ of $\psi_n(\bm{\beta})$ exists with probability tending to 1.

\item[(2)] $\widetilde{\bm{\beta}} \to \bm{\beta}_0$ in probability.

{\color{black} 
\item[(3)] Recalling the matrix $A$ defined in Lemma \ref{lemma-psi-0} Part (3), we have
    \begin{eqnarray}
    \widetilde{\bm{\beta}} - \bm{\beta}_0 &=&
   \frac{  A^{-1} }{n^2} \sum_{i\neq j} \Big[
\left\{X_i - X_j - \varphi_0\left((X_i - X_j)^T \bm{\beta}_0\right)\right\} \nonumber\\
&& \times \left\{I(Y_i>Y_j) - F_0\left((X_i - X_j)^T \bm{\beta}_0\right)\right\} \Big]\nonumber \\  &&+ o_p(\widetilde{\bm{\beta}} - \bm{\beta}_0) + o_p(n^{-1/2}).
 \label{beta-tilde-expansion}
\end{eqnarray}

}

\end{itemize}
\end{theorem}
{\color{black} 
To derive the asymptotic distribution for $\widetilde{\bm{\beta}}$, we need to work on a U-statistic with the kernel:
\begin{eqnarray*}
h(\bm{x}_1, y_1; \bm{x}_2, y_2) = \left(\bm{x}_1 - \bm{x}_2\right)\left\{I(y_1>y_2) - F_0\left((\bm{x}_1 - \bm{x}_2)^T \bm{\beta}_0\right)\right\}.
\end{eqnarray*}
Denote
\begin{eqnarray}
\hbar(\bm{x}_1, y_1) = E\left\{h(\bm{x}_1, y_1; X_2, Y_2)\right\}. \label{def-hbar}
\end{eqnarray}
The theorem above leads to the asymptotic distribution of $\widetilde{\bm{\beta}}$.
}
{\color{black} 
\begin{corollary} \label{corollary-normality}
Assume the conditions of Theorem \ref{thm-normality}. Furthermore, assume that both $\Sigma_X = \mbox{var}(X)$ and $\Sigma_{\hbar} = \mbox{var}\left\{\hbar(X, Y)\right\}$ are of full rank, and
\begin{eqnarray}
\bm{\alpha}^T E\left(X_2 - X_1| (X_2-X_1)^T\bm{\beta}_0\right) = 0, \nonumber \\ \mbox{ for any } \bm{\alpha} \in \mathbb{R}^p, \mbox{ and } \bm{\alpha}^T\Sigma_X \bm{\beta}_0 = 0.  \label{condition-distribution-X}
\end{eqnarray}
Denote $\widetilde A = \psi_0'(\bm{\beta}_0)\Sigma_X^{-1} + {\bar A}${, where $\bar A$ is defined in Lemma \ref{lemma-psi-0} Part (3).} We have
 \begin{eqnarray}
\sqrt{n}(\widetilde{\bm{\beta}} - \bm{\beta}_0) \to 2 A^{-1} \widetilde A^{-1} \psi_0'(\bm{\beta}_0) \Sigma_X^{-1} N(\bm{0}, \Sigma_{\hbar}) \label{eq-coro-normality}
\end{eqnarray}
in distribution, as $n\to \infty$.

\end{corollary}

\begin{remark}
Write $Z_1 = \bm{\alpha}^T (X_2 - X_1)$, $Z_2 = \bm{\beta}_0^T(X_2 - X_1)$. Then  the condition given by (\ref{condition-distribution-X}) essentially requires that $E(Z_1|Z_2) = E(Z_1)$ is implied by $\mbox{cov}(Z_1, Z_2) = 0$, since the latter is straightforwardly verified by $\bm{\alpha}^T\Sigma_X \bm{\beta}_0 = 0$. This condition is satisfied, for example, when $X_i$ follows a multivariate normal distribution.

\end{remark}

}

{\color{black} 
\begin{remark} \label{remark-BPCI}
In the above lemma, we have derived the explicit formula for the asymptotic variance of $\widetilde{\bm{\beta}}$; therefore, plug-in methods can be employed to estimate this variance when the data are available. However, this formula has a complicated form, and plug-in methods may not lead to the desired accuracy. We suggest that one can incorporate bootstrap methods to estimate this variance and perform statistical inference for $\bm{\beta}$; in Section \ref{section-4}, we suggest a bootstrap percentile confidence interval (BPCI) method that can be used for this inference. From our numerical studies, we observe that the coverage probability of the BPCI based on $\widetilde{\bm{\beta}}$ is reasonably close to the nominal level. 
\end{remark}
}

The proof of this corollary uses standard results from the literature for the asymptotic distribution of the U-statistic; the details are given in the supplementary material.

{\color{black} Next, we establish the asymptotic distribution of $\widetilde F(\cdot)=\widehat F_{\widetilde{\bm{\beta}}}(\cdot)$.
\begin{theorem} \label{theorem-F-asymptotics}

Assume that all the conditions of Corollary \ref{corollary-normality} are satisfied, and also assume the following conditions:
\begin{itemize}

\item[] \underline{Condition F1}: Recall that $F_\epsilon(s)$ is the c.d.f. for $\epsilon_i$. Assume that it is continuous for $s\in \mathbb{R}$ and is continuously differentiable and strictly monotone for $s$ in its support.


\item[] \underline{Condition F2}: Denote by $F_{X^T\bm{\beta}}(s)$ the c.d.f. for $X^T \bm{\beta}$. Assume that it is continuously differentiable for $s$ in its support. Let
$f_{X^T\bm{\beta}}(s) \equiv \frac{\partial F_{X^T\bm{\beta}}(s)}{\partial s}$; assume that for any $s$, it is continuous for $\bm{\beta}$ in the neighborhood of $\bm{\beta}_0$; and $f_{X^T\bm{\beta}_0}(s)$ is continuous for $s$ in its support.

\end{itemize}
Then, we have for every $t$,
\begin{eqnarray*}
n^{1/2}\left\{ \widehat F_{\widetilde{\bm{\beta}}}(t) - F_0(t)\right\} \to N\left(0,  \sigma^2(t)/g_0^2(t)\right),
\end{eqnarray*}
in distribution,
where $g_0(t)$ denotes the p.d.f. for $(X_2-X_1)^T \bm{\beta}_0$, $\sigma^2(t) = \mbox{var}\left\{\zeta_t(X, Y)\right\}$,
\begin{eqnarray*}
\zeta_t(\bm{x}, y) &=& \left\{1-F_\epsilon(H_0(y) - \bm{x}^T \bm{\beta}_0-t) - F_0(t)  \right\} f_{X^T\bm{\beta}_0}(\bm{x}^T \bm{\beta}_0+t)\\
&& +\left\{F_\epsilon(H_0(y) - \boldsymbol{x}^T\boldsymbol{\beta}_0+t) - F_0(t)  \right\}f_{X^T\bm{\beta}_0}(\bm{x}^T \bm{\beta}_0-t)\\
&&+2 g_0(t) \dot{F}_0^T(t) A^{-1} \widetilde A^{-1} \psi_0'(\bm{\beta}_0) \Sigma_X^{-1} \hbar(\bm{x}, y),
\end{eqnarray*}
with $\dot{F}_0(t) = \partial F_{\bm{\beta}}(t)/ \partial \bm{\beta}\big|_{\bm{\beta} = \bm{\beta}_0}$ and  $\hbar(\cdot, \cdot)$ defined by (\ref{def-hbar}).

\end{theorem}

With developments similar to but simpler than those of Theorem~\ref{theorem-F-asymptotics}, we are able to establish the asymptotic distribution for $\widehat F_{\bm{\beta}}(t)$ for every given $\bm{\beta}\in \mathB$ and $t\in \mathbb{R}$; we summarize this result in the following corollary.

\begin{corollary} \label{coro-F-beta-asymptotic}
Assume Conditions 0--2 and A1 in the Appendix and Conditions F1' and F2' given below.
\begin{itemize}
\item[] \underline{Condition F1'}: $F_{\epsilon, \bm{\beta}}(y, s)$ defined by (\ref{def-F-eps-beta}) is continuous for $s\in \mathbb{R}$ and is continuously differentiable for $s$ in the support and $\bm{\beta} \in \mathcal{B}$.


\item[] \underline{Condition F2'}: Denote by $F_{X^T\bm{\beta}}(s)$ the c.d.f. for $X^T \bm{\beta}$. Assume that it is continuously differentiable for $s$ in the support. Recall
$f_{X^T\bm{\beta}}(s)$ defined by Condition F2 in Theorem \ref{theorem-F-asymptotics}; assume that it is continuous for $s$ in the support and $\bm{\beta} \in \mathcal{B}$.
\end{itemize}
For every $t\in \mathbb{R}$ and $\bm{\beta}\in \mathcal{B}$, we have
\begin{eqnarray*}
n^{1/2}\left\{ \widehat F_{\bm{\beta}}(t) - F_{\bm{\beta}}(t)\right\} \to N\left(0,  \sigma_{\bm{\beta}}^2(t)/g_{\bm{\beta}}^2(t)\right),
\end{eqnarray*}
in distribution, where $g_{\bm{\beta}}(\cdot)$ denotes the p.d.f. of $(X_2-X_1)^T\bm{\beta}$, and
\begin{eqnarray}
\sigma_{\bm{\beta}}(t) &=& E\Big[\mbox{var}\Big\{f_{X^T\boldsymbol{\beta}}(X_1^T\boldsymbol{\beta}+t)F_{\epsilon, \bm{\beta}}(Y_1, X_1^T\bm{\beta} + t)\\ &&\hspace{0.5in}- f_{X^T\boldsymbol{\beta}}(X_1^T\boldsymbol{\beta}-t) F_{\epsilon, \bm{\beta}}(Y_1, X_1^T\bm{\beta} - t)  \Big| X_1 \Big\} \Big] \nonumber\\
F_{\epsilon, \bm{\beta}}(y, s) &=& E\left\{F_\epsilon(H(y) - X^T\bm{\beta_0})\Big| X^T\bm{\beta} = s\right\}. \label{def-F-eps-beta}
\end{eqnarray}

\end{corollary}

{\color{black} Interestingly, we observe from Theorem \ref{theorem-F-asymptotics} and Corollary \ref{coro-F-beta-asymptotic} that with any fixed $\bm{\beta}$ or our $\widetilde{\bm{\beta}}$ estimate, our $F(\cdot)$ estimate is able to achieve root $n$ consistency and is asymptotically normally distributed. As far as we are aware, we are the first to establish a root $n$ consistent estimate for this $F(\cdot)$. A good estimate may be valuable in both theory and practice. First, we have illustrated in the real-data example that this $F(\cdot)$ estimate can be used to perform diagnostics for the distribution of $\epsilon_i$. Second, this estimate with its desirable theoretical properties may benefit other research problems. For example, in the Harrells C-index measurement in medical applications, $F(\cdot)$ plays an important role because
\begin{eqnarray*}
&& E\left\{I(Y_i<Y_j) I(X_i^T\bm{\beta} < X_j^T\bm{\beta})\Big| X_i, X_j\right\}\\ &=&  F \left( (X_j-X_i)^T \bm{\beta}\right)I\left( (X_i-X_j)^T\bm{\beta}<0 \right).
\end{eqnarray*}

}

}

\section{Simulation Studies} \label{section-4}

In this section, we investigate the performance of our methods. {We compare our {pairwise rank likelihood} estimators from Section \ref{section-2} (referred to as ``Our-PRL'') and the score-function-based estimators from Section \ref{section-alt-beta} (referred to as ``Our-score") with}:
(1) the PDR4 method reviewed in Section \ref{section-PDR};
(2)
 the partial likelihood method based on the Cox PH model reviewed in Section \ref{section-coxPH}; {
 (3) the smoothed rank correlation estimator (Ma and Huang 2005, Lin and Peng 2013) using the normal cumulative density function to approximate the indicator function (denoted ``Smooth-Normal"); (4) the smoothed rank correlation estimator using the sigmoid function to approximate the indicator function (denoted ``Smooth-Sigmoid"; Song et al. 2007);  and (5) the smoothed maximum rank correlation estimator (denoted ``SMRCE"; Zhang et al. 2018).}
 {Clearly, all the methods are built on semiparametric models. However, the Cox PH method requires a stronger model assumption than the others, i.e., it requires that the random error follows the extreme value distribution.}

{\color{black} 
As discussed in Remark \ref{remark-BPCI}, the variance formula for $\widetilde{\bm{\beta}}$ has a complicated form; estimating this variance by a plug-in method may not be practical. 
On the other hand, in many applications it may be of particular interest to estimate the standard error of $\widetilde \beta_k$ for $k=1,\ldots,p$
and to carry out statistical inference for this parameter. For example, it may be of interest to test the hypothesis $H_0: \beta_k = \beta_{k0}$ versus $H_1: \beta_k \neq \beta_{k0}$ for some given $\beta_{k0} \in \bbR$. We propose to estimate the standard error of $\widetilde \beta_k$ and  perform this test via a nonparametric bootstrap method (Efron 1979).}
The procedure is summarized as follows:
\begin{itemize}
\item Generate $B$ bootstrap samples, $\{(X_{i,b}^*,Y_{i,b}^*),i=1,\ldots,n\}$, $b=1,\ldots,B$, with $B$ being a large number.
For each $b=1,\ldots, B$, $\{(X_{i,b}^*,Y_{i,b}^*),i=1,\ldots,n\}$ is a random sample from $\{(X_i,Y_i),i=1,\ldots,n\}$ with replacement.
\item For each $b=1,\ldots, B$, based on {Our-score}, we can estimate $\widetilde \beta_{k,b}^*$ from
the bootstrap sample $\{(X_{i,b}^*,Y_{i,b}^*),i=1,\ldots,n\}$.
The standard error of $\widetilde \beta_k$ can be estimated by the sample standard deviation of these $\{\widetilde \beta_{k,b}^*\}_{b=1}^B$. The level $1-\alpha$ (BPCI) for $\beta_k$ can be constructed as $[L_k^*, U_k^*]$, where $L_k^*$ and $U_k^*$ are respectively the $(\alpha/2)$th and $(1-\alpha/2)$th quantiles of $\{\widetilde \beta_{k,b}^*\}_{b=1}^B$; this can be used to perform the aforementioned test.
\end{itemize}

For the other methods given at the beginning of this section, we can similarly establish
the corresponding standard errors of the $\beta_k$ estimates and the level $1-\alpha$ BPCIs.  In this section, {we will also compare the performance of $\widetilde{\bm{\beta}}$ with that of the other methods based on their BPCIs.
{\color{black} 
Since we have not established the asymptotic normality of $\widehat{\bm \beta}$, 
we  do not include  the simulation results of the BPCI based on $\widehat {\bm\beta}$.}}

{\color{black} 
We have not established the theoretical justification for the BPCI
based on the proposed $\widetilde {\bm\beta}$. The main obstacle is  the nonparametric monotone component $\widehat F_{\bm\beta}(\cdot)$  in $\psi_n(\bm{\beta})$ (Shao and Tu, 1995; Kosorok, 2008). 
We use simulation studies to gain insight into the BPCI based on $\widetilde {\bm\beta}$, and
we observe that the coverage proportions of the BPCIs based on $\widetilde {\bm\beta}$ are reasonably close to the nominal level. We leave the theoretical justification for future research.}

\subsection{Data simulation}\label{section-4-1}

We simulate the responses $Y_i$ from the following model:
$$H(Y_i)=\beta_1 X_{i1}+\beta_2 X_{i2}+\epsilon_i.$$
The covariates $X_{i1}$ and $X_{i2}$ are simulated by $X_{i1}\sim \chi_1^2$ and $X_{i2}\sim N(X_{i1},1)$. We set $\beta_1 = \beta_2 = 1$ as the true value in the data simulation. To evaluate the performance of the methods for different true values of the nonparametric components, we experiment with different combinations of the transformation function $H(\cdot)$ and the distribution of the random errors $\epsilon_i$. In particular,
\begin{itemize}
\item
we consider (1) $H(y) = y$ and (2) $H(y) = \log(y)$;
\item for the distribution of the random errors $\epsilon_i$, we use (1) the extreme value distribution, i.e., $F_\epsilon(t) = 1-\exp\{-\exp(t)\}$; {(2) the $N(0, \pi^2/6)$ distribution; (3) the  logistic distribution with mean $0$ and scale $1/\sqrt{2}$}; {so that all these distributions have the same variance $\pi^2/6$. }
\end{itemize}
For each combination, we consider the sample sizes $n=100$ and $200$. We repeat each simulation 1000 times.

\subsection{Results} \label{section-4-2}

We apply the {seven} methods to the simulated data in Section \ref{section-4-1}, and in the results we normalize {\color{black} $\|{\bm{\beta}}\|_2 = 1$}  for comparison purposes. {\color{black} Tables \ref{extreme}--\ref{chisq} report the relative bias (RB), variance (Var), and MSE for $\beta_1$ and $\beta_2$, over 1000 repetitions; here RB is defined as $(\mbox{estimate} - \mbox{true value})/(\mbox{true value})$. All the values reported in these tables are $100\times\mbox{computed values}$. Table~\ref{table-BPCI} gives the proportions of the level 95\% BPCIs that cover the true values of $\beta_1$ and $\beta_2$, called the coverage proportion (CP), based on
the Our-score, Cox PH, Smooth-Normal, Smooth-Sigmoid, and SMRCE methods over 1000 repetitions. Here we use $B=200$ to compute the BPCI for each method. 
We do not include the BPCIs for Our-PRL because the asymptotic distribution of $\widehat{\bm{\beta}}$ is not available; we also omit those for PDR4 because they are too computationally intensive. Furthermore, since all the methods under comparison are established only on the ranks of the $Y_i$ in the observed sample, they are invariant to the specific function form of $H(\cdot)$ in the estimation of $\bm{\beta}$. We observe the same $\bm{\beta}$ estimates when we simulate the data by $H(y) = y$ or $H(y) = \log(y)$. Therefore, we have reported only the results for data simulated from $H(y) = y$. }

{\color{black} 
Table \ref{extreme} presents the results for the case where the $\epsilon_i$'s follow the extreme value distribution.
We observe in this table, as well as in Tables \ref{normal} and \ref{chisq}, that the kernel-based methods (Smooth-Normal, Smooth-Sigmoid, and SMRCE) usually lead to larger MSE values than those of {our methods (Our-PRL and Our-score)}; hereafter, we focus our discussion on the non-kernel methods ({Our-PRL,  Our-score}, PDR4, and Cox PH). In this setup, the assumptions for all the methods are satisfied. The Cox PH method has the best performance; this is not surprising, since the assumption that the random error follows the extreme value distribution is satisfied, whereas the other methods do not need this assumption. Our-score, Our-PRL, and PDR4 have comparable MSE values for both $\beta_1$ and $\beta_2$ in this example. 


\begin{table}[!http]
\caption{\label{extreme}  Comparison of $\beta_1$ and $\beta_2$ estimation; the error follows the extreme value distribution with c.d.f. $F(t)=1-\exp\{-\exp(t)\}$}
\centering
\fbox{\tabcolsep=2mm
\begin{tabular}{ll ccc ccc }
\noalign{\smallskip}\noalign{\smallskip}
&&\multicolumn{3}{c}{$\beta_1$} & \multicolumn{3}{c}{$\beta_2$}\\
\cline{3-5}\cline{6-8}
$n$ &Method &RB& Var & MSE  &RB& Var & MSE \\
 100 & Our-PRL                 &-1.47&1.64&1.65&-1.61&1.42&1.43\\
       &Our-score           &-1.98&1.48&1.50&-0.80&1.28&1.28\\
     & PDR4                  &-2.62&1.52&1.55&-0.23&1.30&1.30\\
     & Cox PH               &-1.85&0.87&0.89&0.17&0.79&0.79\\
     & Smooth-Normal   &-4.11&1.95&2.03&0.47&1.61&1.61\\
     & Smooth-Sigmoid &-5.52&1.70&1.85&2.33&1.31&1.34\\
     & SMRCE               &-11.37&2.75&3.40&6.07&1.72&1.90\\
 200 & Our-PRL           &0.07&0.76&0.76&-1.58&0.74&0.76\\
      &Our-score            &-0.53&0.73&0.73&-0.91&0.70&0.70\\
     & PDR4                  &-0.84&0.73&0.73&-0.58&0.69&0.69\\
     & Cox PH               &-0.03&0.42&0.42&-0.80&0.41&0.41\\
     & Smooth-Normal   &-1.83&0.96&0.97&0.00&0.86&0.86\\
     & Smooth-Sigmoid &-2.60&0.86&0.89&0.95&0.76&0.76\\
     & SMRCE            &-6.22&1.19&1.39&3.86&0.90&0.98\\
\end{tabular}}
\end{table}

%

Tables \ref{normal} and \ref{chisq} present the results for the cases where the $\epsilon_i$'s follow the normal and logistic distributions respectively. In both cases, especially the latter, the model assumption for the Cox PH method is violated, and this method has the worst performance among the non-kernel methods. In both tables, Our-score has the smallest MSE values. The MSEs of Our-PRL and PDR4 are comparable, but PDR4 is much more computationally intensive.

\begin{table}[!http]
\caption{\label{normal}  Comparison of $\beta_1$ and $\beta_2$ estimation; the error follows the normal distribution}
\centering
\fbox{\tabcolsep=2mm
\begin{tabular}{ll ccc ccc }
\noalign{\smallskip}\noalign{\smallskip}
&&\multicolumn{3}{c}{$\beta_1$} & \multicolumn{3}{c}{$\beta_2$}\\
\cline{3-5}\cline{6-8}
$n$ &Method &RB& Var & MSE  &RB& Var & MSE \\
 100 & Our-PRL         &-2.38&1.66&1.68&-0.76&1.46&1.46\\
       &Our-score         &-2.09&1.45&1.47&-0.68&1.30&1.30\\
     & PDR4                 &-2.77&1.62&1.66&-0.31&1.42&1.42\\
     & Cox PH              &-4.99&1.69&1.81&1.73&1.43&1.45\\
     & Smooth-Normal &-4.64&2.51&2.61&-0.01&2.04&2.04\\
     & Smooth-Sigmoid&-6.01&2.20&2.37&1.93&1.69&1.71\\
     & SMRCE             &-12.73&3.50&4.30&6.08&2.16&2.34\\
 200 & Our-PRL         &-0.91&0.72&0.72&-0.52&0.71&0.71\\
      &Our-score          &-0.81&0.69&0.69&-0.55&0.67&0.67\\
     & PDR4                 &-1.11&0.74&0.75&-0.36&0.73&0.73\\
     & Cox PH              &-3.62&0.83&0.90&1.99&0.72&0.74\\
     & Smooth-Normal &-1.87&1.07&1.09&-0.24&1.02&1.02\\
     & Smooth-Sigmoid&-2.85&1.00&1.04&0.91&0.91&0.91\\
     & SMRCE             &-7.83&1.50&1.80&4.82&1.09&1.21\\
\end{tabular}}
\end{table}

Table \ref{table-BPCI} shows that the CPs from Our-score  are reasonably close to the nominal level (i.e., 0.95), indicating that the suggested bootstrap method controls the type I error well if we apply it to the testing problems $H_0: \beta_j = \beta_{j,0}$ versus $\beta_j \neq \beta_{j,0}$; $j=1$ or $2$.

\begin{table}[!http]
\caption{\label{chisq}  Comparison of $\beta_1$ and  $\beta_2$ estimation; the error follows the logistic distribution}
\centering
\fbox{\tabcolsep=2mm
\begin{tabular}{ll ccc ccc }
\noalign{\smallskip}\noalign{\smallskip}
&&\multicolumn{3}{c}{$\beta_1$} & \multicolumn{3}{c}{$\beta_2$}\\
\cline{3-5}\cline{6-8}
$n$ &Method &RB& Var & MSE  &RB& Var & MSE \\
  100  & Our-PRL           &-1.33&1.58&1.58&-1.72&1.45&1.47\\
    & Our-score         &-1.27&1.40&1.40&-1.43&1.28&1.29\\
     & PDR4                 &-1.97&1.53&1.54&-0.94&1.36&1.37\\
     & Cox PH              &-5.38&2.06&2.21&1.53&1.63&1.64\\
     & Smooth-Normal &-3.81&2.47&2.54&-0.67&1.95&1.95\\
     & Smooth-Sigmoid&-5.22&2.10&2.24&1.40&1.58&1.59\\
     & SMRCE             &-11.44&3.24&3.89&5.36&2.05&2.19\\
 200 & Our-PRL         &-0.55&0.72&0.73&-0.88&0.70&0.71\\
      &Our-score          &-0.42&0.67&0.67&-0.91&0.66&0.66\\
     & PDR4                 &-0.71&0.69&0.70&-0.67&0.68&0.69\\
     & Cox PH              &-5.73&1.13&1.30&3.46&0.91&0.97\\
     & Smooth-Normal &-1.59&1.01&1.02&-0.42&0.99&0.98\\
     & Smooth-Sigmoid&-2.59&0.91&0.95&0.79&0.86&0.86\\
     & SMRCE             &-6.59&1.30&1.52&3.97&1.02&1.10\\
\end{tabular}}
\end{table}

We now examine the numerical performance of our $\widetilde F(\cdot)$ in the estimation of $F(\cdot)$.  We evaluate the mean and the sample standard deviation (SD) of the integrated square errors (ISEs) of  $\widetilde  F(\cdot)$ over 1000 replicates, for all the simulated data in Section \ref{section-4-1}. The results are given in Table \ref{MISE_F}.  We observe that as the sample size $n$ increases, both the mean and SD of the ISEs decrease; this agrees with the asymptotic results we derived in Section \ref{section-3}. To further illustrate the performance of $\widetilde F(\cdot)$ in the estimation of $F(\cdot)$, we construct the mean and the percentile bands. In Figure \ref{fig-F}, we summarize the estimates of $\widetilde F(\cdot)$ from our method for the case where $n=200$ and $F(t)=1-\exp\{-\exp(t)\}$, i.e., the error $\epsilon_i$ follows the extreme value distribution.  We have included the true curve, the 2.5\% percentile, the 97.5\% percentile, and the mean $\widetilde F(\cdot)$ curves over 1000 repetitions. We observe that the mean curve of $\widetilde F(\cdot)$ almost overlaps with the true curve; the  95\% pointwise confidence bands  contain the true curve; and the widths of the bands are quite small. This reinforces the asymptotic results derived for $\widetilde F(\cdot)$ in Section \ref{section-3}.

\begin{table}[!http]
\caption{\label{table-BPCI}   CP (\%)  of BPCI for $\beta_1$ and $\beta_2$}
\centering
\fbox{\tabcolsep=2mm
\begin{tabular}{l l cc cc cc }

&&\multicolumn{2}{c}{Extreme Value} &\multicolumn{2}{c}{Normal} &\multicolumn{2}{c}{Logistic} \\
$n$&Method&$\beta_1$&$\beta_2$ &$\beta_1$&$\beta_2$ &$\beta_1$&$\beta_2$ \\
100&Our-score               & 95.8&95.8&95.3&95.3&94.3&94.3\\
&Cox PH                  & 93.8&93.8&93.2&93.2&91.8&91.8\\
&Smooth-Normal     & 96.1&96.1&96.3&96.3&96.5&96.5\\
&Smooth-Sigmoid   &  95.7&95.7&95.4&95.4&95.2&95.2 \\
&SMRCE                &  97.2&98.1&97.8&99.1&96.8&97.7\\
200&Our-score            &94.2&94.2&93.7&93.7&94.1&94.1\\
&Cox PH               &93.8&93.8&93.7&93.7&92.2&92.2\\
&Smooth-Normal  &97.5&97.5&96.2&96.2&96.8&96.8\\
&Smooth-Sigmoid &96.2&96.2&95.4&95.4&95.9&95.9\\
&SMRCE               &98.1&98.1&97.4&97.4&98.0&98.0\\
\end{tabular}}
\end{table}

\begin{table}[!http]
\caption{\label{MISE_F}  Mean ($\times 1000$ and SD ($\times 1000$) of ISE for $\widetilde F(\cdot)$}
\centering
\fbox{
\tabcolsep=2mm
\begin{tabular}{l  cc cc}
&\multicolumn{2}{c}{$n=100$} & \multicolumn{2}{c}{$n=200$}\\
Distribution & MISE & SD & MISE & SD  \\
\hline
Extreme Value&1.20&0.97&0.52&0.41\\
Normal &1.12&0.94&0.50&0.42\\
Logistic &1.15&0.91&0.53&0.44\\
\end{tabular}
}

\end{table}

\begin{figure}[!http]
\centering{\includegraphics[scale=0.4]{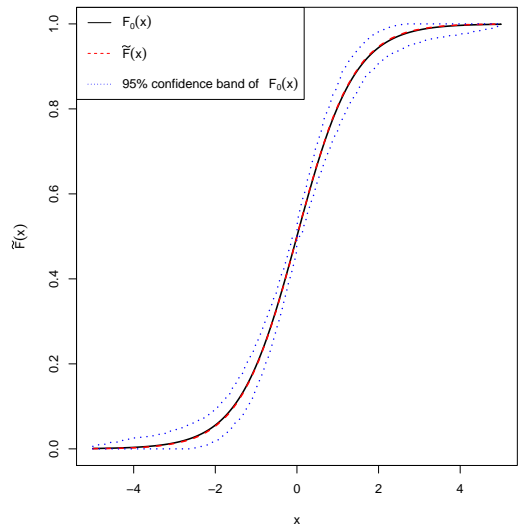}}
  \caption{Summarized $F(\cdot)$ estimation based on 1000 simulations; $n=200$ and  $F(t)=1-\exp\{-\exp(t)\}$.}\label{fig-F}
\end{figure}

In summary, our methods perform well for all the simulation setups. Among the PDR4, Our-PRL, and Our-score methods, Our-score leads to the smallest MSE values in most of the simulation examples.  Our-PRL and PDR4 have similar performance in the $\bm{\beta}$ estimation, but PDR4 is much more computationally intensive.} The Cox PH method performs the best when its model assumptions are satisfied. It fails when these assumptions are (partially) invalid, so these assumptions are clearly restrictive in practice. We therefore recommend Our-score unless there is strong prior scientific evidence that the assumptions of the Cox PH model are satisfied.

\section{Application to an Alzheimer's Disease Study} \label{section-5}

In this section, we analyze  the National Alzheimer's
Coordinating Center (NACC) uniform data set (UDS) (Beekly et al.~2007). Approximately five million people in the United States and more than thirty-seven million people worldwide suffer from Alzheimer's disease. This disease gradually destroys the patient's memory and ability to learn and to carry out daily activities such as talking and eating. As the disease progresses, there may also be changes in personality and behavior.
Unfortunately, to date there is no cure and no effective way to predict how quickly an individual will progress through the stages of the disease. However, early diagnosis and appropriate treatment can slow its progression.

A popular test for Alzheimer's disease is the mini-mental state examination (MMSE) test (Folstein et~al.~1975). This is a 30-point test that gives a numerical measurement of the cognitive impairment. We use the MMSE score as our response ($Y$) and study the association between this score and the covariates of interest, including the age (in years; $X_1$) and education level (in years; $X_2$) of the corresponding individual. In particular, we consider the linear transformation model, given by
\begin{eqnarray*}
H(Y) = \beta_1X_1 + \beta_2X_2 + \epsilon.
\end{eqnarray*}
A sample of 100 subjects enrolled in 2005 was used to study the relationship between the covariates and the responses.

{\small
\begin{table}[!http]
\caption{\label{table-NACCUDS} Estimates and 95\% BPCI of $\bm{\beta}$ estimates for the NACC UDS data; the Length columns report the length of the corresponding BPCIs}
\centering
\fbox{\tabcolsep=2mm
\begin{tabular}{l ccc| ccc }
&\multicolumn{3}{c|}{$\beta_1$} & \multicolumn{3}{c}{$\beta_2$}\\
Method & Estimate & 95\% BPCI&Length& Estimate & 95\% BPCI&Length\\ \hline
Our-Score&0.931&[0.796, 0.977]&0.181&-0.365&[-0.605, -0.212]&0.393\\
PDR4&0.934&[0.736, 0.979]&0.243&-0.359&[-0.677, -0.205]&0.472\\
Smooth-Normal&0.922&[0.669, 0.984]&0.315&-0.388&[-0.743, -0.180]&0.563\\
Smooth-Sigmoid&0.924&[0.672, 0.984]&0.312&-0.383&[-0.740, -0.179]&0.561\\
SMRCE&0.905&[0.627, 0.981]&0.354&-0.426&[-0.779, -0.194]&0.585\\
Cox PH&0.957&[0.812, 0.988]&0.176&-0.289&[-0.584, -0.157]&0.427\\
LR&0.934&[0.648, 0.983]&0.335&-0.358&[-0.761, -0.185]&0.576\\
\end{tabular}}

\end{table}
}

\begin{figure}[!ht]
  \centering{\includegraphics[scale=0.39]{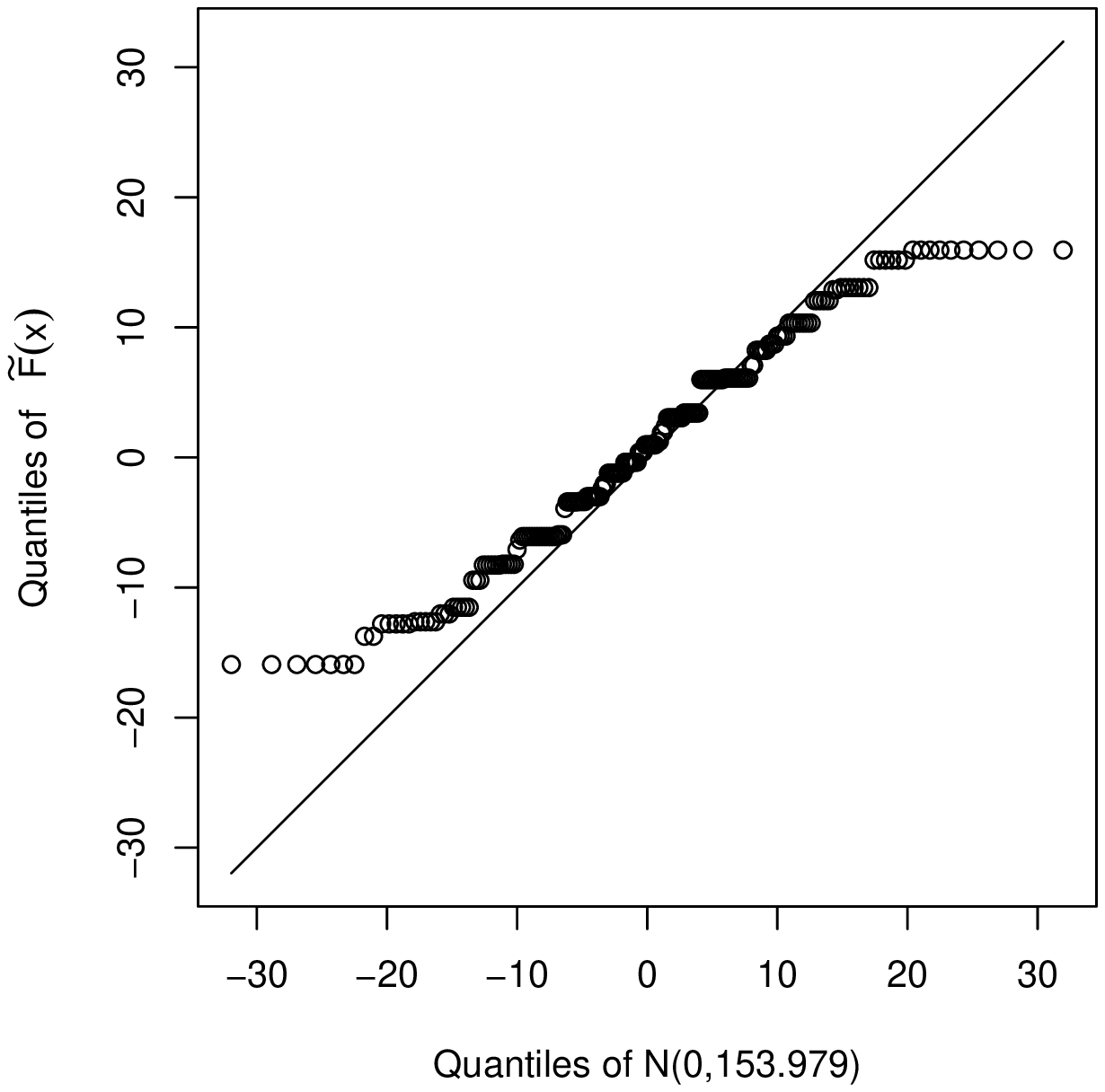}}
  \centering{\includegraphics[scale=0.39]{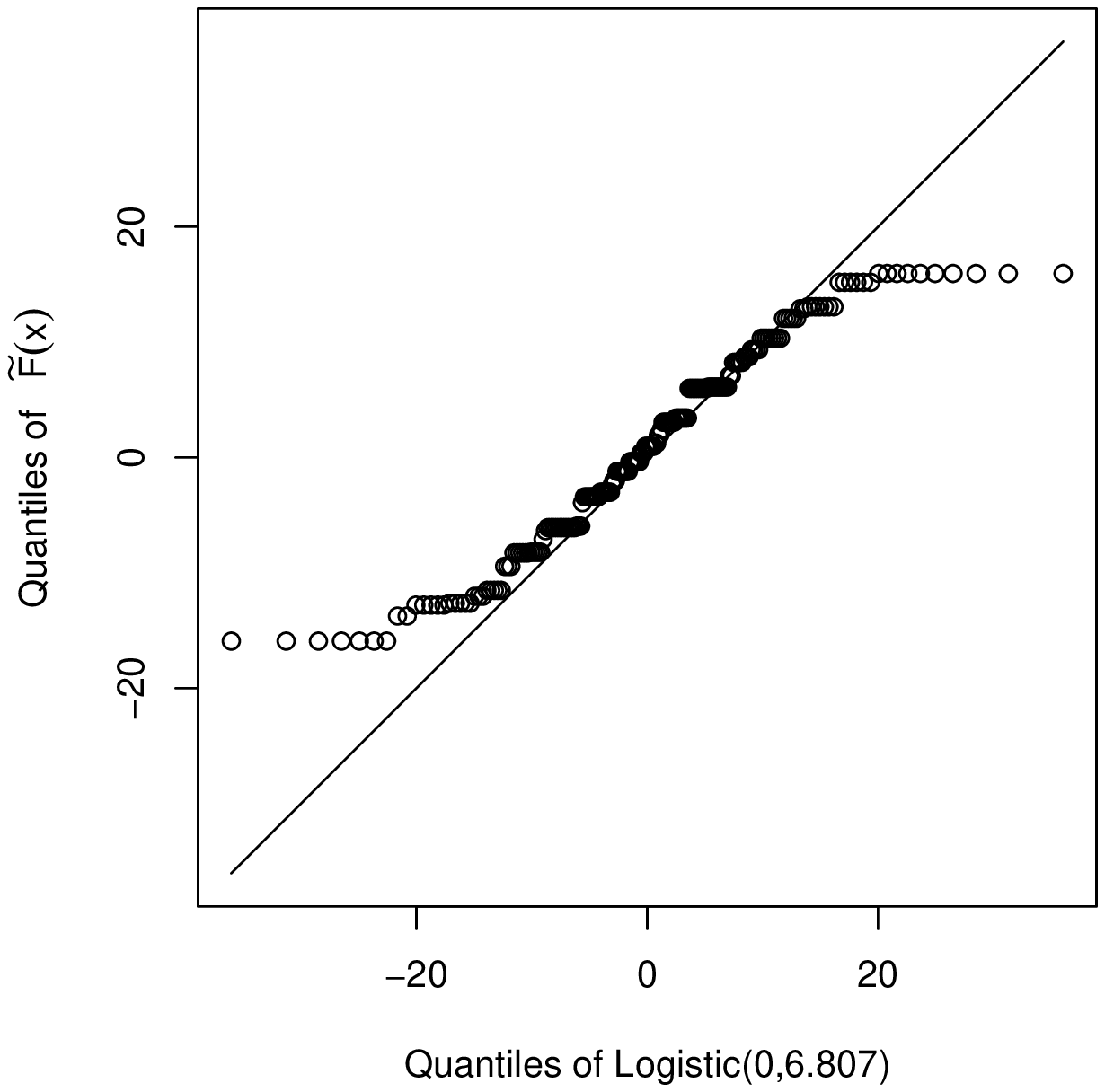}}
  \caption{Quantile-quantile plots of the estimated $\widetilde F(\cdot)$ from Our-Score versus those from the LR method (left panel) and the Cox PH method (right panel) for the NACC UDS data.}\label{PP-plot}
\end{figure}

{\color{black} 
As we have observed in the simulation studies, Our-score has better performance than Our-PRL; in this real-data example, we do not include the results from Our-PRL. We apply the following methods:
(1) Our-score; (2) PDR4;  (3) Smooth-Normal; (4) Smooth-Sigmoid; (5) SMRCE; (6) Cox PH; and (7) classical linear regression (LR). We summarize the $\beta$ estimates in
Table \ref{table-NACCUDS}, where we regularize $\|\bm{\beta}\|_2 = 1$ for comparison purposes;  the 95\% BPCIs are based on the method given in Section \ref{section-4} with {\bf $B=200$}. We observe that the $\bm{\beta}$ estimates from PDR4, Smooth-Normal, Smooth-Sigmoid, and LR are close to those from our methods; whereas those from Cox PH and SMRCE are slightly different. The lengths of the BPCIs from Our-score are comparable to those from Cox PH and much shorter than those from the other methods. 

The estimates of $F(\cdot)$ from our methods can be used to check the model assumptions of LR and Cox PH. Specifically,  note that $\widetilde F(\cdot)$ is the estimate for $F(\cdot)$, which is the c.d.f. of $\epsilon_i - \epsilon_j$. This c.d.f. can also be estimated under the LR model with normal errors or the Cox PH model; the estimated $F(\cdot)$'s are respectively $N(0, 153.979)$ and $\mbox{Logistic}(0, 6.807)$, the logistic distribution with mean 0 and scale 6.807. Figure~\ref{PP-plot} shows quantile-quantile plots of the $\widetilde F(\cdot)$ from Our-Score versus those of $N(0, 153.979)$ (left panel) and  $\mbox{Logistic}(0, 6.807)$ (right panel). This figure indicates that the distribution of the errors may deviate from the normal and extreme value distributions.
}

Combining the observations above with those of Section \ref{section-4-2}, we expect that in this example, the bias from Cox PH could be larger than that from the other methods; our methods might have produced {\color{black} good} estimates of the unknown parameters. However, we are unable to compare the estimates with their true values.


\section{Discussion} \label{section-6}

The linear transformation model has been widely studied in statistics and econometrics. It includes as special cases many important and popular models that are commonly applied in practice. The development of appropriate methodology, fast algorithms, and solid statistical theory is an important but challenging task. In the literature, there are two general strategies: kernel-smoothing techniques and rank-based methods. {\color{black} However, the former approach needs a tuning parameter,} and {\color{black} the latter may not make full use of the data information} or may be computationally expensive.

We have developed two methods for analyzing the linear transformation model: (1) a pairwise rank likelihood method; and (2) a score-function-based method. Our methods  do not need a tuning parameter and {make effective use of} the information carried in the original data. For the pairwise rank likelihood estimator, we have established a theoretical upper bound on the asymptotic convergence rate; for the score-function-based estimator, we have established asymptotic normality. Furthermore, via extensive numerical studies, we have demonstrated that our methods are more appealing than existing methods because they are not only robust to the distribution of the random errors but also in many cases {\color{black} lead to comparable or smaller MSEs} in the estimation of the model parameters.

We expect that the methodology, theoretical results, and technical tools of this paper will benefit the study of similar models, e.g., the popular Box--Cox model (Box and Cox 1964) has a similar structure. {\color{black} We have assumed that the effect of the covariates on the transformed response is linear; future work will include relaxing and performing hypothesis testing on this assumption by considering more complicated models. The pairwise rank likelihood method could therefore be extended to these models. Some of the technical tools and theoretical results in this paper could facilitate these studies. }Furthermore, the development of the asymptotic properties of the estimators for the pairwise rank likelihood (\ref{pseudo-lik}) is challenging; see the more detailed discussion in Remark \ref{remark-1}. Thus far, we have proved only $O_p(n^{-1/3})$ convergence. We conjecture that this rate may not be sharp for the estimator $\widehat{\bm{\beta}}$; there may be room for improvement. In addition, we have developed the theory only for the random design, i.e., where the $X_i$'s are random variables. We conjecture that for the fixed design, these theoretical results are still valid under appropriate regularity conditions. We leave these important and interesting topics for future research.

\section*{Acknowlegement}
Dr. Yu's research is supported in part by Singapore Ministry Education Academic Research Fund Tier 1: R-155-000-157-112, R-155-000-202-114, and Ministry of Education of Singapore: MOE2014-T2-1-072.
Dr. Li's research is supported in part by NSERC Grant RGPIN-2020-04964.
Dr. Chen was supported in part by National Institute on Aging grant U01AG016976.
The NACC database is funded by NIA/NIH Grant U01 AG016976. The NACC data are contributed by the NIA-funded ADCs: P30 AG019610 (PI Eric Reiman, MD), P30 AG013846 (PI Neil Kowall, MD), P50 AG008702 (PI Scott Small, MD), P50 AG025688 (PI Allan Levey, MD, PhD), P30 AG010133 (PI Andrew Saykin, PsyD), P50 AG005146 (PI Marilyn Albert, PhD), P50 AG005134 (PI Bradley Hyman, MD, PhD), P50 AG016574 (PI Ronald Petersen, MD, PhD), P50 AG005138 (PI Mary Sano, PhD), P30 AG008051 (PI Steven Ferris, PhD), P30 AG013854 (PI M. Marsel Mesulam, MD), P30 AG008017 (PI Jeffrey Kaye, MD), P30 AG010161 (PI David Bennett, MD), P30 AG010129 (PI Charles DeCarli, MD), P50 AG016573 (PI Frank LaFerla, PhD), P50 AG016570 (PI David Teplow, PhD), P50 AG005131 (PI Douglas Galasko, MD), P50 AG023501 (PI Bruce Miller, MD), P30 AG035982 (PI Russell Swerdlow, MD), P30 AG028383 (PI Linda Van Eldik, PhD), P30 AG010124 (PI John Trojanowski, MD, PhD), P50 AG005133 (PI Oscar Lopez, MD), P50 AG005142 (PI Helena Chui, MD), P30 AG012300 (PI Roger Rosenberg, MD), P50 AG005136 (PI Thomas Montine, MD, PhD), P50 AG033514 (PI Sanjay Asthana, MD, FRCP), and P50 AG005681 (PI John Morris, MD).

\section*{Appendix: Technical Conditions}

We impose the following regularity conditions to establish the theoretical results in Section \ref{section-asym-prle}. They are not necessarily the weakest possible.

\begin{itemize}
\item[] \underline{\bf Condition 0}: $H(y)$ is a strictly increasing function on the support $\mathY$ of $Y$, and $\epsilon_i$ and $X_i$ are independent.
\item[] \underline{\bf Condition 1}: $\bm{\beta}\in \mathB$ and $F_X(\bm{x})$ is supported on $\mathX$, where both $\mathB$ and $\mathX$ are compact subspaces of $\bbR^{p}$.
\item[] \underline{\bf Condition 2}: For any $x_1, x_2 \in \bbR$, we have
$$\sup_{\bm{\beta}\in \mathB}\left| F_{\bm{\beta}^TX}(x_1) - F_{\bm{\beta}^TX}(x_2) \right|\lesssim |x_1-x_2|.$$
\item[] \underline{\bf Condition 3}: $\inf_{\bm{x_1},\bm{x}_2 \in \mathX, \bm{\beta}\in \mathB} F_0((\bm{x}_1-\bm{x}_2)^T\bm{\beta})>0$.
\item[] \underline{\bf Condition 4}: If $F(x)$ is a c.d.f. such that $F((\bm{x}_1 - \bm{x_2})^T\bm{\beta}) = F_0((\bm{x}_1 - \bm{x_2})^T\bm{\beta}_0)$ for all $\bm{x}_1, \bm{x}_2 \in \mathX$, then $\bm{\beta} = \bm{\beta}_0$ and $F(\cdot) = F_0(\cdot)$.
\item[] \underline{\bf Condition 5}: $F_0(u)$ is second-order differentiable with bounded second-order derivatives and $\inf_{\bm{v}\in \mathX-\mathX, \bm{\beta}\in \mathB} F_0'(\bm{v}^T\bm{\beta})>0$.
\item[] {\color{black} \underline{\bf Condition 6}: There exists an $\eta_0>0$ such that
\begin{eqnarray*}
\inf_{\bm{\beta}\in \mathO_{\eta_0}}\lambda_2 \Big(\left[\mbox{var}\big\{ (X_1 -X_2)-E((X_1-X_2)|(X_1 - X_2)^T\bm{\beta}) \big\}\right] \Big)>0,
\end{eqnarray*}
where $\mathO_{\eta_0} = \{\bm{\beta}: \|\bm{\beta}-\bm{\beta}_0\|_2\le\eta_0, \|\bm{\beta}\|_2 = 1\}$;  $\lambda_2(B)$ denotes the second smallest eigenvalue of matrix $B$.}

\end{itemize}

Furthermore, we need the following additional conditions to establish the theoretical results for $\widetilde{\bm{\beta}}$ in Section \ref{section-asym-score}.

\begin{itemize}

\item[] \underline{\bf Condition A1}: $F_{\bm{\beta}}(t)$ is continuously differentiable in $\bm{\beta}$ and $t$, and
    \begin{eqnarray*}
    \inf_{t\in \left\{\bm{v}^T\bm{\beta}: \bm{v}\in \mathX-\mathX, \bm{\beta}\in \mathB \right\}, \bm{\beta}\in \mathB} \frac{\partial F_{\bm{\beta}}(t)}{\partial t} > 0.
    \end{eqnarray*}

\item[] \underline{\bf Condition A2}: $f_{X_1-X_2|(X_1-X_2)^T\bm{\beta} = t} (\bm{v})$ is continuously differentiable in $\bm{\beta}\in \mathB$ and $t\in\{\bm{x}^T\bm{\beta}: \bm{x}\in \mathX, \bm{\beta}\in \mathB\}$.

\item[] \underline{\bf Condition A3}: We have $\bm{\beta} = \bm{\beta}_0$ if \begin{eqnarray*}
    \mbox{cov}\left\{ \bm{\beta}_0^T(X_1-X_2), F_0(\bm{\beta}_0^T(X_1-X_2)) \Big| \bm{\beta}^T(X_1-X_2)\right\} = 0
    \end{eqnarray*}
    almost surely.

\end{itemize}

{\color{black} 
\begin{remark}
Some of the conditions above are not intuitive and not easily checked in practice.  In the supplementary document, we provide more discussion and give stronger but more intuitive conditions.
\end{remark}
}

{\color{black} 
\begin{remark} \label{remark-identifiablity}
Condition 4 is required to ensure the identifiability of the model; this is because based on this condition and 
\eqref{eq-pseudo-l-added}, we can conclude that $\bm{\beta}$ and $F(\cdot)$ are uniquely determined by the transformation model, and so is 
$H(\cdot)$. Based on our derivation in Section 3.2 of the supplementary document, this condition can be replaced with a more intuitive condition: $\|\bm{\beta}\|_2 = 1$, $F_0(\cdot)$ is strictly increasing, and $\mathX$ contains at least one interior point in $\bbR^p$. 
Furthermore, we observe that
 Condition A3 replaces Condition 4 when we ensure the identifiability of $\bm{\beta}$ in the proof of Theorem \ref{thm-normality}. 
\end{remark}

}



\newpage

{\centering {\Large {\bf Supplementary Material for \\ ``Maximum pairwise-rank-likelihood-based inference for the semiparametric transformation model"}}}
\bigskip

\noindent
This supplementary document   contains
technical details for the theoretical results in Section 4 of the main article (Sections \ref{section-review}--\ref{sec-proof-of-normality-F-beta}), some details for the numerical algorithms of our $\widehat{\bm{\beta}}$ and $\widetilde{\bm{\beta}}$ estimates (Section \ref{details-numerical-imp}),
an extension of our methods to data with ties in the responses (Section \ref{section-ties}), and an extension of our methods to right-censored data (Section \ref{section-right-censor}).

\setcounter{equation}{0}
\setcounter{section}{0}
\renewcommand{\theequation} {S.\arabic{equation}}
\setcounter{theorem}{0}
\setcounter{lemma}{0}
\setcounter{corollary}{0}
\setcounter{definition}{0}
\setcounter{remark}{0}

\section{Review of Technical Conditions and Theorems in the Main Article} \label{section-review}

In this section, we review the technical conditions in the Appendix and theorems in Section 4 of the main article.

\subsection{Technical conditions}

We have imposed the following technical conditions in the Appendix of the main article.

\begin{itemize}
\item[] \underline{\bf Condition 0}: $H(y)$ is a strictly increasing function on the support $\mathY$ of $Y$, and $\epsilon_i$ and $X_i$ are independent.
\item[] \underline{\bf Condition 1}: $\bm{\beta}\in \mathB$ and $F_X(\bm{x})$ is supported on $\mathX$, where both $\mathB$ and $\mathX$ are compact subspaces of $\bbR^{p}$.
\item[] \underline{\bf Condition 2}: For any $x_1, x_2 \in \bbR$, we have
$$\sup_{\bm{\beta}\in \mathB}\left| F_{\bm{\beta}^TX}(x_1) - F_{\bm{\beta}^TX}(x_2) \right|\lesssim |x_1-x_2|.$$
\item[] \underline{\bf Condition 3}: $\inf_{\bm{x_1},\bm{x}_2 \in \mathX, \bm{\beta}\in \mathB} F_0((\bm{x}_1-\bm{x}_2)^T\bm{\beta})>0$.
\item[] \underline{\bf Condition 4}: If $F(x)$ is a c.d.f. such that $F((\bm{x}_1 - \bm{x_2})^T\bm{\beta}) = F_0((\bm{x}_1 - \bm{x_2})^T\bm{\beta}_0)$ for all $\bm{x}_1, \bm{x}_2 \in \mathX$, then $\bm{\beta} = \bm{\beta}_0$ and $F(\cdot) = F_0(\cdot)$.
\item[] \underline{\bf Condition 5}: $F_0(u)$ is second-order differentiable with bounded second-order derivatives and $\inf_{\bm{v}\in \mathX-\mathX, \bm{\beta}\in \mathB} F_0'(\bm{v}^T\bm{\beta})>0$.
\item[] {\color{black} \underline{\bf Condition 6}: There exists an $\eta_0>0$ such that
\begin{eqnarray*}
\inf_{\bm{\beta}\in \mathO_{\eta_0}}\lambda_2 \Big(\left[\mbox{var}\big\{ (X_1-X_2)-E((X_1 - X_2)|(X_1 - X_2)^T\bm{\beta}) \big\}\right] \Big)>0,
\end{eqnarray*}
where $\mathO_{\eta_0} = \{\bm{\beta}: \|\bm{\beta}-\bm{\beta}_0\|_2\le\eta_0, \|\bm{\beta}\|_2 = 1\}$;  $\lambda_2({\color{black} B})$ denotes the second smallest eigenvalue of matrix ${B}$.}

\end{itemize}

Furthermore, we need the following additional conditions to establish the asymptotic distribution for $\widetilde{\bm{\beta}}$.

\begin{itemize}

\item[] \underline{\bf Condition A1}: $F_{\bm{\beta}}(t)$ is continuously differentiable in $\bm{\beta}$ and $t$, and
    \begin{eqnarray*}
    \inf_{t\in \left\{\bm{v}^T\bm{\beta}: \bm{v}\in \mathX-\mathX, \bm{\beta}\in \mathB \right\}, \bm{\beta}\in \mathB} \frac{\partial F_{\bm{\beta}}(t)}{\partial t} > 0.
    \end{eqnarray*}

\item[] \underline{\bf Condition A2}: $f_{X_1-X_2|(X_1-X_2)^T\bm{\beta} = t} (\bm{v})$ is continuously differentiable in $\bm{\beta}\in \mathB$ and $t\in\{\bm{x}^T\bm{\beta}: \bm{x}\in \mathX, \bm{\beta}\in \mathB\}$.

\item[] \underline{\bf Condition A3}: We have $\bm{\beta} = \bm{\beta}_0$ if \begin{eqnarray*}
    \mbox{cov}\left\{ \bm{\beta}_0^T(X_1-X_2), F_0(\bm{\beta}_0^T(X_1-X_2)) \Big| \bm{\beta}^T(X_1-X_2)\right\} = 0
    \end{eqnarray*}
    almost surely.

\end{itemize}

\subsection{Theorem in Section 4.1 of the main article} \label{section-4.1-main}

In Section 4.1 of the main article, we have presented the following theorem, which establish the asymptotic properties of $\widehat{\bm{\beta}}$ and $\widehat F(\cdot)$.
 Let
\begin{eqnarray}
 \label{def-D} &&D\Big(\bm{\beta}_2, F_2; \bm{\beta}_1, F_1\Big)\\
  &=& \left[\int \int \left\{F_1((\bm{x}_1-\bm{x}_2)^T \bm{\beta}_1) - F_2((\bm{x}_1-\bm{x}_2)^T \bm{\beta}_2) \right\}^2 d F_{X}(\bm{x}_1)d F_{X}(\bm{x}_2)\right]^{1/2}. \nonumber
\end{eqnarray}

The following is Theorem 1 in the main article.

\begin{theorem}
\label{theorem-1}
Assume Conditions 0--6. We have
\begin{itemize}
\item[(a)]
$
D\Big( \widehat{\bm{\beta}}, \widehat F; \bm{\beta}_0, F_0\Big) = O_p(n^{-1/3}),
$
\item[(b)] $\widehat{\bm{\beta}}-\bm{\beta}_0 = O_p(n^{-1/3})$,
\end{itemize}
where $\bm{\beta}_0$ and $F_0$ are the true values of $\bm{\beta}$ and $F$.
\end{theorem}

\subsection{Theorems in Section 4.2 of the main article}

In Section 4.2 of the main article, we have presented the following theorems, which establish the asymptotic distributions for $\widetilde {\bm{\beta}}$, {\color{black} $\widehat F_{\bm{\beta}}$, and $\widehat F_{\widetilde{\bm{\beta}}}$}.

Recall that we have the following notation. For any $\bm{\beta}\in \mathB$, define
\begin{eqnarray}
F_{\bm{\beta}}(t) &=& E\left\{I(Y_1>Y_2)\Big|(X_1-X_2)^T\bm{\beta} = t\right\} \nonumber \\
&=& \int F_0(t + \bm{v}^T (\bm{\beta}_0 - \bm{\beta})) f_{(X_1-X_2)|(X_1-X_2)^T\bm{\beta} = t} (\bm{v}) d\bm{v} \label{def-F-beta}
\end{eqnarray}
and
\begin{eqnarray}
\varphi_{\bm{\beta}}(t) = E(X_1 - X_2|(X_1-X_2)^T \bm{\beta} = t), \quad \varphi_0(t) =  \varphi_{\bm{\beta}_0}(t). \label{def-varphi}
\end{eqnarray}
$\widetilde{\bm{\beta}}$ is defined to be the zero-crossing of $\psi_n(\bm{\beta})$, where
\begin{eqnarray*}
&&\psi_n(\bm{\beta}) =\frac{1}{n^2} \sum_{i\neq j}(X_i-X_j)\left\{I(Y_i>Y_j)- \widehat F_{\bm{\beta}}\left((X_i-X_j)^T \bm{\beta} \right)\right\}\\
&=& \int \int (\bm{x}_1 - \bm{x}_2)\left\{I(y_1>y_2) - \widehat F_{\bm{\beta}}((\bm{x}_1 - \bm{x}_2)^T \bm{\beta})\right\} d \bbF_{X,Y}(\bm{x}_1, y_1) d \bbF_{X,Y}(\bm{x}_2, y_2),
\end{eqnarray*}
and the zero-crossing of a function (or a mapping) is defined below.
\begin{definition}
For a function $\psi$: $\mathX \to \bbR$, $x$ is called the zero-crossing of $\psi$ if every open neighborhood of $x$ contains $x_1, x_2$ such that $\psi(x_1)\psi(x_2) \leq 0$.
For a mapping $\widetilde \psi$: $\mathX \to \bbR^d$, $x$ is called the zero-crossing of $\widetilde \psi$ if $x$ is the zero-crossing of each component of $\widetilde \psi$.
\end{definition}
The population version of $\psi_n(\cdot)$ is given by
\begin{eqnarray}
 \label{def-psi-0} \psi_0(\bm{\beta}) &=& \int \int (\bm{x}_1 - \bm{x}_2)\left\{I(y_1>y_2) -  F_{\bm{\beta}}((\bm{x}_1 - \bm{x}_2)^T \bm{\beta})\right\} \\ &&
 \hspace{1in}\times d F_{X,Y}(\bm{x}_1, y_1) d F_{X,Y}(\bm{x}_2, y_2). \nonumber
\end{eqnarray}
We have the following lemma for $\psi_0(\cdot)$; it is Lemma 1 in the main article.  The proof is given in Section \ref{proof-lemma-psi-0}.

\begin{lemma}\label{lemma-psi-0}
Assume Conditions 6, A1, and A2. We have the following:

\begin{itemize}
\item[(1)] $\psi_0(\bm{\beta}_0) = 0$;

{\color{black} 
\item[(2)] $\psi_0'(\bm{\beta}_0) = \frac{\partial \psi_0(\bm{\beta})}{\partial \bm{\beta}}\Big|_{\bm{\beta} = \bm{\beta}_0}$ exists with rank $p-1$.

\item[(3)] Since $\|\bm{\beta}_0\|_2 = 1$, there exists an $i$ such that the $i$th component of $\bm{\beta}_0$ is nonzero. Define $A = \psi_0'(\bm{\beta}_0) + \bar A$, where $\bar A$ is a $p\times p$ matrix with $i$th row $\bm{\beta}_0^T$ and all other entries $0$. Then $A$ is of full rank, and
\begin{eqnarray*}
\psi_0(\bm{\beta}) = A\cdot (\bm{\beta}-\bm{\beta}_0) + o(\bm{\beta} - \bm{\beta}_0).
\end{eqnarray*}

}

\end{itemize}

\end{lemma}

We have the following theorem, which is Theorem 2 in the main article; it establishes the asymptotic properties of $\widetilde{\bm{\beta}}$, and the proof is given in Sections \ref{sec-pf-part-1}--\ref{sec-pf-part-3}.

\begin{theorem} \label{thm-normality}
Assume Conditions 0--2, 6, and A1--A3. Denote by $\widetilde{\bm{\beta}}$ the zero-crossing of $\psi_n(\bm{\beta})$ (if it exists). We have the following:
\begin{itemize}

\item[(1)] When $n\to \infty$, a zero-crossing $\widetilde{\bm{\beta}}$ of $\psi_n(\bm{\beta})$ exists with probability tending to 1.

\item[(2)] $\widetilde{\bm{\beta}} \to \bm{\beta}_0$ in probability.

{\color{black} 
\item[(3)] Recalling the matrix $A$ defined in Lemma \ref{lemma-psi-0} Part (3), we have
    \begin{eqnarray}
    \widetilde{\bm{\beta}} - \bm{\beta}_0 &=&
   \frac{  A^{-1} }{n^2} \sum_{i\neq j} \Big[
\left\{X_i - X_j - \varphi_0\left((X_i - X_j)^T \bm{\beta}_0\right)\right\} \nonumber\\
&& \times \left\{I(Y_i>Y_j) - F_0\left((X_i - X_j)^T \bm{\beta}_0\right)\right\} \Big]\nonumber \\  &&+ o_p(\widetilde{\bm{\beta}} - \bm{\beta}_0) + o_p(n^{-1/2}).
 \label{beta-tilde-expansion}
\end{eqnarray}

}

\end{itemize}
\end{theorem}

{\color{black} 
To derive the asymptotic distribution for $\widetilde{\bm{\beta}}$, we need to work on a U-statistic with the kernel:
\begin{eqnarray*}
h(\bm{x}_1, y_1; \bm{x}_2, y_2) = \left(\bm{x}_1 - \bm{x}_2\right)\left\{I(y_1>y_2) - F_0\left((\bm{x}_1 - \bm{x}_2)^T \bm{\beta}_0\right)\right\}.
\end{eqnarray*}
Denote
\begin{eqnarray}
\hbar(\bm{x}_1, y_1) = E\left\{h(\bm{x}_1, y_1; X_2, Y_2)\right\}. \label{def-hbar}
\end{eqnarray}
The theorem above leads to the asymptotic distribution of $\widetilde{\bm{\beta}}$.
}
{\color{black} 
\begin{corollary} \label{corollary-normality}
Assume the conditions of Theorem \ref{thm-normality}. Furthermore, assume that both $\Sigma_X = \mbox{var}(X)$ and $\Sigma_{\hbar} = \mbox{var}\left\{\hbar(X, Y)\right\}$ are of full rank, and
\begin{eqnarray}
\bm{\alpha}^T E\left(X_2 - X_1| (X_2-X_1)^T\bm{\beta}_0\right) = 0, \nonumber \\ \mbox{ for any } \bm{\alpha} \in \mathbb{R}^p, \mbox{ and } \bm{\alpha}^T\Sigma_X \bm{\beta}_0 = 0.  \label{condition-distribution-X}
\end{eqnarray}
Denote $\widetilde A = \psi_0'(\bm{\beta}_0)\Sigma_X^{-1} + \bar A$, where $\bar A$ is defined in Lemma \ref{lemma-psi-0}. 
 We have
 \begin{eqnarray}
\sqrt{n}(\widetilde{\bm{\beta}} - \bm{\beta}_0) \to 2 A^{-1} \widetilde A^{-1} \psi_0'(\bm{\beta}_0) \Sigma_X^{-1} N(\bm{0}, \Sigma_{\hbar}) \label{eq-coro-normality}
\end{eqnarray}
in distribution, as $n\to \infty$.

\end{corollary}
}

{\color{black} 

Theorem \ref{theorem-F-asymptotics} below is the Theorem 3 in the main article; it establishes the asymptotic distribution of $\widetilde F(\cdot) = \widehat F_{\widetilde{\bm{\beta}}}(\cdot)$; the proof is given in Section \ref{sec-proof-normality-F-betahat}.

\begin{theorem} \label{theorem-F-asymptotics}

Assume that all the conditions of Corollary \ref{corollary-normality} are satisfied, and also assume the following conditions:
\begin{itemize}

\item[] \underline{Condition F1}: Recall that $F_\epsilon(s)$ is the c.d.f. for $\epsilon_i$. Assume that it is continuous for $s\in \mathbb{R}$ and is continuously differentiable and strictly monotone for $s$ in its support.


\item[] \underline{Condition F2}: Denote by $F_{X^T\bm{\beta}}(s)$ the c.d.f. for $X^T \bm{\beta}$. Assume that it is continuously differentiable for $s$ in its support. Let
$f_{X^T\bm{\beta}}(s) \equiv \frac{\partial F_{X^T\bm{\beta}}(s)}{\partial s}$; assume that for any $s$, it is continuous for $\bm{\beta}$ in the neighborhood of $\bm{\beta}_0$; and $f_{X^T\bm{\beta}_0}(s)$ is continuous for $s$ in its support.

\end{itemize}
Then, we have for every $t$,
\begin{eqnarray*}
n^{1/2}\left\{ \widehat F_{\widetilde{\bm{\beta}}}(t) - F_0(t)\right\} \to N\left(0,  \sigma^2(t)/g_0^2(t)\right),
\end{eqnarray*}
in distribution,
where $g_0(t)$ denotes the p.d.f. for $(X_2-X_1)^T \bm{\beta}_0$, $\sigma^2(t) = \mbox{var}\left\{\zeta_t(X, Y)\right\}$,
\begin{eqnarray*}
\zeta_t(\bm{x}, y) &=& \left\{1-F_\epsilon(H_0(y) - \bm{x}^T \bm{\beta}_0-t) - F_0(t)  \right\} f_{X^T\bm{\beta}_0}(\bm{x}^T \bm{\beta}_0+t)\\
&& +\left\{F_\epsilon(H_0(y) - \boldsymbol{x}^T\boldsymbol{\beta}_0+t) - F_0(t)  \right\}f_{X^T\bm{\beta}_0}(\bm{x}^T \bm{\beta}_0-t)\\
&&+2 g_0(t) \dot{F}_0^T(t) A^{-1} \widetilde A^{-1} \psi_0'(\bm{\beta}_0) \Sigma_X^{-1} \hbar(\bm{x}, y),
\end{eqnarray*}
with $\dot{F}_0(t) = \partial F_{\bm{\beta}}(t)/ \partial \bm{\beta}\big|_{\bm{\beta} = \bm{\beta}_0}$ and  $\hbar(\cdot, \cdot)$ defined by (\ref{def-hbar}).

\end{theorem}

\begin{remark} \label{remark-condition-F}
Based on the definition
\begin{eqnarray*}
F_0(s) = P(\epsilon_1-\epsilon_2\leq s) = E\left\{F_\epsilon(\epsilon_2 + s)\right\},
\end{eqnarray*}
the dominant convergence theorem, and Condition F1, we can verify that
$F_0'(s) \equiv \partial F_0(s)/\partial s$ exists and continuous, and $F_0'(t) > 0$ for any $t$ in the support of $F_0(\cdot)$.
Similarly,  based on the definition
\begin{eqnarray*}
G_{\bm{\beta}}(s) &=& P((X_1-X_2)^T\bm{\beta}<s)\\
&=& E\left\{ F_{X^T\bm{\beta}}(X_2^T\bm{\beta} + s) \right\},
\end{eqnarray*}
Condition F2 implies that
$G_{\bm{\beta}}'(s) \equiv \partial G_{\bm{\beta}}(s)/\partial s$ exists and continuous for $(s, \bm{\beta})$ in a neighbourhood of $(t, \bm{\beta}_0)$,  and $G_0'(t) > 0$. We denote $g_{\bm{\beta}}(s) = G_{\bm{\beta}}'(s)$ and $g_0(s) = g_{\bm{\beta}_0}(s)$.

\end{remark}

With developments similar to but simpler than those of Theorem \ref{theorem-F-asymptotics}, we are able to establish the asymptotic distribution for $\widehat F_{\bm{\beta}}(t)$ for every given $\bm{\beta}\in \mathB$ and $t\in \mathbb{R}$; we summarise this result in the following corollary; a sketched proof is given in Section \ref{sec-proof-of-normality-F-beta}.

\begin{corollary} \label{coro-F-beta-asymptotic}
Assume Conditions 0--2 and A1, and Conditions F1' and F2' given below.
\begin{itemize}
\item[] \underline{Condition F1'}: $F_{\epsilon, \bm{\beta}}(y, s)$ defined by (\ref{def-F-eps-beta}) is continuous for $s\in \mathbb{R}$ and is continuously differentiable for $s$ in the support and $\bm{\beta} \in \mathcal{B}$.


\item[] \underline{Condition F2'}: Denote by $F_{X^T\bm{\beta}}(s)$ the c.d.f. for $X^T \bm{\beta}$. Assume that it is continuously differentiable for $s$ in the support. Recall
$f_{X^T\bm{\beta}}(s)$ defined by Condition F2 in Theorem \ref{theorem-F-asymptotics}; assume that it is continuous for $s$ in the support and $\bm{\beta} \in \mathcal{B}$.
\end{itemize}
For every $t\in \mathbb{R}$ and $\bm{\beta}\in \mathcal{B}$, we have
\begin{eqnarray*}
n^{1/2}\left\{ \widehat F_{\bm{\beta}}(t) - F_{\bm{\beta}}(t)\right\} \to N\left(0,  \sigma_{\bm{\beta}}^2(t)/g_{\bm{\beta}}^2(t)\right),
\end{eqnarray*}
in distribution,
where $g_{\bm{\beta}}(\cdot)$ denotes the p.d.f. of $(X_2-X_1)^T\bm{\beta}$, and
\begin{eqnarray}
\sigma_{\bm{\beta}}(t) &=& E\Big[\mbox{var}\Big\{f_{X^T\boldsymbol{\beta}}(X_1^T\boldsymbol{\beta}+t)F_{\epsilon, \bm{\beta}}(Y_1, X_1^T\bm{\beta} + t)\\ &&\hspace{0.5in}-f_{X^T\boldsymbol{\beta}}(X_1^T\boldsymbol{\beta}-t) F_{\epsilon, \bm{\beta}}(Y_1, X_1^T\bm{\beta} - t)  \Big| X_1 \Big\} \Big] \nonumber\\
F_{\epsilon, \bm{\beta}}(y, s) &=& E\left\{F_\epsilon(H(y) - X^T\bm{\beta_0})\Big| X^T\bm{\beta} = s\right\}. \label{def-F-eps-beta}
\end{eqnarray}

\end{corollary}

}

{\color{black} 
\section{Some Discussion and Special Cases for Conditions 2, 4, and 6} \label{section-special-condition}

In this section, we discuss some special cases under which Conditions 2, 4, and 6 are satisfied.

\subsection{Special cases for Condition 2}

Recall that Condition 2 requires that for any $x_1, x_2 \in \bbR$, we have
\begin{eqnarray*}
\sup_{\bm{\beta} \in \mathB} \left| F_{\bm{\beta}^T X} (x_1) - F_{\bm{\beta}^T X}(x_2) \right| \lesssim |x_1 - x_2|. \label{eq-Cond-2-1}
\end{eqnarray*}
Clearly, a sufficient condition for this is the following Condition 2':

\begin{itemize}

\item[] \underline{\bf Condition 2'}: The p.d.f. of $\bm{\beta}^T X$, denoted by $f_{\bm{\beta}^T X}(x)$, exists and $$\sup_{\bm{\beta} \in \mathB, x \in \bbR} f_{\bm{\beta}^T X}(x) \lesssim 1.$$

\end{itemize}

Moreover, Condition 2' is easily satisfied for many popular distributions of $X$; for example,
\begin{itemize}
\item if $X\sim N(\mu, \Sigma_X)$ with $\lambda_{\mbox{min}}(\Sigma_X) > 0$, then Condition 2' is satisfied, since $\bm{\beta}^T X\sim N(\bm{\beta}^T \mu, \bm{\beta}^T \Sigma_X \bm{\beta})$, and the variance
    \begin{eqnarray*}
    \inf_{\bm{\beta} \in \mathB}\bm{\beta}^T \Sigma_X \bm{\beta} \geq \lambda_{\mbox{min}}(\Sigma_X) \inf_{\bm{\beta} \in \mathB} \|\bm{\beta}\|_2^2 = \lambda_{\mbox{min}}(\Sigma_X) > 0,
    \end{eqnarray*}
    where we have used the fact that $\|\bm{\beta}\|_2 = 1$;

\item if $X \sim \mbox{Uniform}(\Omega)$, where $\Omega$ is a bounded subset of $\bbR^p$, then Condition 2' is satisfied;

\item more generally, it is satisfied if the distribution of $X$ belongs to a location-scale family, whose density has the structure $f_X(\bm{x}) = \mbox{det}(\Sigma_X^{-1/2}) f(\Sigma_X^{-1/2} (\bm{x} - \bm{\mu}_X))$, where $\mbox{det}(\cdot)$ denotes the determinant of a matrix,  $\Sigma_X$ and $\bm{\mu}_X$ are unknown parameters satisfying $\lambda_{\mbox{min}}(\Sigma_X)>0$, and $f(\cdot)$ is some given density function such that $\sup_{x\in \bbR} f(x) \lesssim 1$.

\end{itemize}

\subsection{Special cases for Condition 4}

Condition 4 is to ensure that $F(\cdot)$ and $\bm{\beta}$ are identifiable. The following lemma gives a sufficient, but more intuitive condition under which Condition 4 is satisfied. We need the following Condition 4':
\begin{itemize}

\item \underline{\bf Condition 4'}: There exists an $\widetilde{\bm{x}} \in \mathX$ and $\eta >0$ such that $\{\bm{x}: \|\bm{x}- \widetilde{\bm{x}}\|_2\leq \eta\} \subset \mathX$.

\end{itemize}

\begin{remark}

Condition 4' is satisfied if we assume that $\lambda_{\min}\left\{\mbox{var}(X)\right\} > 0$ and $f_X(\bm{x})$, the p.d.f. of $X$, is continuous.

\end{remark}

\begin{lemma}

Assume $\|\bm{\beta}\|_2 = 1$, $F_0(\cdot)$ is strictly increasing, and Condition 4' given above. Then Condition 4 is satisfied.

\end{lemma}

{\it Proof}: We need to show only that if $F(\bm{v}^T \bm{\beta}) = F_0(\bm{v}^T \bm{\beta}_0)$ for all $\bm{v} \in \mathX - \mathX$, then
$\bm{\beta} = \bm{\beta}_0$.
We prove this by contradiction. Suppose otherwise, then there exists a $\bm{\beta}_1 \neq \bm{\beta}_0$ such that $F(\bm{v}^T \bm{\beta}_1) = F_0(\bm{v}^T \bm{\beta}_0)$ for all $\bm{v} \in \mathX - \mathX$. Since $F\in \mathF$ is nondecreasing and $F_0(\cdot)$ is strictly increasing, we must have for any $\bm{v}_1, \bm{v}_2 \in \mathX-\mathX$,
\begin{itemize}

\item if $\bm{v}_1^T \bm{\beta}_0 > \bm{v}_2^T \bm{\beta}_0$, then $\bm{v}_1^T \bm{\beta}_1 \geq \bm{v}_2^T \bm{\beta}_1$; otherwise, since $F(\cdot)$ is nondecreasing,
\begin{eqnarray*}
F_0(\bm{v}_1^T \bm{\beta}_0) = F(\bm{v}_1^T \bm{\beta}_1) \leq F(\bm{v}_2^T \bm{\beta}_1) = F_0(\bm{v}_2^T \bm{\beta}_0),
\end{eqnarray*}
which contradicts $\bm{v}_1^T \bm{\beta}_0 > \bm{v}_2^T \bm{\beta}_0$ and $F_0(\cdot)$ is strictly increasing;

\item likewise if $\bm{v}_1^T \bm{\beta}_0 < \bm{v}_2^T \bm{\beta}_0$, then $\bm{v}_1^T \bm{\beta}_1 \leq \bm{v}_2^T \bm{\beta}_1$.

\end{itemize}
In summary, we must have
\begin{eqnarray}
(\bm{v}_1-\bm{v}_2)^T \bm{\beta}_1 \bm{\beta}_0^T (\bm{v}_1-\bm{v}_2)\geq 0. \label{eq-condition-4-special-1}
\end{eqnarray}

On the other hand, since $\bm{\beta}_1 \neq \bm{\beta}_0$ and $\|\bm{\beta}_1 \|_2 = \|\bm{\beta}_0\|_2 = 1$, we have $|\bm{\beta}_1^T \bm{\beta}_0| < 1$. Because of Condition 4', we can verify that there exists an $\widetilde \eta > 0$ such that
$0\neq \widetilde{\bm{v}} \in \{\bm{v}_1- \bm{v}_2: \bm{v}_1 \in \mathX - \mathX, \bm{v}_2 \in \mathX - \mathX\}$, where
$\widetilde{\bm{v}} = \widetilde \eta\left(\bm{\beta_1} - \bm{\beta}_0\right).$
As a consequence, noting that $\bm{\beta}_1^T \bm{\beta}_1 = \bm{\beta}_0^T \bm{\beta}_0 = 1$, we have
\begin{eqnarray*}
\widetilde{\bm{v}}^T \bm{\beta}_1 \bm{\beta}_0^T \widetilde{\bm{v}}  = -\widetilde \eta^2 \left(1-\bm{\beta}_0^T \bm{\beta}_1\right)^2 < 0,
\end{eqnarray*}
which contradicts (\ref{eq-condition-4-special-1}). This completes the proof of this lemma. \epf

\subsection{Special cases for Condition 6}

In this section, we consider some stronger but more intuitive conditions than Condition 6.

 Let $\bm{\beta} \in \mathO_{\eta_0}$ and $\Sigma(\bm{\beta}) = \mbox{var}\left(V - E(V|V^T \bm{\beta})\right)$. Then $\Sigma(\bm{\beta})$ is positive semidefinite; there exists an orthonormal matrix $P(\bm{\beta})$ and a diagonal matrix $\Lambda(\bm{\beta}) = \mbox{diag}(\lambda_1(\bm{\beta}), \ldots, \lambda_p(\bm{\beta}))$ with $0\leq \lambda_1(\bm{\beta}) \leq \ldots \leq \lambda_p(\bm{\beta})$ such that $\Sigma(\bm{\beta}) = P(\bm{\beta}) \Lambda(\bm{\beta}) P(\bm{\beta})^T$.
Furthermore, since
\begin{eqnarray}
\bm{\beta}^T \Sigma(\bm{\beta}) \bm{\beta} &=& \bm{\beta}^T\mbox{var}\left(V - E(V|V^T \bm{\beta})\right) \bm{\beta}  \label{eq-beta-Sigma-beta}\\
 &=& \mbox{var}\left( \bm{\beta}^T V - E(\bm{\beta}^TV|V^T \bm{\beta})\right) = \mbox{var}(0) = 0, \nonumber
\end{eqnarray}
meaning that $\Sigma(\bm{\beta})$ is singular, we have $\lambda_1(\bm{\beta}) = 0$. As a consequence,
\begin{eqnarray*}
\Lambda(\bm{\beta}) = \mbox{diag}(0, \lambda_2(\bm{\beta}), \ldots, \lambda_p(\bm{\beta})),
\end{eqnarray*}
and Condition 6 is equivalent to
\begin{eqnarray*}
\inf_{\bm{\beta}\in \mathO_{\eta_0}}\lambda_2(\bm{\beta}) > 0.
\end{eqnarray*}

We shall show that the following  Condition 6' is a sufficient condition for Condition 6.

\begin{itemize}

\item[] \underline{\bf Condition 6'}: There exists an $\eta_1>0$ such that all entries of $\Sigma(\bm{\beta})$ are continuous functions for $\bm{\beta}\in \mathO_{\eta_1}$; and  for any vector $\bm{0} \neq \bm{\alpha} \in \bbR^p$ such that $\bm{\alpha}^T \bm{\beta}_0 = 0$, we have $$\mbox{var}\left\{\bm{\alpha}^TV - E(\bm{\alpha}^T V| \bm{\beta}_0^T V)\right\} > 0.$$

\end{itemize}

\begin{lemma}

Condition 6 holds if Condition 6' holds.

\end{lemma}

{\it Proof}: We need the following notational convention for matrices: for any matrix $B$, $B_i$ denotes the $i$th row, and $B_{i:j}$ denotes the $i$th to $j$th rows of $B$; $B_{i,j}$ denotes the $(i,j)$th entry of $B$.

We first show that $\lambda_2(\bm{\beta}_0) > 0$. In fact, there exists an orthonormal matrix $U(\bm{\beta}_0)$ whose first row is given by $U_1(\bm{\beta}_0) = \bm{\beta}_0^T/\|\bm{\beta}_0\|_2$, and since $U_{2:p}(\bm{\beta}_0) \Sigma (\bm{\beta}_0) U_{2:p}(\bm{\beta}_0)$ is positive semidefinite, there exists a $(p-1)\times(p-1)$ orthonormal matrix $R(\bm{\beta}_0)$ such that
$$R^T(\bm{\beta}_0) U_{2:p}(\bm{\beta}_0) \Sigma(\bm{\beta}_0) U_{2:p}^T(\bm{\beta}_0) R(\bm{\beta}_0) = \mbox{diag}(\tau_2(\bm{\beta}_0), \ldots, \tau_p(\bm{\beta}_0)),$$
where $0\leq \tau_2(\bm{\beta}_0)\leq \ldots\leq \tau_p(\bm{\beta}_0)$.
It is straightforward to verify that $\widetilde R(\bm{\beta}_0) = (U_1^T(\bm{\beta_0}), U_{2:p}^T (\bm{\beta}_0) R(\bm{\beta}_0))^T$ is a $p\times p$ orthonormal matrix, and
\begin{eqnarray}
\label{eq-cond-6-0}&&\widetilde R(\bm{\beta}_0) \Sigma(\bm{\beta}_0) \widetilde R^T(\bm{\beta}_0) \\
&=& \widetilde R(\bm{\beta}_0)\mbox{var}\left(V - E(V|V^T \bm{\beta}_0)\right) \widetilde R^T(\bm{\beta}_0)(\bm{\beta}_0) \nonumber\\
&=& \mbox{var}\left(\left(\begin{matrix}
  \bm{\beta}_0^T/\|\bm{\beta}_0\|_2 \\ R^T(\bm{\beta}_0) U_{2:p}(\bm{\beta}_0)
\end{matrix} \right)V - E\left(\left(\begin{matrix}
  \bm{\beta}_0^T/\|\bm{\beta}_0\|_2 \\ R^T(\bm{\beta}_0) U_{2:p}(\bm{\beta}_0)
\end{matrix} \right)V\Bigg|V^T \bm{\beta}_0\right)\right) \nonumber\\
&=& \mbox{var}\left( \begin{matrix} 0 \\ R^T(\bm{\beta}_0) U_{2:p}(\bm{\beta}_0) (V - E(V|V^T \bm{\beta}_0) \end{matrix} \right)\nonumber\\
&=& \left( \begin{matrix} 0 & \bm{0} \\ \bm{0} & R^T(\bm{\beta}_0) U_{2:p}(\bm{\beta}_0) \Sigma(\bm{\beta}_0) U_{2:p}^T(\bm{\beta}_0) R(\bm{\beta}_0) \end{matrix} \right) \nonumber\\
&=& \left( \begin{matrix} 0 & \bm{0} \\ \bm{0} & \mbox{diag}(\tau_2(\bm{\beta}_0), \ldots, \tau_p(\bm{\beta}_0))  \end{matrix} \right), \nonumber
\end{eqnarray}
indicating that $0 = \tau_1(\bm{\beta}_0) \leq \tau_2(\bm{\beta}_0)\leq \ldots\leq \tau_p(\bm{\beta}_0)$ are the eigenvalues of $\Sigma(\bm{\beta}_0)$. Based on the uniqueness of eigenvalues, $\lambda_i(\bm{\beta}_0) = \tau_i(\bm{\beta}_0)$ for $i=2, \ldots, p$. If $\lambda_2(\bm{\beta}_0) = 0$, then $\tau_2(\bm{\beta}_0) = \lambda_2(\bm{\beta}_0) = 0$. Consider $\widetilde R_2(\bm{\beta}_0)$, the second row of the orthonormal matrix $\widetilde R(\bm{\beta}_0)$; we have
\begin{itemize}
  \item[(i)] $\|\widetilde R_2^T(\bm{\beta}_0)\|_2 = 1\neq 0$, since $\widetilde R(\bm{\beta}_0)$ is orthonormal;

  \item[(ii)] $\widetilde R_2(\bm{\beta}_0) \bm{\beta}_0 = 0$, since $\bm{\beta}_0^T/\|\bm{\beta}_0\|_2$ is the first row of $\widetilde R(\bm{\beta}_0)$;

  \item[(iii)] based on (\ref{eq-cond-6-0}),
  \begin{eqnarray*}
\mbox{var}\left(\widetilde R_2(\bm{\beta}_0)V - E(\widetilde R_2(\bm{\beta}_0) V| \bm{\beta}_0^T V)\right) = \widetilde R_2(\bm{\beta}_0) \Sigma(\bm{\beta}_0) \widetilde R_2^T(\bm{\beta}_0) = \tau_2(\bm{\beta}_0) = 0.
\end{eqnarray*}
\end{itemize}

We observe that (i)--(iii) introduce a contradiction to Condition 6'. Therefore, we must have $\lambda_2(\bm{\beta}_0) > 0$. Furthermore, as assumed by Condition 6', we must have that $\Sigma(\bm{\beta})$ is continuous for $\bm{\beta}\in \mathO_{\eta_1}$; and its eigenvalues are the solutions of the polynomial $\det(\lambda I_p - \Sigma(\bm{\beta}))$. Therefore, $\lambda_2(\bm{\beta})$ is a continuous function for $\bm{\beta}\in \mathO_{\eta_1}$; see Zedek (1965). As a consequence, there exists $0< \eta_0 \leq \eta_1$ such that $\inf_{\bm{\beta} \in \mathO_{\eta_0}}\lambda_2(\bm{\beta}) > 0$.
Similarly, $\lambda_{\min}(\Sigma_{-1,-1}(\bm{\beta}))$ is a continuous function for $\bm{\beta} \in \mathO_{\eta_0}$, and therefore there exists
a $\bm{\beta}' \in \mathO_{\eta_0}$ such that
\begin{eqnarray*}
\lambda_{\min}(\Sigma_{-1,-1}(\bm{\beta}')) = \inf_{\bm{\beta}\in \mathO_{\eta_0}} \lambda_{\min}(\Sigma_{-1,-1}(\bm{\beta})).
\end{eqnarray*}
Clearly,
$$\lambda_2(\bm{\beta}') \geq \inf_{\bm{\beta} \in \mathO_{\eta_0}}\lambda_2(\bm{\beta}) > 0.$$
We complete the proof of this lemma. \epf
\\

Finally, we give one special example under which Condition 6' is satisfied.

\begin{example}
Assume $X_i$, $i=1,\ldots, n$, are i.i.d.~and follow the multivariate normal distribution with a variance matrix $\Sigma_X$ that is of full rank. Then Condition 6' is satisfied.
\end{example}

{\it Proof}: Clearly, $\Sigma_V = \mbox{var}(X_i-X_j) = 2 \Sigma_X$, which is strictly positive definite; thus, $\Sigma_V = P_V^T \Lambda_V P_V$ with  $P_V$ being an orthonormal matrix and $\Lambda_V = \mbox{diag}(\nu_1, \ldots, \nu_p)$, $0< \nu_1 \leq \nu_2 \leq \ldots \leq \nu_p$. For any $\bm{\beta} \in \mathO_{\eta_1}$,
let $U_V(\bm{\beta})$ be an orthonormal matrix whose first row is given by
$$U_{V,1} = \frac{\bm{\beta}^T P_V^T \Lambda_V^{1/2}}{\sqrt{\bm{\beta}^T\Sigma_V \bm{\beta}}}.$$
Let $Q_V = U_V(\bm{\beta}) \Lambda_V^{-1/2} P_V$. Clearly, the first row of $Q_V$ is $Q_{V,1} = \frac{\bm{\beta}^T}{\sqrt{\bm{\beta}^T \Sigma_V \bm{\beta}}}$, and
\begin{eqnarray*}
\mbox{var}(Q_V V) = Q_V \Sigma_V Q_V^T = U_V(\bm{\beta}) \Lambda_V^{-1/2} P_V (P_V^T \Lambda_V P_V) P_V^T \Lambda_V^{-1/2} U_V^T(\bm{\beta}) = I_p,
\end{eqnarray*}
which together with the normality assumption indicates that $Q_{V, 2:p} V$ is independent of $\bm{\beta}^T V$. Furthermore, note that $Q_V^{-1} = P_V^T \Lambda_V^{1/2}U_V^T(\bm{\beta})$. Therefore,
\begin{eqnarray*}
E(V|\bm{\beta}^T V) &=& Q_V^{-1} E(Q_VV|\bm{\beta}^T V) = P_V^T \Lambda_V^{1/2}U_V^T(\bm{\beta}) \left( \begin{matrix}
  \bm{\beta}^T V/\sqrt{\bm{\beta}^T \Sigma_V \bm{\beta}} \\ \bm{0}
\end{matrix}  \right)\\
&=& P_V^T \Lambda_V P_V \bm{\beta}\bm{\beta}^T V/(\bm{\beta}^T \Sigma_V \bm{\beta}) = \Sigma_V\bm{\beta}\bm{\beta}^T V/(\bm{\beta}^T \Sigma_V \bm{\beta}),
\end{eqnarray*}
and thus
\begin{eqnarray}
\Sigma(\bm{\beta}) &=& \mbox{var}\left(V - E(V|V^T \bm{\beta})\right) \nonumber \\
&=& \left( I_p - \frac{\Sigma_V \bm{\beta}\bm{\beta}^T}{\bm{\beta}^T \Sigma_V \bm{\beta}}\right)\Sigma_V\left( I_p - \frac{  \bm{\beta}\bm{\beta}^T \Sigma_V}{\bm{\beta}^T \Sigma_V \bm{\beta}}\right), \label{eq-example-1-1}
\end{eqnarray}
which is clearly a continuous function of $\bm{\beta}$.

It remains to show that for any vector $\bm{0} \neq \bm{\alpha} \in \bbR^p$ that satisfies $\bm{\alpha}^T \bm{\beta}_0 = 0$, we must have
\begin{eqnarray}
\mbox{var}\left(\bm{\alpha}^TV - E(\bm{\alpha}^T V| \bm{\beta}_0^T V)\right) > 0. \label{eq-example-1-2}
\end{eqnarray}
In fact, noting (\ref{eq-example-1-1}), we have
\begin{eqnarray*}
\mbox{var}\left(\bm{\alpha}^TV - E(\bm{\alpha}^T V| \bm{\beta}_0^T V)\right) &=& \bm{\alpha}^T \Sigma(\bm{\beta}_0) \bm{\alpha} \\
&=& \bm{\alpha}^T \left( I_p - \frac{\Sigma_V \bm{\beta}_0\bm{\beta}_0^T}{\bm{\beta}_0^T \Sigma_V \bm{\beta}_0}\right)\Sigma_V\left( I_p - \frac{  \bm{\beta}_0\bm{\beta}_0^T \Sigma_V}{\bm{\beta}_0^T \Sigma_V \bm{\beta}_0}\right) \bm{\alpha}.
\end{eqnarray*}
Therefore, because $\Sigma_V$ is of full rank, (\ref{eq-example-1-2}) is verified if
\begin{eqnarray*}
\bm{\alpha}^T \left( I_p - \frac{\Sigma_V \bm{\beta}_0\bm{\beta}_0^T}{\bm{\beta}_0^T \Sigma_V \bm{\beta}_0}\right) \neq \bm{0}.
\end{eqnarray*}
Suppose otherwise; then
\begin{eqnarray*}
0 &=& \bm{\alpha}^T \left( I_p - \frac{\Sigma_V \bm{\beta}_0\bm{\beta}_0^T}{\bm{\beta}_0^T \Sigma_V \bm{\beta}_0}\right)  \bm{\alpha}\\
&=& \|\bm{\alpha}\|_2^2 - \bm{\alpha}^T  \frac{\Sigma_V \bm{\beta}_0(\bm{\beta}_0^T\bm{\alpha})}{\bm{\beta}_0^T \Sigma_V \bm{\beta}_0} = \|\bm{\alpha}\|_2^2,
\end{eqnarray*}
which contradicts $\bm{\alpha}\neq \bm{0}$. This completes the proof of this example. \epf

}

\section{Notations and Some Preliminary Results}

We first introduce some notation used throughout the technical development. Let ``$\lesssim$" (``$\gtrsim$") denote smaller (greater) than, up to a universal constant. If not otherwise stated, for any arbitrary random variable (vector) $Z$, denote by $F_Z(\cdot)$ the c.d.f.\ of $Z$; likewise for $Z_1, \ldots, Z_n$, denote by $\bbF_Z(\cdot)$ the empirical c.d.f. We use $\|\cdot\|_q$
to  denote the $l_q$ norm in Euclidean space;  for any probability measure $P$, we use $\|f(\cdot)\|_{q, P}$ to denote the $L_q(P)$ norm of $f$. {\color{black} We need the following definitions of covering number, bracketing number, and entropy for a class of functions; these concepts play key roles in modern empirical process theory. These definitions are adapted from Definitions 2.1 and 2.2 in van de Geer (2000).

\begin{definition}
Let $\Im$ be a class of functions. For any $\delta>0$ and $q>0$, let $N_q(\delta, \Im, P)$ be the smallest value of $N$ for which there exists a collection of functions $\{g_1, \ldots, g_N\}$ such that for any $g\in \Im$, there exists a $j=j(g)\in \{1, \ldots, N\}$, such that $\|g-g_j\|_{q,P} \leq \delta$. $N_q(\delta, \Im, P)$ is called the $\delta$-covering number of $\Im$, and $$H_q(\delta, \Im, P) = \log N_q(\delta, \Im, P)$$ is called the  $\delta$-entropy of $\Im$ (for the $L_q(P)$-metric).
\end{definition}
}

\begin{definition}
Let $\Im$ be a class of functions. For any $\delta>0$ and $q>0$, let
$N_{q,B}(\delta,\Im,P)$ be the smallest value of $N$
for which there exists a set of pairs of functions
$\{(g_j^L,g_j^U)\}_{j=1}^N$ such that (i) $\|g_j^U-g_j^L\|_{q,P} \le \delta$, and (ii) for any $g\in \Im$, there exists a $j=j(g)$ such that
$$g_j^L\leq g\leq g_j^U.$$
$N_{q,B}(\delta,\Im,P)$ is called the $\delta$-bracketing number of $\Im$, and $$H_{q,B}(\delta,\Im,P)=\log N_{q,B}(\delta,\Im,P)$$
is called the $\delta$-entropy with bracketing of $\Im$.
\end{definition}

{\color{black} We reiterate the following notational convention for matrices: for any matrix $B$, $B_i$ denotes the $i$th row, and $B_{i:j}$ denotes the $i$th to $j$th rows of $B$; $B_{i,j}$ denotes the $(i,j)$th entry of $B$; $B_{-i,-j}$ denotes the matrix by removing the $i$th row and $j$th column $B$.}

We first introduce two lemmas.
Lemma \ref{lemma-p-1} is from Lemma 5.13 in van de Geer (2000).
\begin{lemma} \label{lemma-p-1}
Let $\Im$ be a class of functions and  $Z_1,\ldots, Z_n$ be an i.i.d.~sample.  Assume that
\begin{eqnarray*}
\sup_{g\in \Im}|g-g_0|_{\infty} \le 1, \\
H_{2,B}(\delta, \Im, F_Z) \le A \delta^{-\alpha},
\end{eqnarray*}
for every $\delta >0$ and some $0<\alpha<2$ and some constant $A$. Then, for some constants $c$ and $n_0$ depending on $\alpha$ and $A$, we have for all $T\ge c$ and $n\ge n_0$,
\begin{eqnarray*}
&& P\left(\sup_{g\in \Im, \|g-g_0\|_{2,F_Z} \le n^{-1/(2+\alpha)}}\left| \int (g-g_0) d(\bbF_Z - F_Z) \right| \ge T n^{-2/(2+\alpha)}\right) \\ &&\hspace{2.8in} \leq c \exp\left\{-\frac{T n^{\alpha/(2+\alpha)}}{c^2}\right\},\\
&&P\left(\sup_{g\in \Im, \|g-g_0\|_{2,F_Z} > n^{-1/(2+\alpha)}}\frac{\sqrt{n}\left| \int (g-g_0) d(\bbF_Z - F_Z) \right|}{\|g-g_0\|_{2,F_Z}^{1-\alpha/2}} \ge T\right) \\&&\hspace{3in} \le c \exp\left(-\frac{T}{c^2}\right).
\end{eqnarray*}
\end{lemma}

To facilitate our subsequent development, we define the following function classes and derive their $\delta$-entropies:
\begin{eqnarray*}
\mathF &=& \left\{ F(\cdot): F(x) \in [0,1] \mbox{ and is monotonically increasing}  \right\}, \\
\mathF_1 &=& \left\{F((\bm{x}_1-\bm{x}_2)^T\bm{\beta}): F \in \mathF, \bm{\beta}\in \mathB \right\},\\
\mathF_2 &=& \left\{F((\bm{x}_1-\bm{x}_2)^T\bm{\beta}): F \in \mathF, \bm{\beta}\in \mathB, \bm{x}_2 \in \mathX \right\},\\
\mathD &=& \bigg\{d_{\bm{\beta}, F} (\bm{x}_2) = \left[ \int \left\{F_0((\bm{x}_1-\bm{x}_2)^T\bm{\beta}_0) - F((\bm{x}_1-\bm{x}_2)^T\bm{\beta})\right\}^2 dF_X(\bm{x}_1) \right]^{1/2}: \\ && \hspace{3.8 in}F \in \mathF, \bm{\beta}\in \mathB  \bigg\}.
\end{eqnarray*}
 The following lemma establishes the $\delta$-entropy of $\mathF_1,\mathF_2$, and $\mathD$.

\begin{lemma} \label{lemma-1}
Assume Conditions 1 and 2. For any arbitrary $1\le q<\infty$, we have
\begin{eqnarray}
H_{q, B} (\delta, \mathF_1, F_{X_1,X_2})&\lesssim& 1/\delta, \label{lem-1-1-1}\\
H_{q, B} (\delta, \mathF_2, F_X)&\lesssim& 1/\delta, \label{lem-1-1} \\
H_{q, B} (\delta, \mathD, F_X)&\lesssim& 1/\delta, \label{lem-1-1-2}
\end{eqnarray}
where ``$\lesssim$" is up to a universal constant depending only on $q$.
\end{lemma}

\proof We first prove a preliminary result, which is helpful in the proof of Lemma \ref{lemma-1}.

\begin{lemma} \label{lemma-abs-entropy}
Let $\Im$ be an arbitrary class of function such that $H_{q, B}(\delta, \Im, P) < \infty$. Then
\begin{eqnarray*}
H_{q,B}(2\delta, |\Im|, P) \le H_{q, B}(\delta, \Im, P).
\end{eqnarray*}
\end{lemma}
\proof By the definition of $H_{q,B}(\delta, \Im, P)$, there exists a set of brackets $\{[l_i, u_i]\}_{i=1}^N$ that covers $\Im$, where $N = \exp\left\{H_{q, B}(\delta, \Im, P)\right\}$. Let $|g|$ be an arbitrary function in $|\Im|$ and hence $g\in \Im$. Let $[l_i, u_i]$ be the bracket such that $l_i \le g \le u_i$. We immediately have $l_i^+ + u_i^- \le g^+ + g^- \le l_i^- + u_i^+ $, that is $l_i^+ + u_i^- \le |g| \le l_i^- + u_i^+ $. Furthermore,
\begin{eqnarray*}
|l_i^- + u_i^+ - l_i^+ -u_i^-| \le |l_i^--u_i^-| + |u_i^+ - l_i^+| \le 2|u_i-l_i|,
\end{eqnarray*}
and hence
\begin{eqnarray*}
\|l_i^- + u_i^+ - l_i^+ -u_i^-\|_{q,P} \le 2 \|u_i-l_i\|_{q,P}.
\end{eqnarray*}
This indicates, every $\delta$-bracket under $L_q(P)$ in $\Im$ leads to a $2\delta$-bracket under $L_q(P)$ in $|\Im|$. This completes our proof of this lemma. \epf

We now move back to the proof of Lemma \ref{lemma-1}.
First, we show (\ref{lem-1-1}); (\ref{lem-1-1-1}) can be obtained with very similar but simpler arguments. For any $\delta>0$, let $\delta_1 = \delta^q$.  Because of Condition 1, we have the following:
\begin{itemize}
\item[(i)] There exist $\bm{\beta_1}, \ldots, \bm{\beta}_{N_1} \in \bbR^p$, such that $\mathB = \cup_{j_1=1}^{N_1} \{\bm{\beta}: \|\bm{\beta} - \bm{\beta}_{j_1}\|_1 < \delta_1, \bm{\beta} \in \mathB\}$, where $N_1\lesssim 1/\delta_1^p$.
\item[(ii)] There exist $\{[a_{j_2}, b_{j_2}]: a_{j_2} \in \bbR, b_{j_2} \in \bbR\}_{j_2 = 1}^{N_2}$ with $N_2 \lesssim 1/\delta_1^{pq}$ and $|b_{j_2}-a_{j_2}|< \delta_1$, such that for every $\bm{\beta}\in \mathB$ and $\bm{x}_2\in \mathX$, we have $\bm{x}_2^T \bm{\beta} \in [a_{j_2}, b_{j_2}]$ for some $j_2\in \{1, \ldots, N_2\}$.
\end{itemize}
Next, we define a set of brackets $\{[l_j(\bm{x}_1), u_j(\bm{x}_1)]: j=1\ldots, N\}$ that covers $\{(\bm{x}_1-\bm{x}_2)^T\bm{\beta}: \bm{\beta} \in \mathB, \bm{x}_2 \in \mathX\}$. Here $N= N_1 N_2 \lesssim 1/\delta_1^{p(1+q)}$. To this end, for $j_1 = 1 \ldots, N_1$ let
\begin{eqnarray*}
\bm{\beta}_{j_1+} = \bm{\beta}_{j_1} + (\delta_1, \ldots, \delta_1)^T; \qquad \bm{\beta}_{j_1-} = \bm{\beta}_{j_1} - (\delta_1, \ldots, \delta_1)^T.
\end{eqnarray*}
Now, for every $j_1 = 1, \ldots, N_1$ and $j_2 = 1,\ldots, N_2$, we define
\begin{eqnarray*}
l_{j_1,j_2}(\bm{x}_1) = \bm{\beta}_{j_1-}^T \bm{x}_1^+ - \bm{\beta}_{j_1+}^T \bm{x}_1^- - b_{j_2}; \quad u_{j_1,j_2}(\bm{x}_1) = \bm{\beta}_{j_1+}^T \bm{x}_1^+ - \bm{\beta}_{j_1-}^T \bm{x}_1^- - a_{j_2},
\end{eqnarray*}
where $\bm{x}_1^+$ and $\bm{x}_1^-$ respectively denote the positive and negative parts of $\bm{x}_1$. Based on (i) and (ii) above, we immediately observe that for every $\bm{\beta} \in \mathB$ and $\bm{x}_2 \in \mathX$, there exist $j_1 \in \{1,\ldots, N_1\}$ and $j_2\in \{ 1, \ldots, N_2\}$ such that
$l_{j_1, j_2}(\bm{x}_1) \le (\bm{x}_1-\bm{x}_2)^T \bm{\beta} \le u_{j_1, j_2}(\bm{x}_1)$. This indicates $\{[l_{j_1, j_2}(\bm{x}_1), u_{j_1,j_2}(\bm{x}_1)]: j_1=1\ldots, N_1, j_2 = 1,\ldots, N_2\}$ covers $\{(\bm{x}_1-\bm{x}_2)^T\bm{\beta}: \bm{\beta} \in \mathB, \bm{x}_2 \in \mathX\}$. For notational convenience, with a little abuse of notation, we denote this set to be $\{[l_j(\bm{x}_1), u_j(\bm{x}_1)]: j=1\ldots, N\}$, where $N= N_1N_2 \lesssim 1/\delta_1^{p(q+1)}$, $l_j(\bm{x}_1) =  \bm{\beta}_{j-}^T \bm{x}_1^+ - \bm{\beta}_{j+}^T \bm{x}_1^- - b_{j}$, and $u_j(\bm{x}_1) = \bm{\beta}_{j+}^T \bm{x}_1^+ - \bm{\beta}_{j-}^T \bm{x}_1^- - a_{j}$.

Next, we establish a set of brackets that cover $\mathF_2$. Since $\mathF$ is the subset of the class of monotonically increasing functions. Based on the well known results in the empirical process literature (see for example Theorem 9.24 in Kosorok 2008), there exists a set $\{[L_{i,j}(u), U_{i,j}(u)]\}_{i=1}^{M_j}$ of brackets that cover $\mathF$, $\log M_j\lesssim 1/\delta$, and for every $i=1,\ldots,M_j$,
\begin{eqnarray}
\| L_{i,j} - U_{i,j} \|_{q, F_j} \le  \delta, \label{lem-1-2}
\end{eqnarray}
where $F_j$ is the c.d.f. of $u_j(X)$. Without loss of generality, we assume that $L_{i,j}$ and $U_{i,j}$ are monotonically increasing functions, $L_{i,j}(u)\in [0,1]$, and $U_{i,j}(u)\in [0,1]$. We consider the set of brackets
\begin{eqnarray}
\left\{\left[L_{i,j}(l_j(\bm{x}_1)), U_{i,j}(u_j(\bm{x}_1)) \right], i = 1,\ldots,M_j; \quad j=1,\ldots,N \right\}, \label{lem-1-3}
\end{eqnarray}
which contains the number of brackets
\begin{eqnarray*}
\sum_{j=1}^N M_j; \quad \mbox{ and } \log\sum_{j=1}^N M_j \lesssim \log \left(Ne^{1/\delta} \right)\lesssim 1/\delta.
\end{eqnarray*}
Finally, (\ref{lem-1-1}) follows if we can verify (\ref{lem-1-4}) and (\ref{lem-1-5}) given below:
\begin{eqnarray}
 \label{lem-1-4}
L_{i,j}(l_j(\bm{x}_1)) \le F((\bm{x}_1-\bm{x}_2)^T \bm{\beta}) \le U_{i,j}(u_j(\bm{x}_2)), \\ \mbox{for every }F((\bm{x}_1-\bm{x}_2)^T \bm{\beta}) \in \mathF_2; \nonumber
\end{eqnarray}
and uniformly in $i\in\{1,\ldots,M\}$ and $j\in\{1,\ldots,N\}$,
\begin{eqnarray}
\|U_{ij}(u_j) - L_{i,j}(l_j)\|_{q, F_X} \lesssim \delta.   \label{lem-1-5}
\end{eqnarray}
We first verify (\ref{lem-1-4}). Based on the definition of $\{[l_j(\bm{x}_1), u_j(\bm{x}_2)]: j=1\ldots, N\}$, for every $\bm{\beta} \in \mathB$ and $\bm{x}_2 \in \mathX$, there exists $j\in \{1,\ldots,N\}$ such that $l_j(\bm{x}_1) \le (\bm{x}_1-\bm{x}_2)^T \bm{\beta} \le u_j(\bm{x}_1)$ for all $\bm{x}_1\in \mathX$. For this $j$, based on (\ref{lem-1-2}), for every $F \in \mathF$ there exists an $i\in \{1,\ldots,M\}$ such that $L_{i,j}(u)\le F(u) \le U_{i,j}(u)$. Therefore, we have
\begin{eqnarray*}
L_{i,j}(l_j(\bm{x}_1)) \le F(l_j(\bm{x}_1)) \le F((\bm{x}_1-\bm{x}_2)^T \bm{\beta}) \le F(u_j(\bm{x}_1)) \le U_{i,j}(u_j(\bm{x}_1)),
\end{eqnarray*}
which proves (\ref{lem-1-4}). We proceed to show (\ref{lem-1-5}). Consider
\begin{eqnarray}
&&\|U_{ij}(u_j) - L_{i,j}(l_j)\|_{q, F_X} \nonumber \\ &\le& \|U_{ij}(u_j) - L_{i,j}(u_j)\|_{q, F_X} + \|L_{i,j}(u_j) - L_{i,j}(l_j)\|_{q, F_X} \nonumber\\
&=& \|U_{i,j} - L_{i,j}\|_{q, F_j} +\|L_{i,j}(u_j) - L_{i,j}(l_j)\|_{q, F_X} \nonumber\\
&\le& \delta + \|L_{i,j}(u_j) - L_{i,j}(l_j)\|_{q, F_X}. \label{lem-1-6}
\end{eqnarray}
Therefore, it is left to bound $\|L_{i,j}(u_j) - L_{i,j}(l_j)\|_{q, F_X}$. Since $L_{i,j}(u) \in [0,1]$ is a monotonically increasing function, there exists a monotonically increasing function $S(u) = \sum_{l}c_l I(d_l < u)$ such that $0\le c_l\le 1$, $\sum_{l} c_l \le 1$, and
\begin{eqnarray*}
\sup_{u}|S(u) - L_{i,j}(u)| \lesssim \delta.
\end{eqnarray*}
Now that
{\small
\begin{eqnarray}
&&\|L_{i,j}(u_j) - L_{i,j}(l_j)\|_{q, F_X} \le 2 \delta + \|S(u_j) - S(l_j)\|_{q, F_X} \nonumber \\
&=& 2\delta + \left[\int \left\{\sum_l c_l I(l_j(\bm{x}_1) \le d_l < u_j(\bm{x}_1))\right\}^q d F_X(\bm{x}_1)\right]^{1/q} \nonumber\\
&\le& 2\delta + \sum_l c_l \left[ \int I(l_j(\bm{x}_1) \le d_l < u_j(\bm{x}_1)) dF_X(\bm{x}_1) \right]^{1/q} \nonumber\\
&=& 2\delta + \sum_l c_l \left[ P(l_j(\bm{x}_1) \le d_l) - P(u_j(\bm{x}_1)\le d_l) \right]^{1/q} \nonumber\\
&=& 2\delta + \sum_l c_l \left[ P\left( \bm{\beta}_j^T X \le \delta_1 \|X\|_1 + b_j + d_l\right) - P\left(\bm{\beta}_j^T X \le -\delta_1 \|X\|_1 + a_j + d_l\right) \right]^{1/q}\nonumber \\
&\le& 2\delta + \sum_l c_l \bigg[ P\left( \bm{\beta}_j^T X \le \delta_1 \sup_{\bm{x}_1\in \mathX}\|\bm{x}_1\|_1 + b_j + d_l\right) \nonumber  \\
&& \hspace{0.7in} - P\left(\bm{\beta}_j^T X \le -\delta_1 \sup_{\bm{x}_1 \in \mathX}\|\bm{x}_1\|_1 + a_j + d_l\right) \bigg]^{1/q}\nonumber \\
&=& 2\delta +  \sum_l c_l \Big[ F_{\bm{\beta}_j^T X}(\delta_1 \sup_{\bm{x}_1 \in \mathX}\|\bm{x}_1\|_1 + b_j + d_l) -  F_{\bm{\beta}_j^TX}(-\delta_1 \sup_{\bm{x}_1 \in \mathX}\|\bm{x}_1\|_1 + a_j + d_l) \Big]^{1/q}. \nonumber
\end{eqnarray} }
By Condition 2, we further have
\begin{eqnarray}
\|U_{ij}(u_j) - L_{i,j}(l_j)\|_{q, F_X}
&\lesssim & 2\delta + \sum_l c_l \left( 2\delta_1 \sup_{\bm{x}_1 \in \mathX}\|\bm{x}_1\|_1 + |a_j - b_j|\right)^{1/q}\nonumber\\
&\lesssim & 2\delta + \delta_1^{1/q} \lesssim \delta. \label{lem-1-7}
\end{eqnarray}
Combining (\ref{lem-1-7}) with (\ref{lem-1-6}), (\ref{lem-1-5}) follows, and we finish the proof of (\ref{lem-1-1}).

We proceed to show (\ref{lem-1-1-2}). Based on (\ref{lem-1-1-1}) and Lemma \ref{lemma-abs-entropy}, we immediately have
\begin{eqnarray*}
H_{q_1,B}(\delta, |F_0((\bm{x}_1-\bm{x}_2)^T \bm{\beta}_0 - \mathF_1|, F_{X_1, X_2}) \lesssim 1/\delta,
\end{eqnarray*}
for any arbitrary $1\le q_1<\infty$. Now for any arbitrary $q$, we set $q_1=2$ when $1\le q\le 2$ and $q_1 = q$ when $1\le q\le 2$ in above. Then,
there exist a set of $\delta$-brackets $\{[l_i(\bm{x}_1, \bm{x}_2), u_i(\bm{x}_1, \bm{x}_2)]\}_{i=1}^{N_3}$ to cover $|F_0((\bm{x}_1-\bm{x})^T \bm{\beta}_0 - \mathF_1|$ under the $L_{q_1}(F_{X_1,X_2})$ norm, where $\log N_3\lesssim 1/\delta$. Without loss of generality, we assume $l_i(\bm{x}_1, \bm{x}_2)\ge 0$ and $u_i(\bm{x}_1, \bm{x}_2)\ge 0$. For $i=1,\ldots, N_3$, define
\begin{eqnarray}
L_i(\bm{x}_2) &=& \left[ \int l_i^2(\bm{x}_1, \bm{x}_2) dF_X(\bm{x}_1) \right]^{1/2} \nonumber \\
U_i(\bm{x}_2) &=& \left[ \int u_i^2(\bm{x}_1, \bm{x}_2) dF_X(\bm{x}_1) \right]^{1/2}. \label{lem-1-8}
\end{eqnarray}
Now for any $d_{\bm{\beta}, F} (\bm{x}_2) = \left[ \int \left\{F_0((\bm{x}_1-\bm{x}_2)^T\bm{\beta}_0) - F((\bm{x}_1-\bm{x}_2)^T\bm{\beta})\right\}^2 dF_X(\bm{x}_1) \right]^{1/2} \in \mathD$, there exists $1\le i\le N_3$, such that
\begin{eqnarray*}
0\le l_i(\bm{x}_1, \bm{x}_2) \le \left|F_0((\bm{x}_1-\bm{x}_2)^T\bm{\beta}_0) - F((\bm{x}_1-\bm{x}_2)^T\bm{\beta})\right| \le u_i(\bm{x}_1, \bm{x}_2),
\end{eqnarray*}
which immediately implies
\begin{eqnarray*}
L_i(\bm{x}_2) \le d_{\bm{\beta}, F} (\bm{x}_2) \le U_i(\bm{x}_2).
\end{eqnarray*}
This indicates that the set of brackets $\{[L_i, U_i]\}_{i=1}^{N_3}$ with $L_i, U_i$ defined by (\ref{lem-1-8}) covers $\mathD$. Furthermore, for every $i=1,\ldots,N_3$, we have
\begin{eqnarray*}
0&\le& U_i(\bm{x}_2)-L_i(\bm{x}_2) = \left[ \int u_i^2(\bm{x}_1, \bm{x}_2) dF_X(\bm{x}_1) \right]^{1/2} - \left[ \int l_i^2(\bm{x}_1, \bm{x}_2) dF_X(\bm{x}_1) \right]^{1/2}\\
&=& \frac{\int \left\{u_i(\bm{x}_1, \bm{x}_2)-l_i(\bm{x}_1, \bm{x}_2)\right\}\left\{u_i(\bm{x}_1, \bm{x}_2)+l_i(\bm{x}_1, \bm{x}_2)\right\} dF_X(\bm{x}_1) }{\left[ \int u_i^2(\bm{x}_1, \bm{x}_2) dF_X(\bm{x}_1) \right]^{1/2} + \left[ \int l_i^2(\bm{x}_1, \bm{x}_2) dF_X(\bm{x}_1) \right]^{1/2}}\\
&\le& \frac{\left[\int \left\{u_i(\bm{x}_1, \bm{x}_2)-l_i(\bm{x}_1, \bm{x}_2)\right\}^2 dF_X(\bm{x}_1)\right]^{1/2}   \left[\int \left\{u_i(\bm{x}_1, \bm{x}_2)+l_i(\bm{x}_1, \bm{x}_2)\right\}^2 dF_X(\bm{x}_1)\right]^{1/2} }{\left[ \int u_i^2(\bm{x}_1, \bm{x}_2) dF_X(\bm{x}_1) \right]^{1/2} + \left[ \int l_i^2(\bm{x}_1, \bm{x}_2) dF_X(\bm{x}_1) \right]^{1/2}}\\
&\le&\left[\int \left\{u_i(\bm{x}_1, \bm{x}_2)-l_i(\bm{x}_1, \bm{x}_2)\right\}^2 dF_X(\bm{x}_1)\right]^{1/2},
\end{eqnarray*}
where the last ``$\le$" follows from the fact that
\begin{eqnarray*}
&& \int \left\{u_i(\bm{x}_1, \bm{x}_2)+l_i(\bm{x}_1, \bm{x}_2)\right\}^2 dF_X(\bm{x}_1) \\ && \hspace{0.2in} - \left[\left\{ \int u_i^2(\bm{x}_1, \bm{x}_2) dF_X(\bm{x}_1) \right\}^{1/2} + \left\{ \int l_i^2(\bm{x}_1, \bm{x}_2) dF_X(\bm{x}_1) \right\}^{1/2} \right]^2 \\
&=&  2\int u_i(\bm{x}_1, \bm{x}_2)l_i(\bm{x}_1, \bm{x}_2) dF_X(\bm{x}_1)\\ && \hspace{0.2in} -2 \left\{ \int u_i^2(\bm{x}_1, \bm{x}_2) dF_X(\bm{x}_1)\int l_i^2(\bm{x}_1, \bm{x}_2) dF_X(\bm{x}_1) \right\}^{1/2} \le 0.
\end{eqnarray*}
Therefore
\begin{eqnarray*}
&&\|U_i-L_i\|_{q, F_X} \nonumber \\ &\le& \left[ \int \left[\int \left\{u_i(\bm{x}_1, \bm{x}_2)-l_i(\bm{x}_1, \bm{x}_2)\right\}^2 dF_X(\bm{x}_1)\right]^{q/2} d F_X(\bm{x}_2) \right]^{1/q} \\
&\le& \left\{ \begin{array}{ll}\left[ \int \int \left\{u_i(\bm{x}_1, \bm{x}_2)-l_i(\bm{x}_1, \bm{x}_2)\right\}^q dF_X(\bm{x}_1) d F_X(\bm{x}_2) \right]^{1/q} & \mbox{when } q>2 \\
\left[ \int \int \left\{u_i(\bm{x}_1, \bm{x}_2)-l_i(\bm{x}_1, \bm{x}_2)\right\}^2 dF_X(\bm{x}_1) d F_X(\bm{x}_2) \right]^{1/2} & \mbox{when } 1\le q \le 2 \end{array} \right. \\
&=& \|u_i-l_i\|_{q_1, F_{X_1,X_2}} \le \delta,
\end{eqnarray*}
based on the fact that $[l_i({\bm{x}_1, \bm{x}_2}), u_i(\bm{x}_1, \bm{x}_2) ], i=1,\ldots,N_3$ are $\delta$-brackets that cover $\mathF_1$.
This completes our proof of (\ref{lem-1-1-2}). The proof of Lemma \ref{lemma-1} is completed. \epf

%

\section{Proof of Theorem \ref{theorem-1}}

We first show Part (a) of Theorem \ref{theorem-1}. It is structured as Lemmas \ref{lemma-basic} and \ref{lemma-u-stat} below.

Lemma \ref{lemma-basic} establishes a basic inequality that plays a key role in developing the asymptotic property for $D\Big( \widehat{\bm{\beta}}, \widehat F; \bm{\beta}_0, F_0\Big)$. We need the following notation:
\begin{eqnarray*}
\gamma_1(\bm{v}; \widehat F, \widehat{\bm{\beta}})&=&4\left\{ \sqrt{\frac{\widehat F(\bm{v}^T \widehat{\bm{\beta}})}{F_0(\bm{v}^T \bm{\beta}_0)}} -1 \right\}, \\
\gamma_2(\bm{v}; \widehat F, \widehat{\bm{\beta}})&=&4\left\{\sqrt{\frac{1-\widehat F(\bm{v}^T \widehat{\bm{\beta}})}{1-F_0(\bm{v}^T \bm{\beta}_0)}} -1 \right\}.
\end{eqnarray*}

\begin{lemma}\label{lemma-basic}
Recall the definition of $D\Big( \widehat{\bm{\beta}}, \widehat F; \bm{\beta}_0, F_0\Big)$ in (\ref{def-D}):
\begin{eqnarray*}
D\Big( \widehat{\bm{\beta}}, \widehat F; \bm{\beta}_0, F_0\Big)
= \left[\int \int \left\{F_0((\bm{x}_1-\bm{x}_2)^T \bm{\beta}) - \widehat F((\bm{x}_1-\bm{x}_2)^T \widehat{\bm{\beta}}) \right\}^2 d F_{X}(\bm{x}_1)d F_{X}(\bm{x}_2)\right]^{1/2}.
\end{eqnarray*}
 We have
\begin{eqnarray}
\label{lem-2-0} && D^2\Big( \widehat{\bm{\beta}}, \widehat F; \bm{\beta}_0, F_0\Big)
\leq 2 \int \left\{I(u>0) \gamma_1(\bm{v}; \widehat F, \widehat{\bm{\beta}}) + I(u<0) \gamma_2(\bm{v}; \widehat F, \widehat{\bm{\beta}}) \right\} \\
&& \hspace{2.5in} \times d\left\{\bbF_{U,V}(u,\bm{v}) - F_{U,V}(u,\bm{v})) \right\},  \nonumber
\end{eqnarray}
where $F_{U,V}$ is the joint c.d.f. of $(U_{i,j},V_{i,j})$ and $\bbF_{U,V}(u,\bm{v})$ is the empirical c.d.f. of $\{(U_{i,j},V_{i,j})\}_{i,j=1}^n$, $U_{i,j} = Y_i-Y_j$, $V_{i,j} = X_i-X_j$.
\end{lemma}

\proof Since $P(Y_i = Y_j) = 0$ for $i\neq j$ by Conditions 0 and 5, without loss of generality, we assume $Y_i \neq Y_j$  when $i\neq j$. Then, we can write 
\begin{eqnarray*}
\ell(\bm{\beta}, F) = \sum_{i,j}[I(Y_i>Y_j)\log\{ F((X_i-X_j)^T\bm{\beta})\}+I(Y_i < Y_j)\log \{1-F((X_i-X_j)^T\bm{\beta})\}].
\end{eqnarray*}
We have
\begin{eqnarray}
0 &\ge& \ell(\bm{\beta}_0, F_0) - \ell(\widehat{\bm{\beta}}, \widehat F) \nonumber \\
&=& -\sum_{i,j} I(U_{i,j}>0) \log \frac{\widehat F(V_{i,j}^T \widehat{\bm{\beta}})}{F_0(V_{i,j}^T\bm{\beta}_0)} - \sum_{i,j} I(U_{i,j}<0) \log \frac{1-\widehat F(V_{i,j}^T \widehat{\bm{\beta}})}{1-F_0(V_{i,j}^T\bm{\beta}_0)} \nonumber\\
&=& -n^2\int I(u>0) \log \frac{\widehat F(\bm{v}^T \widehat{\bm{\beta}})}{F_0(\bm{v}^T \bm{\beta}_0)} d\bbF_{U,V}(u,\bm{v}) \nonumber\\ && \quad -n^2\int I(u<0) \log \frac{1-\widehat F(\bm{v}^T \widehat{\bm{\beta}})}{1-F_0(\bm{v}^T \bm{\beta}_0)} d\bbF_{U,V}(u,\bm{v}). \label{lem-2-1}
\end{eqnarray}
On the other hand, using the inequality $0.5 \log x \le \sqrt{x} -1$ for any $x>0$, we have
\begin{eqnarray}
-\log \frac{\widehat F(\bm{v}^T \widehat{\bm{\beta}})}{F_0(\bm{v}^T \bm{\beta}_0)}  \ge 2\left\{ 1 - \sqrt{\frac{\widehat F(\bm{v}^T \widehat{\bm{\beta}})}{F_0(\bm{v}^T \bm{\beta}_0)}} \right\} = -0.5 \gamma_1(\bm{v}; \widehat F, \widehat{\bm{\beta}}), \label{lem-2-2}
\end{eqnarray}
and likewise
\begin{equation}
-\log \frac{1-\widehat F(\bm{v}^T \widehat{\bm{\beta}})}{1-F_0(\bm{v}^T \bm{\beta}_0)}  \ge 2\left\{ 1 - \sqrt{\frac{1-\widehat F(\bm{v}^T \widehat{\bm{\beta}})}{1-F_0(\bm{v}^T \bm{\beta}_0)}} \right\} = -0.5 \gamma_2(\bm{v}; \widehat F, \widehat{\bm{\beta}}). \label{lem-2-3}
\end{equation}
Combining (\ref{lem-2-1})--(\ref{lem-2-3}), we have
{\small
\begin{eqnarray}
\label{lem-2-4} 0&\ge&  -\int \left\{I(u>0) \gamma_1(\bm{v}; \widehat F, \widehat{\bm{\beta}}) + I(u<0) \gamma_2(\bm{v}; \widehat F, \widehat{\bm{\beta}}) \right\}d\bbF_{U,V}(u,\bm{v}) \\
&=& -\int \left\{I(u>0) \gamma_1(\bm{v}; \widehat F, \widehat{\bm{\beta}}) + I(u<0) \gamma_2(\bm{v}; \widehat F, \widehat{\bm{\beta}}) \right\} \nonumber \\ && \hspace{1.5 in} \times d\left\{\bbF_{U,V}(u,\bm{v}) - F_{U,V}(u, \bm{v}) \right\} \nonumber\\
&&-\int \left\{I(u>0) \gamma_1(\bm{v}; \widehat F, \widehat{\bm{\beta}}) + I(u<0) \gamma_2(\bm{v}; \widehat F, \widehat{\bm{\beta}}) \right\} d F_{U,V}(u, \bm{v}).  \nonumber
\end{eqnarray}
}
Therefore, to show (\ref{lem-2-0}), we need to show only that
\begin{eqnarray}
-\int \left\{I(u>0) \gamma_1(\bm{v}; \widehat F, \widehat{\bm{\beta}}) + I(u<0) \gamma_2(\bm{v}; \widehat F, \widehat{\bm{\beta}}) \right\} d F_{U,V}(u, \bm{v})
\nonumber \\\ge 0.5 D^2\Big( \widehat{\bm{\beta}}, \widehat F; \bm{\beta}_0, F_0\Big). \label{lem-2-4-1}
\end{eqnarray}
To this end,  note that
\begin{eqnarray}
&&-\int \left\{I(u>0) \gamma_1(\bm{v}; \widehat F, \widehat{\bm{\beta}}) + I(u<0) \gamma_2(\bm{v}; \widehat F, \widehat{\bm{\beta}}) \right\} d F_{U,V}(u, \bm{v}) \nonumber\\
&=& -\int \Big\{ \gamma_1(\bm{v}; \widehat F, \widehat{\bm{\beta}}) \int I(u>0) dF_{U|\bm{v}}(u)  \nonumber \\ && \hspace{0.5in} + \gamma_2(\bm{v}; \widehat F, \widehat{\bm{\beta}}) \int I(u<0) dF_{U|\bm{v}}(u) \Big\} d F_V(\bm{v}),\label{lem-2-5}
\end{eqnarray}
where $F_{U|\bm{v}}(u)$ denotes the conditional c.d.f. of $U_{i,j}|V_{i,j}=\bm{v}$ and $F_V(\bm{v})$ is the marginal c.d.f. of $V_{i,j}$. Note that by Condition 0,
\begin{eqnarray}
\int I(u>0) dF_{U|\bm{v}}(u) &=& E\left\{I(U_{i,j}>0)\Big| V_{i,j}=\bm{v} \right\} \nonumber\\
&=& E\left\{I(Y_i-Y_j>0)\Big| X_i-X_j=\bm{v}\right\} \nonumber\\
&=& P(Y_i>Y_j|X_i-X_j=\bm{v}) = F_0(\bm{v}^T\bm{\beta}_0), \label{lem-2-6}
\end{eqnarray}
and likewise
\begin{eqnarray}
\int I(u<0) dF_{U|\bm{v}}(u) = 1-F_0(\bm{v}^T\bm{\beta}_0). \label{lem-2-7}
\end{eqnarray}
Combining (\ref{lem-2-2}), (\ref{lem-2-3}), (\ref{lem-2-5}), (\ref{lem-2-6}), and (\ref{lem-2-7}), we have
\begin{eqnarray}
&&-\int \left\{I(u>0) \gamma_1(\bm{v}; \widehat F, \widehat{\bm{\beta}}) + I(u<0) \gamma_2(\bm{v}; \widehat F, \widehat{\bm{\beta}}) \right\} d F_{U,V}(u, \bm{v}) \nonumber\\
&=& 4 \int  \bigg\{ F_0(\bm{v}^T \bm{\beta}_0) - \sqrt{F_0(\bm{v}^T \bm{\beta}_0)}\sqrt{\widehat F(\bm{v}^T \widehat{\bm{\beta}}) } \nonumber\\
&& +  1- F_0(\bm{v}^T \bm{\beta}_0) - \sqrt{1-F_0(\bm{v}^T \bm{\beta}_0)}\sqrt{1-\widehat F(\bm{v}^T \widehat{\bm{\beta}}) }\bigg\}            d F_V(\bm{v}) \nonumber\\
&=& 4 \int  \Big\{ 1 - \sqrt{F_0(\bm{v}^T \bm{\beta}_0)}\sqrt{\widehat F(\bm{v}^T \widehat{\bm{\beta}}) } \nonumber \\ &&  \hspace{0.5in}- \sqrt{1-F_0(\bm{v}^T \bm{\beta}_0)}\sqrt{1-\widehat F(\bm{v}^T \widehat{\bm{\beta}}) }\Big\}            d F_V(\bm{v}) \nonumber\\
&=& 2 \int \left\{ \sqrt{F_0(\bm{v}^T \bm{\beta}_0)} -    \sqrt{\widehat F(\bm{v}^T \widehat{\bm{\beta}}) }      \right\}^2 d F_V(\bm{v}) \nonumber \\ &&+ \int \left\{ \sqrt{1-F_0(\bm{v}^T \bm{\beta}_0)} -    \sqrt{1-\widehat F(\bm{v}^T \widehat{\bm{\beta}}) }      \right\}^2 d F_V(\bm{v}) \nonumber\\
&\ge& 0.5 \int \left\{F_0(\bm{v}^T \bm{\beta}_0) - \widehat F(\bm{v}^T \widehat{\bm{\beta}}) \right\}^2 d F_V(\bm{v}),
 \label{lem-2-8}
\end{eqnarray}
where to derive the last ``$\ge$" we used the fact that $F_0(\cdot) \in [0,1]$ and $\widehat F(\cdot) \in [0,1]$.

On the other hand,
\begin{eqnarray*}
F_{V}(\bm{v}) &=& P(X_i-X_j\le \bm{v}) = E\left\{ P(X_i\le X_j+ \bm{v} | X_j) \right\}\\
&=& E\left\{ F_X(X_j+\bm{v}) \right\} = \int F_X(\bm{x}_2+\bm{v}) d F_X(\bm{x}_2).
\end{eqnarray*}
Therefore,
\begin{eqnarray}
&&\int \left\{F_0(\bm{v}^T \bm{\beta}_0) - \widehat F(\bm{v}^T \widehat{\bm{\beta}}) \right\}^2 d F_V(\bm{v}) \nonumber\\
&=&\int \int \left\{F_0(\bm{v}^T \bm{\beta}_0) - \widehat F(\bm{v}^T \widehat{\bm{\beta}}) \right\}^2 d F_X(\bm{x}_2+\bm{v}) d F_X(\bm{x}_2). \label{lem-2-9}
\end{eqnarray}
Setting $\bm{x}_1 = \bm{x}_2+\bm{v}$, (\ref{lem-2-9}) immediately implies
\begin{eqnarray*}
&&\int \left\{F_0(\bm{v}^T \bm{\beta}_0) - \widehat F(\bm{v}^T \widehat{\bm{\beta}}) \right\}^2 d F_V(\bm{v}) \nonumber\\
&=& \int \int \left\{F_0((\bm{x}_1-\bm{x}_2)^T \bm{\beta}_0) - \widehat F((\bm{x}_1-\bm{x}_2)^T \widehat{\bm{\beta}}) \right\}^2 d F_X(\bm{x}_1) d F_X(\bm{x}_2) \nonumber\\
&=& D^2\Big( \widehat{\bm{\beta}}, \widehat F; \bm{\beta}_0, F_0\Big),
\end{eqnarray*}
which together with (\ref{lem-2-8}) leads to (\ref{lem-2-4-1}), and therefore proves (\ref{lem-2-0}). \epf

 Lemma \ref{lemma-u-stat} below establishes the asymptotic order of the right-hand side of (\ref{lem-2-0}).

\begin{lemma} \label{lemma-u-stat}
Assume Conditions 1--3. We have
\begin{eqnarray}
&&\int \left\{I(u>0) \gamma_1(\bm{v}; \widehat F, \widehat{\bm{\beta}}) + I(u<0) \gamma_2(\bm{v}; \widehat F, \widehat{\bm{\beta}}) \right\}\nonumber\\
&& \hspace{2in}\times d\left\{\bbF_{U,V}(u,\bm{v}) - F_{U,V}(u,\bm{v}) \right\} \nonumber\\
&=&O_p(n^{-2/3})\vee\left\{O_p(n^{-1/2})D^{1/2}\Big( \widehat{\bm{\beta}}, \widehat F; \bm{\beta}_0, F_0\Big)  \right\}.
 \label{lem-3-1}
\end{eqnarray}
\end{lemma}

\proof
Recall that we have defined the following notations in Lemma \ref{lemma-basic}.
$F_{U,V}$ denotes  the joint c.d.f. of $(U_{i,j},V_{i,j})$ and likewise, $\bbF_{U,V}(u,\bm{v})$ denotes the empirical c.d.f. of $\{(U_{i,j},V_{i,j})\}_{i,j=1}^n$, where $U_{i,j}=Y_i-Y_j$ and $V_{i,j}=X_i-X_j$.

In the subsequent proof of this lemma, we need Lemma 3.4.2 in van der Vaart (1996), which is reviewed as follows.
\begin{lemma} \label{lemma-p-2}
Let $\Im$ be a class of functions such that for every $g\in \Im$, $\int g^2 dP< \widetilde \delta^2$ and $\|g\|_\infty < M$ for some $M>0$, and $Z_1, \ldots,Z_n$ be i.i.d. sample.  Then
\begin{eqnarray*}
\sup_{g\in \Im} \left|\int g d (\bbF_Z - F_Z) \right| \lesssim n^{-1/2} \widetilde J_{[]}(\widetilde \delta, \Im, L_2(F_Z))\left\{1+\frac{\widetilde J_{[]}(\widetilde \delta, \Im, L_2(F_Z))}{\widetilde \delta^2\sqrt{n}}M\right\},
\end{eqnarray*}
where
\begin{eqnarray*}
\widetilde J_{[]}(\widetilde \delta, \Im, L_2(F_Z)) = \int_0^{\widetilde \delta} \sqrt{1+H_{2,B}(\delta, \Im, F_Z)}.
\end{eqnarray*}
\end{lemma}

We now move back to the proof of Lemma \ref{lemma-u-stat}.  Note that we only need to show
\begin{eqnarray}
&&\int I(u>0) \gamma_1(\bm{v}; \widehat F, \widehat{\bm{\beta}})d\left\{\bbF_{U,V}(u,\bm{v}) - F_{U,V}(u,\bm{v}) \right\} \nonumber\\
&=&O_p(n^{-2/3})\vee\left\{O_p(n^{-1/2})D^{1/2}\Big( \widehat{\bm{\beta}}, \widehat F; \bm{\beta}_0, F_0\Big)  \right\}. \label{lem-3-1-1}
\end{eqnarray}
The same arguments can be applied to show
\begin{eqnarray}
&&\int I(u<0) \gamma_2(\bm{v}; \widehat F, \widehat{\bm{\beta}}) d\left\{\bbF_{U,V}(u,\bm{v}) - F_{U,V}(u,\bm{v})) \right\} \nonumber\\
&=&O_p(n^{-2/3})\vee\left\{O_p(n^{-1/2})D^{1/2}\Big( \widehat{\bm{\beta}}, \widehat F; \bm{\beta}_0, F_0\Big)  \right\}. \label{lem-3-1-2}
\end{eqnarray}
Then, combining (\ref{lem-3-1-1}) and (\ref{lem-3-1-2}) leads to (\ref{lem-3-1}).

To this end, note that
by the definition  of $\bbF_{U,V}(u,\bm{v})$,
\begin{eqnarray}
&&\int I(u>0) \gamma_1(\bm{v}; \widehat F, \widehat{\bm{\beta}}) d\bbF_{U,V}(u,\bm{v})\nonumber\\
&=& \frac{1}{n}\sum_{j=1}^n \frac{1}{n} \sum_{i=1}^n I(Y_i>Y_j) \gamma_1(X_i-X_j; \widehat F, \widehat{\bm{\beta}}) \nonumber \\
&=& \int \int I(y_1>y_2)\gamma_1(\bm{x}_1-\bm{x}_2; \widehat F, \widehat{\bm{\beta}}) d \bbF_{X,Y}(\bm{x}_1,y_1) d\bbF_{X,Y}(\bm{x}_2,y_2), \label{lem-3-3}
\end{eqnarray}
where $\bbF_{X,Y}(\bm{x},y)$ denotes the empirical c.d.f. of $\{(X_i,Y_i)\}_{i=1}^n$. Furthermore by the definition of $F_{U,V}(u,\bm{v})$, we have
\begin{eqnarray*}
F_{U,V}(u,\bm{v}) &=& P(Y_1-Y_2\le u, X_1-X_2 \le \bm{v}) \nonumber\\
&=& E\left\{ P(Y_1-Y_2\le u, X_1-X_2 \le \bm{v}| Y_2, X_2) \right\} \nonumber\\
&=& E\left\{ F_{X,Y}(X_2 + \bm{v}, Y_2 + u) \right\} \nonumber\\
&=& \int F_{X,Y}(\bm{x}_2 + \bm{v}, y_2 + u)  d F_{X,Y}(\bm{x}_2, y_2),
\end{eqnarray*}
which together with the transformation $u=y_1-y_2$, $\bm{v} = \bm{x}_1-\bm{x}_2$ immediately leads to
\begin{eqnarray}
&&\int I(u>0) \gamma_1(\bm{v}; \widehat F, \widehat{\bm{\beta}})dF_{U,V}(u,\bm{v}) \nonumber\\
&=&\int I(u>0) \gamma_1(\bm{v}; \widehat F, \widehat{\bm{\beta}})\int d F_{X,Y}(\bm{x}_2 + \bm{v}, y_2 + u)  d F_{X,Y}(\bm{x}_2, y_2) \nonumber \\
&=& \int\int I(y_1-y_2>0) \gamma_1(\bm{x}_1-\bm{x}_2; \widehat F, \widehat{\bm{\beta}})d F_{X,Y}(\bm{x}_1, y_1)  d F_{X,Y}(\bm{x}_2, y_2). \label{lem-3-3-1}
\end{eqnarray}
Combining (\ref{lem-3-3}) with (\ref{lem-3-3-1}), to show (\ref{lem-3-1-1}), we only need to show
\begin{eqnarray}
&&\int \int I(y_1>y_2)\gamma_1(\bm{x}_1-\bm{x}_2; \widehat F, \widehat{\bm{\beta}}) d \bbF_{X,Y}(\bm{x}_1,y_1) d\bbF_{X,Y}(\bm{x}_2,y_2) \nonumber\\
&& -  \int\int I(y_1-y_2>0) \gamma_1(\bm{x}_1-\bm{x}_2; \widehat F, \widehat{\bm{\beta}}) \nonumber \\  && \hspace{2in} \times d F_{X,Y}(\bm{x}_1, y_1)  d F_{X,Y}(\bm{x}_2, y_2) \nonumber\\
&=&O_p(n^{-2/3})\vee\left\{O_p(n^{-1/2})D^{1/2}\Big( \widehat{\bm{\beta}}, \widehat F; \bm{\beta}_0, F_0\Big)  \right\}. \label{lem-3-3-2}
\end{eqnarray}

We will use the following two steps to establish (\ref{lem-3-3-2}).
\begin{itemize}
\item In {\it step 1}, we show
\begin{eqnarray}
 \label{lem-3-3-3} &&\int \left|\int I(y_1>y_2)\gamma_1(\bm{x}_1-\bm{x}_2; \widehat F, \widehat{\bm{\beta}}) d \left\{\bbF_{X,Y}(\bm{x}_1,y_1)- F_{X,Y}(\bm{x}_1, y_1) \right\} \right|\\ && \hspace{2in} \times d\bbF_{X,Y}(\bm{x}_2,y_2) \nonumber\\
&=&O_p(n^{-2/3})\vee\left[O_p(n^{-1/2})\left\{D\Big( \widehat{\bm{\beta}}, \widehat F; \bm{\beta}_0, F_0\Big)+O_p(n^{-1/2})\right\}^{1/2}  \right].\nonumber
\end{eqnarray}
\item In {\it step 2}, we show
\begin{eqnarray}
\label{lem-3-3-4} &&\int\left\{\int I(y_1-y_2>0) \gamma_1(\bm{x}_1-\bm{x}_2; \widehat F, \widehat{\bm{\beta}})   d F_{X,Y}(\bm{x}_1, y_1) \right\} \\ && \hspace{2in} \times d( \bbF_{X,Y}(\bm{x}_1, y_1)- F_{X,Y}(\bm{x}_2, y_2) )\nonumber \\
&=& O_p(n^{-2/3})\vee\left\{O_p(n^{-1/2})D^{1/2}\Big( \widehat{\bm{\beta}}, \widehat F; \bm{\beta}_0, F_0\Big)  \right\}.\nonumber
\end{eqnarray}
\end{itemize}

We start with \underline{\it step 1}.
 Consider the class of functions
\begin{eqnarray}
\widetilde \mathG &=& \Big\{I(y_1>y_2)\gamma_1(\bm{x}_1-\bm{x}_2;  F, \bm{\beta}): \nonumber \\ && \hspace{0.5in} F\in \mathF, \bm{\beta} \in \mathB, y_2\in \mathY, \bm{x}_2\in \mathX \Big\} = \mathG_1\cdot \mathG_2,
\label{lem-3-4}
\end{eqnarray}
where
\begin{eqnarray*}
\mathG_1 = \{I(y_1>y_2): y_2 \in \mathY\}; \quad \mathG_2 = \{\gamma_1(\bm{x}_1-\bm{x}_2;  F, \bm{\beta}): F\in \mathF, \bm{\beta} \in \mathB,  \bm{x}_2\in \mathX\}. \nonumber \\
\label{lem-3-5}
\end{eqnarray*}
It is straightforward to check that there exists a universal constant $M>0$, such that
\begin{eqnarray}
\sup_{g\in \widetilde \mathG}|g|_\infty \le M. \label{lem-3-9}
\end{eqnarray}
We can also check
\begin{eqnarray}
H_{q,B}(\delta, \mathG_1, F_{X,Y})\lesssim -\log\delta. \label{lem-3-6}
\end{eqnarray}
 Furthermore, applying Lemma \ref{lemma-1} and Condition 3, we can easily check that
\begin{eqnarray}
H_{q,B}(\delta, \mathG_2, F_{X,Y})\lesssim 1/\delta. \label{lem-3-7}
\end{eqnarray}
Combining (\ref{lem-3-4}), (\ref{lem-3-6}), and (\ref{lem-3-7}), and applying Lemma 9.25 in Kosorok (2008), we can conclude that
\begin{eqnarray}
H_{q,B}(\delta, \widetilde \mathG, F_{X,Y})\lesssim 1/\delta.  \label{lem-3-8}
\end{eqnarray}
Now, let $\mathG = \frac{1}{M} \widetilde \mathG$. Then, we have
\begin{eqnarray*}
\sup_{g\in \mathG}|g|_\infty \le 1 \label{lem-3-10} \\
H_{q,B}(\delta, \mathG, F_{X,Y})\lesssim 1/\delta, \label{lem-3-11}
\end{eqnarray*}
which together with Lemma \ref{lemma-p-1} lead to
{\small
\begin{eqnarray}
P\left( \sup_{g\in \mathG(n^{-1/3})}\left| \int g(\bm{x}_1,y_1) d\left\{ \bbF_{X,Y}(\bm{x}_1,y_1)-F_{X,Y}(\bm{x}_1,y_1)\right\} \right|   \ge T n^{-2/3} \right) \nonumber \\ \le c \exp\left(-Tn^{1/3}/c^2\right), \label{lem-3-12}
\end{eqnarray}
and
\begin{eqnarray}
P\left( \sup_{g\in \mathG, \|g\|_{2, F_{X,Y}} > n^{-1/3}} \frac{\left| \int g(\bm{x}_1,y_1) d\left\{ \bbF_{X,Y}(\bm{x}_1,y_1)-F_{X,Y}(\bm{x}_1,y_1)\right\} \right|}{\|g\|_{2, F_{X,Y}}^{1/2}}  \ge T n^{-1/2}\right) \nonumber \\ \le c\exp\left(-T/c^2 \right), \label{lem-3-13}
\end{eqnarray}
}
for all $T\ge c$ and $n\ge n_0$, where $c$ and $n_0$ are constants.

Note that (\ref{lem-3-13}) leads to
{\small
\begin{eqnarray}
\label{lem-3-14}\\
&&1- c\exp\left(-T/c^2 \right) \nonumber \\
&<& P\left( \sup_{g\in \mathG, \|g\|_{2, F_{X,Y}} > n^{-1/3}} \frac{\left| \int g(\bm{x}_1,y_1) d\left\{ \bbF_{X,Y}(\bm{x}_1,y_1)-F_{X,Y}(\bm{x}_1,y_1)\right\} \right|}{\|g\|_{2, F_{X,Y}}^{1/2}}< T n^{-1/2}\right)\nonumber\\
&\leq&P\left( \bigcap_{i=1}^n \left\{ \sup_{g\in \mathG_{X_i,Y_i}, \|g\|_{2, F_{X,Y}} > n^{-1/3}} \frac{\left| \int g(\bm{x}_1,y_1) d\left\{ \bbF_{X,Y}(\bm{x}_1,y_1)-F_{X,Y}(\bm{x}_1,y_1)\right\} \right|}{\|g\|_{2, F_{X,Y}}^{1/2}}< T n^{-1/2} \right\} \right)\nonumber\\
&\leq& P\left( \bigcap_{i=1}^n \left\{ \frac{\left| \int \widehat g_i(\bm{x}_1,y_1) d\left\{ \bbF_{X,Y}(\bm{x}_1,y_1)-F_{X,Y}(\bm{x}_1,y_1)\right\} \right|I\left( \|\widehat g_i\|_{2, F_{X,Y}} > M n^{-1/3}\right)}{\|\widehat g_i\|_{2, F_{X,Y}}^{1/2}}< T n^{-1/2} \right\} \right) \nonumber\\
&\le& P\left(\frac{\frac{1}{n} \sum_{i=1}^n \left| \int \widehat g_i(\bm{x}_1,y_1) d\left\{ \bbF_{X,Y}(\bm{x}_1,y_1)-F_{X,Y}(\bm{x}_1,y_1)\right\} \right|I\left( \|\widehat g_i\|_{2, F_{X,Y}} > M n^{-1/3}\right)}{\frac{1}{n} \sum_{i=1}^n\|\widehat g_i\|_{2, F_{X,Y}}^{1/2}}< T n^{-1/2}  \right),\nonumber
\end{eqnarray}
}
where
\begin{eqnarray}
\mathG_{X_i,Y_i} &=& \frac{1}{M} \Big\{I(y_1>Y_i)\gamma_1(\bm{x}_1-X_i;  F, \bm{\beta}): F\in \mathF, \bm{\beta} \in \mathB \Big\} \nonumber\\
\widehat g_i(\bm{x},y) &=& \frac{1}{M}  I(y>Y_i)\gamma_1(\bm{x}-X_i;  \widehat F, \widehat{\bm{\beta}}). \label{lem-3-15}
\end{eqnarray}

Applying the similar arguments on (\ref{lem-3-12}) leads to
\begin{eqnarray}
 \label{lem-3-16}&&1-c \exp\left(-Tn^{1/3}/c^2\right)
\\&<& P\bigg( \frac{1}{n}\sum_{i=1}^n\left| \int \widehat g_i(\bm{x}_1,y_1) d\left\{ \bbF_{X,Y}(\bm{x}_1,y_1)-F_{X,Y}(\bm{x}_1,y_1)\right\}\right| \nonumber \\ && \hspace{1in} \times I\left( \|\widehat g_i\|_{2, F_{X,Y}} \le M n^{-1/3}\right) < T n^{-2/3}  \bigg).\nonumber
\end{eqnarray}

Combining (\ref{lem-3-14}) and (\ref{lem-3-16}), we have
\begin{eqnarray*}
&&P\bigg(\frac{1}{n}\sum_{i=1}^n\left| \int \widehat g_i(\bm{x}_1,y_1) d\left\{ \bbF_{X,Y}(\bm{x}_1,y_1)-F_{X,Y}(\bm{x}_1,y_1)\right\}\right| \nonumber \\ && \hspace{ 2in}   < Tn^{-2/3} + Tn^{-1/2}\frac{1}{n}\sum_{i=1}^n \|\widehat g_i\|_{2, F_{X,Y}}^{1/2}\bigg)\\
&>& 1-c \exp\left(-Tn^{1/3}/c^2\right) - c\exp\left(-T/c^2 \right),
\end{eqnarray*}
which together the definition of $\widehat g_i$ in (\ref{lem-3-15}) indicates
\begin{eqnarray}
\label{lem-3-17} && \int \left| \int I(y_1>y_2)\gamma_1(\bm{x}_1-\bm{x}_2;  \widehat F, \widehat{\bm{\beta}}) d\left\{ \bbF_{X,Y}(\bm{x}_1,y_1)-F_{X,Y}(\bm{x}_1,y_1)\right\}\right|  \\ &&\hspace{3.2in} \times d\bbF_{X, Y}(\bm{x}_2, y_2) \nonumber \\ &=& O_p(n^{-2/3})\vee O_p(n^{-1/2}) \frac{1}{n}\sum_{i=1}^n \|I(y_1>Y_i)\gamma_1(\bm{x}_1-X_i;  \widehat F, \widehat{\bm{\beta}})\|_{2, F_{X,Y}}^{1/2}. \nonumber
\end{eqnarray}

Next, we shall study the convergence of
\begin{eqnarray}
J_0 = \frac{1}{n}\sum_{i=1}^n \|I(y_1>Y_i)\gamma_1(\bm{x}_1-X_i;  \widehat F, \widehat{\bm{\beta}})\|_{2, F_{X,Y}}^{1/2}. \label{lem-3-18}
\end{eqnarray}
Applying Condition 3,  we have
\begin{eqnarray*}
&&\|I(y_1>Y_i)\gamma_1(\bm{x}_1-X_i;  \widehat F, \widehat{\bm{\beta}})\|_{2, F_{X, Y}}^2   \nonumber \\ &=& \int I(y_1>Y_i)\gamma_1^2(\bm{x}_1-X_i;  \widehat F, \widehat{\bm{\beta}}) dF_{X, Y}(\bm{x}_1, y_1)\nonumber\\
&\le&\int \gamma_1^2(\bm{x}_1-X_i;  \widehat F, \widehat{\bm{\beta}}) dF_X(\bm{x}_1) = 16 \int \left\{ 1 - \sqrt{\frac{\widehat F((\bm{x}_1-X_i)^T \widehat{\bm{\beta}})}{F_0((\bm{x}_1-X_i)^T \bm{\beta}_0)}} \right\}^2 dF_X(\bm{x}_1)\\
&\lesssim& \int \left\{ \sqrt{F_0((\bm{x}_1-X_i)^T \bm{\beta}_0)} - \sqrt{\widehat F((\bm{x}_1-X_i)^T \widehat{\bm{\beta}})} \right\}^2 dF_X(\bm{x}_1)\\
&=& \int \left\{ \frac{F_0((\bm{x}_1-X_i)^T \bm{\beta}_0) - \widehat F((\bm{x}_1-X_i)^T \widehat{\bm{\beta}})}{\sqrt{F_0((\bm{x}_1-X_i)^T \bm{\beta}_0)} + \sqrt{\widehat F((\bm{x}_1-X_i)^T \widehat{\bm{\beta}})}} \right\}^2dF_X(\bm{x}_1)\nonumber\\
&\lesssim& \int \left\{ F_0((\bm{x}_1-X_i)^T \bm{\beta}_0) - \widehat F((\bm{x}_1-X_i)^T \widehat{\bm{\beta}}) \right\}^2dF_X(\bm{x}_1) \\
&=& d_{\widehat{\bm{\beta}}, \widehat F}^2(X_i), \label{lem-3-19}
\end{eqnarray*}
which together with the fact that $x^{1/2}$ for any $x>0$ is a strictly concave function leads to
\begin{eqnarray}
0\le J_0 &\le&  \left\{ \frac{1}{n}\sum_{i=1}^n\|I(y_1>Y_i)\gamma_1(\bm{x}_1-X_i;  \widehat F, \widehat{\bm{\beta}})\|_{2, F_{X,Y}}\right\}^{1/2}\nonumber\\
&\lesssim& \left\{ \frac{1}{n}\sum_{i=1}^n d_{\widehat{\bm{\beta}}, \widehat F}(X_i) \right\}^{1/2}. \label{lem-3-20}
\end{eqnarray}
Note that $d_{\widehat{\bm{\beta}}, \widehat F}(\bm{x}) \in \mathD$. Incorporating (\ref{lem-1-1-2}) with Lemma \ref{lemma-p-2} leads to
\begin{eqnarray}
\label{lem-3-21} &&\frac{1}{n}\sum_{i=1}^n d_{\widehat{\bm{\beta}}, \widehat F}(X_i) \\ &=& \int d_{\widehat{\bm{\beta}}, \widehat F}(\bm{x}_2) dF_X(\bm{x}_2) + O_p(n^{-1/2})\nonumber\\
&=& \int \left[\int \left\{ F_0((\bm{x}_1-\bm{x}_2)^T \bm{\beta}_0) - \widehat F((\bm{x}_1-\bm{x}_2)^T \widehat{\bm{\beta}}) \right\}^2dF_X(\bm{x}_1)\right]^{1/2} d F_X(\bm{x}_2)\nonumber\\
&& +O_p(n^{-1/2}) \nonumber \\
&\le& \left[ \int \int \left\{ F_0((\bm{x}_1-\bm{x}_2)^T \bm{\beta}_0) - \widehat F((\bm{x}_1-\bm{x}_2)^T \widehat{\bm{\beta}}) \right\}^2dF_X(\bm{x}_1) d F_X(\bm{x}_2)\right]^{1/2}\nonumber\\
&& +O_p(n^{-1/2}) \nonumber \\
&=& D\Big( \widehat{\bm{\beta}}, \widehat F; \bm{\beta}_0, F_0\Big) + O_p(n^{-1/2}). \nonumber
\end{eqnarray}

Combining (\ref{lem-3-17})--(\ref{lem-3-21}) leads to
\begin{eqnarray*}
&&\int \left|\int I(y_1>y_2)\gamma_1(\bm{x}_1-\bm{x}_2; \widehat F, \widehat{\bm{\beta}}) d \left\{\bbF_{X,Y}(\bm{x}_1,y_1)- F_{X,Y}(\bm{x}_1, y_1) \right\} \right|d\bbF_{X,Y}(\bm{x}_2,y_2) \nonumber\\
&=&O_p(n^{-2/3})\vee\left[O_p(n^{-1/2})\left\{D\Big( \widehat{\bm{\beta}}, \widehat F; \bm{\beta}_0, F_0\Big)+O_p(n^{-1/2})\right\}^{1/2}  \right],
\end{eqnarray*}
which proves (\ref{lem-3-3-3}) and therefore completes our proof in \underline{\it step 1}.

We proceed to show (\ref{lem-3-3-4}) claimed in \underline{\it step 2}.
Let \begin{eqnarray}
 \label{lem-3-23} \\
\widetilde \mathG_I &=& \left\{\int g_{\bm{\beta}, F}(\bm{x}_1, \bm{x}_2, y_1, y_2) d F_{X,Y}(\bm{x}_1, y_1): g_{\bm{\beta}, F}(\bm{x}_1, \bm{x}_2, y_1, y_2)\in \mathG_{I,1}  \right\}\nonumber
\end{eqnarray}
where
\begin{eqnarray*}
\mathG_{I,1} &=& \left\{g_{\bm{\beta}, F}(\bm{x}_1, \bm{x}_2, y_1, y_2) =I(y_1>y_2)\gamma_1(\bm{x}_1-\bm{x}_2;  F, \bm{\beta}): F\in \mathF, \bm{\beta} \in \mathB \right\}\\
&=& I(y_1>y_2)\cdot \mathG_{I,2},\\
\mathG_{I,2} &=& \{\gamma_1(\bm{x}_1-\bm{x}_2;  F, \bm{\beta}): F\in \mathF, \bm{\beta} \in \mathB\}.\nonumber \\
 \label{lem-3-5}
\end{eqnarray*}
Based on Condition 3, it is easy to check that there exists a universal constant $M>0$, such that
\begin{eqnarray}
\sup_{g\in \widetilde \mathG_I} |g|_\infty <M. \label{lem-3-24}
\end{eqnarray}
Define $\mathG_I = \frac{1}{M} \widetilde \mathG_I$. With (\ref{lem-1-1-1}) in Lemma \ref{lemma-1}
and Condition 3, it is straightforward to show that
\begin{eqnarray*}
H_{q,B}\left(\delta, \frac{1}{M}\mathG_{I,1}, F_{X_1, X_2, Y_1, Y_2}\right) \lesssim 1/\delta,
\end{eqnarray*}
where $F_{X_1, X_2, Y_1, Y_2}(\bm{x_1}, \bm{x_2}, y_1, y_2)$ denotes the joint c.d.f. of $(X_1^T, X_2^T, Y_1, Y_2)^T$. Therefore there exists a set of $\delta$-brackets $\{[l_i(\bm{x}_1, \bm{x}_2, y_1, y_2), u_i(\bm{x}_1, \bm{x}_2, y_1, y_2)]_{i=1}^{N_4}\}$ under $L_q(F_{X_1, X_2, Y_1, Y_2})$ that cover $\frac{1}{M}\mathG_{I,1}$, where $N_4 \lesssim 1/\delta$. Consider the set of brackets
\begin{eqnarray}
\label{lem-3-25}&&\bigg\{[L_i(\bm{x}_2, y_2), U_i(\bm{x}_2, y_2)]: \\ && \begin{array}{l} L_i(\bm{x}_2, y_2) = \int l_i(\bm{x}_1, \bm{x}_2, y_1, y_2) d F_{X,Y}(\bm{x}_1,y_1); \\
U_i(\bm{x}_2, y_2) = \int u_i(\bm{x}_1, \bm{x}_2, y_1, y_2) d F_{X,Y}(\bm{x}_1,y_1);\end{array}  i=1,\ldots,N_4 \bigg\}. \nonumber
\end{eqnarray}
Clearly, for every $\frac{1}{M}\int g_{\bm{\beta}, F}(\bm{x}_1, \bm{x}_2, y_1, y_2) d F_{X,Y}(\bm{x}_1, y_1)\in \mathG_I$, there exists an $i\in \{1,\ldots,N_4\}$ such that
\begin{equation}
L_i(\bm{x}_2, y_2)\le \frac{1}{M}\int g_{\bm{\beta}, F}(\bm{x}_1, \bm{x}_2, y_1, y_2) d F_{X,Y}(\bm{x}_1, y_1) \le U_i(\bm{x}_2, y_2),
 \label{lem-3-26}
\end{equation}
which indicates the set of brackets given in (\ref{lem-3-25}) covers $\mathG_I$, with bracket length
{\small
\begin{eqnarray}
\label{lem-3-27} &&\\
&&\left[\int \left\{U_i(\bm{x}_2, y_2) - L_i(\bm{x}_2, y_2)\right\}^q d F_{X,Y}(\bm{x}_2, y_2)\right]^{1/q} \nonumber\\
&=& \left[\int \left\{\int \left(u_i(\bm{x}_1, \bm{x}_2, y_1, y_2)  - l_i(\bm{x}_1, \bm{x}_2, y_1, y_2)\right) d F_{X,Y}(\bm{x}_1,y_1)\right\}^q d F_{X,Y}(\bm{x}_2, y_2)\right]^{1/q} \nonumber\\
&\le& \left[\int \int \left\{u_i(\bm{x}_1, \bm{x}_2, y_1, y_2)  - l_i(\bm{x}_1, \bm{x}_2, y_1, y_2)\right\}^q d F_{X,Y}(\bm{x}_1,y_1) d F_{X,Y}(\bm{x}_2, y_2)\right]^{1/q} \nonumber\\
&=& \|u_i-l_i\|_{q, F_{X_1, X_2, Y_1, Y_2}} \le \delta.\nonumber
\end{eqnarray}
}
(\ref{lem-3-26}) and (\ref{lem-3-27}) imply that the set of brackets given in (\ref{lem-3-25}) are $\delta$-brackets under $L_q(F_{X,Y})$ that cover $\mathG_I$. Therefore
\begin{eqnarray*}
H_{q,B}(\delta, \mathG_I, F_{X,Y}) \lesssim 1/\delta, \label{lem-3-28}
\end{eqnarray*}
which together with the fact in (\ref{lem-3-24}) and Lemma \ref{lemma-p-1} leads to
\begin{eqnarray*}
P\left( \sup_{g\in \mathG_I(n^{-1/3})}\left| \int g(\bm{x}_2,y_2) d\left\{ \bbF_{X,Y}(\bm{x}_2,y_2)-F_{X,Y}(\bm{x}_2,y_2)\right\} \right| \ge T n^{-2/3} \right) \\ \le c \exp\left(-Tn^{1/3}/c^2\right),
\end{eqnarray*}
and
\begin{eqnarray*}
P\left( \sup_{g\in \mathG_I, \|g\|_{2, F_{X,Y}} > n^{-1/3}} \frac{\left| \int g(\bm{x}_2,y_2) d\left\{ \bbF_{X,Y}(\bm{x}_2,y_2)-F_{X,Y}(\bm{x}_2,y_2)\right\} \right|}{\|g\|_{2, F_{X,Y}}^{1/2}} \ge T n^{-1/2}\right) \\ \le c\exp\left(-T/c^2 \right),
\end{eqnarray*}
for all $T\ge c$ and $n\ge n_0$, where $c$ and $n_0$ are constants. This together with the fact that
\begin{eqnarray*}
\int I(y_1>y_2)\gamma_1(\bm{x}_1-\bm{x}_2;  \widehat F, \widehat{\bm{\beta}}) d F_{X,Y}(\bm{x}_1,y_1) \in \mathG_I
\end{eqnarray*}
immediately implies
\begin{eqnarray}
\label{lem-3-29} &&\int \left\{\int I(y_1>y_2)\gamma_1(\bm{x}_1-\bm{x}_2;  \widehat F, \widehat{\bm{\beta}}) d F_{X,Y}(\bm{x}_1,y_1)\right\} \\ && \hspace{2in} \times d\left\{\bbF_{X,Y}(\bm{x}_2, y_2)-F_{X,Y}(\bm{x}_2, y_2) \right\} \nonumber \\
&=& O_p(n^{-2/3})\vee O_p(n^{-1/2}) \nonumber \\ && \hspace{0.5in}\times  \left\|\int I(y_1>y_2)\gamma_1(\bm{x}_1-\bm{x}_2;  \widehat F, \widehat{\bm{\beta}}) d F_{X,Y}(\bm{x}_1,y_1)\right\|_{2, F_{X,Y}}^{1/2}. \nonumber
\end{eqnarray}
On the other hand, with Condition 3,
\begin{eqnarray}
 \label{lem-3-29-1} &&\left\|\int I(y_1>y_2)\gamma_1(\bm{x}_1-\bm{x}_2;  \widehat F, \widehat{\bm{\beta}}) d F_{X,Y}(\bm{x}_1,y_1)\right\|_{2, F_{X,Y}}^2  \\
&=&\int \left\{\int I(y_1>y_2)\gamma_1(\bm{x}_1-\bm{x}_2;  \widehat F, \widehat{\bm{\beta}}) d F_{X,Y}(\bm{x}_1,y_1) \right\}^2 dF_{X,Y}(\bm{x}_2, y_2)\nonumber\\
&\le& \int \int I(y_1>y_2)\gamma_1^2(\bm{x}_1-\bm{x}_2;  \widehat F, \widehat{\bm{\beta}}) d F_{X,Y}(\bm{x}_1,y_1) dF_{X,Y}(\bm{x}_2, y_2)\nonumber\\
&\le& \int \int \gamma_1^2(\bm{x}_1-\bm{x}_2;  \widehat F, \widehat{\bm{\beta}}) d F_X(\bm{x}_1) dF_X(\bm{x}_2) \nonumber\\
&=&16 \int \int \left\{ 1 - \sqrt{\frac{\widehat F((\bm{x}_1-\bm{x}_2)^T \widehat{\bm{\beta}})}{F_0((\bm{x}_1-\bm{x}_2)^T \bm{\beta}_0)}} \right\}^2 dF_X(\bm{x}_1) dF_X(\bm{x}_2) \nonumber \\
&\lesssim& \int \int \left\{ \sqrt{F_0((\bm{x}_1-\bm{x}_2)^T \bm{\beta}_0)} - \sqrt{\widehat F((\bm{x}_1-\bm{x}_2)^T \widehat{\bm{\beta}})} \right\}^2 dF_X(\bm{x}_1)dF_X(\bm{x}_2) \nonumber \\
&=& \int \int \left\{ \frac{F_0((\bm{x}_1-\bm{x}_2)^T \bm{\beta}_0) - \widehat F((\bm{x}_1-\bm{x}_2)^T \widehat{\bm{\beta}})}{\sqrt{F_0((\bm{x}_1-\bm{x}_2)^T \bm{\beta}_0)} + \sqrt{\widehat F((\bm{x}_1-\bm{x}_2)^T \widehat{\bm{\beta}})}} \right\}^2dF_X(\bm{x}_1)dF_X(\bm{x}_2)\nonumber\\
&\lesssim& \int \int \left\{ F_0((\bm{x}_1-\bm{x}_2)^T \bm{\beta}_0) - \widehat F((\bm{x}_1-\bm{x}_2)^T \widehat{\bm{\beta}}) \right\}^2dF_X(\bm{x}_1)dF_X(\bm{x}_2)\nonumber \\
&=& D^2\Big( \widehat{\bm{\beta}}, \widehat F; \bm{\beta}_0, F_0\Big), \nonumber
\end{eqnarray}
which together with (\ref{lem-3-29}) leads to (\ref{lem-3-3-4}), and therefore completes our proof of \underline{\it step 2}. The proof of Lemma \ref{lemma-u-stat} is completed. \epf

If we combine Lemmas \ref{lemma-basic} and \ref{lemma-u-stat}, the result claimed in Part (a) follows immediately. We proceed to show Part (b). We first establish the consistency of $\widehat{\bm{\beta}}$.

\begin{lemma} \label{lemma-7}
Assume Conditions 1--4. We have
\begin{eqnarray}
\widehat{\bm{\beta}}-\bm{\beta}_0 = o_p(1). \label{thm-1-1-1}
\end{eqnarray}
\end{lemma}

\proof Let
\begin{eqnarray*}
M^*(\bm{\beta}) = \inf_{F\in\mathF} D(\bm{\beta}; F, \bm{\beta}_0, F_0).
\end{eqnarray*}
Then, from the arguments in Wald (1949), to show (\ref{thm-1-1-1}), we need to show only that
\begin{itemize}
\item[(i)] $M^*(\widehat{\bm{\beta}}) = o_p(1)$;
\item[(ii)] $M^*(\bm{\beta}) = 0$ implies that $\bm{\beta} = \bm{\beta}_0$;
\item[(iii)] $M^*(\bm{\beta})$ is continuous in $\mathB$.
\end{itemize}
Part (i) is easily obtained from Part (a) of this theorem, since
\begin{eqnarray*}
0\le M^*(\widehat{\bm{\beta}}) \le D(\widehat{\bm{\beta}}, \widehat F; \bm{\beta}_0, F_0) =o_p(1).
\end{eqnarray*}
Part (ii) holds because of Condition 4. We now show Part (iii).  Similarly to the proof of (\ref{lem-1-1}) in Lemma \ref{lemma-1}, for an arbitrary $\delta_1>0$ and $\bm{\beta}^*$ there exists a set of brackets $[L_i(\bm{x}_1-\bm{x}_2), U_i(\bm{x}_1-\bm{x}_2)]_{i=1}^{N^*}$ satisfying
\begin{equation}
\left[\int\int\left\{U_i(\bm{x}_1-\bm{x}_2) - L_i(\bm{x}_1-\bm{x}_2)\right\}^2 d F_X(\bm{x}_1) d F_X(\bm{x}_2)\right]^{1/2} \le C^* \delta_1,
\label{thm-1-1-2}
\end{equation}
where $C^*$ is a universal constant.
Moreover, for any $F\in \mathF$, there exists an $i \in \{1, \ldots, N^*\}$ such that
\begin{eqnarray}
&&0\le L_i(\bm{x}_1-\bm{x}_2) \le F((\bm{x}_1-\bm{x}_2)^T \bm{\beta}) \le U_i(\bm{x}_1-\bm{x}_2)\le 1, \quad \mbox{and} \nonumber\\
&& 0\le L_i(\bm{x}_1-\bm{x}_2) \le F((\bm{x}_1-\bm{x}_2)^T \bm{\beta}^*) \le U_i(\bm{x}_1-\bm{x}_2)\le 1, \label{thm-1-1-3}
\end{eqnarray}
for any
$\|\bm{\beta}-\bm{\beta}^*\|_1\le \delta_1$. Now for any $\epsilon>0$, we set $\delta_1 = \epsilon/C^*$. Then when $\|\bm{\beta}-\bm{\beta}^*\|_1\le \delta_1$, we have
\begin{eqnarray}
D(\bm{\beta}, F; \bm{\beta}_0, F_0) \ge D(\bm{\beta}^*, F; \bm{\beta}_0, F_0) - D(\bm{\beta}^*, F; \bm{\beta}, F), \label{thm-1-1-4}
\end{eqnarray}
from the triangle inequality. On the other hand, from (\ref{thm-1-1-2}), (\ref{thm-1-1-3}), and the definition of $D(\bm{\beta}^*, F; \bm{\beta}, F)$, we have
\begin{eqnarray*}
&&D(\bm{\beta}^*, F; \bm{\beta}, F) \nonumber \\  &=&\left[\int \int \left\{F((\bm{x}_1-\bm{x}_2)^T \bm{\beta}) -F((\bm{x}_1-\bm{x}_2)^T \bm{\beta}^*) \right\}^2 d F_{X}(\bm{x}_1)d F_{X}(\bm{x}_2)\right]^{1/2} \nonumber\\
&\le &\left[\int \int\left\{U_i(\bm{x}_1-\bm{x}_2) - L_i(\bm{x}_1-\bm{x}_2)\right\}^2 d F_X(\bm{x}_1) d F_X(\bm{x}_2)\right]^{1/2} \nonumber \\
&\le& C^* \delta_1 =\epsilon,
\end{eqnarray*}
which together with (\ref{thm-1-1-4}) implies
\begin{eqnarray*}
D(\bm{\beta}, F; \bm{\beta}_0, F_0) \ge D(\bm{\beta}^*, F; \bm{\beta}_0, F_0) - \epsilon.
\end{eqnarray*}
Therefore,
\begin{eqnarray}
M^*(\bm{\beta}) &=& \inf_{F\in\mathF} D(\bm{\beta}; F, \bm{\beta}_0, F_0) \nonumber\\ &\ge& \inf_{F\in\mathF}D(\bm{\beta}^*, F; \bm{\beta}_0, F_0) - \epsilon = M^*(\bm{\beta}^*)-\epsilon.
 \label{thm-1-1-5}
\end{eqnarray}
Using similar arguments, we can show that
\begin{eqnarray}
M^*(\bm{\beta})\le M^*(\bm{\beta}^*)+\epsilon. \label{thm-1-1-6}
\end{eqnarray}
Combining (\ref{thm-1-1-5}) and (\ref{thm-1-1-6}), we prove Part (iii). This completes the proof of this lemma.
\epf

 Recall that in the proof of Lemma \ref{lemma-basic}, we have shown that
\begin{eqnarray}
D^2(\bm{\beta}; F, \bm{\beta}_0, F_0) =  \int \left\{F_0(\bm{v}^T \bm{\beta}_0) - \widehat F(\bm{v}^T \widehat{\bm{\beta}}) \right\}^2 d F_V(\bm{v}), \label{thm-1-1}
\end{eqnarray}
where $F_V(\bm{v})$ is the c.d.f. of $X_i-X_j$ for $i\neq j$. Let $V$ be a random variable independent of but sharing the same distribution with $X_i-X_j$, and let $\psi(\bm{v}^T \bm{\beta}) = E(F_0(V^T \bm{\beta}_0)|V^T\bm{\beta} = \bm{v}\bm{\beta})$.  Combining (\ref{thm-1-1}) with the results in Part (a) leads to
\begin{eqnarray}
 \label{thm-1-2} &&O_p(n^{-2/3}) = \int \left\{F_0(\bm{v}^T \bm{\beta}_0) - \widehat F(\bm{v}^T \widehat{\bm{\beta}}) \right\}^2 d F_V(\bm{v}) \\
&=& \int \left\{F_0(\bm{v}^T\bm{\beta}_0) - \psi(\bm{v}^T \widehat{\bm{\beta}})\right\}^2 dF_V(\bm{v}) + \int \left\{ \psi(\bm{v}^T \widehat{\bm{\beta}}) - \widehat F(\bm{v}^T \widehat{\bm{\beta}})\right\}^2 dF_V(\bm{v}) \nonumber \\
&& +  \int \left\{F_0(\bm{v}^T\bm{\beta}_0) - \psi(\bm{v}^T \widehat{\bm{\beta}})\right\} \left\{ \psi(\bm{v}^T \widehat{\bm{\beta}}) - \widehat F(\bm{v}^T \widehat{\bm{\beta}})\right\} dF_V(\bm{v})\nonumber\\
&\geq& \int \left\{F_0(\bm{v}^T\bm{\beta}_0) - \psi(\bm{v}^T \widehat{\bm{\beta}})\right\}^2 dF_V(\bm{v})\nonumber\\
&=& E\left[\left\{F_0(V^T\bm{\beta}_0) - \psi(V^T \widehat{\bm{\beta}})\right\}^2\Big| \widehat{\bm{\beta}}\right],  \nonumber
\end{eqnarray}
where the ``$\geq$" is because
{\small
\begin{eqnarray*}
&&\int \left\{F_0(\bm{v}^T\bm{\beta}_0) - \psi(\bm{v}^T \widehat{\bm{\beta}})\right\} \left\{ \psi(\bm{v}^T \widehat{\bm{\beta}}) - \widehat F(\bm{v}^T \widehat{\bm{\beta}})\right\} dF_V(\bm{v}) \nonumber\\
&=& E\left[\left\{F_0(V^T\bm{\beta}_0) - \psi(V^T \widehat{\bm{\beta}})\right\} \left\{ \psi(V^T \widehat{\bm{\beta}}) - \widehat F(V^T \widehat{\bm{\beta}})\right\}\bigg| \{(X_i,Y_i)\}_{i=1}^n \right] \nonumber\\
&=& E\left[E\left[\left\{F_0(V^T\bm{\beta}_0) - \psi(V^T \widehat{\bm{\beta}})\right\} \left\{ \psi(V^T \widehat{\bm{\beta}}) - \widehat F(V^T \widehat{\bm{\beta}})\right\}\bigg| \{(X_i,Y_i)\}_{i=1}^n, V^T\widehat{\bm{\beta}} \right]\bigg| \{(X_i,Y_i)\}_{i=1}^n\right] \\
&=& E\left[\left\{ \psi(V^T \widehat{\bm{\beta}}) - \widehat F(V^T \widehat{\bm{\beta}})\right\}E\left[\left\{F_0(V^T\bm{\beta}_0) - \psi(V^T \widehat{\bm{\beta}})\right\} \bigg| \{(X_i,Y_i)\}_{i=1}^n, V^T\widehat{\bm{\beta}} \right]\bigg| \{(X_i,Y_i)\}_{i=1}^n\right]\\
&=& E\left[\left\{ \psi(V^T \widehat{\bm{\beta}}) - \widehat F(V^T \widehat{\bm{\beta}})\right\}E\left[\left\{F_0(V^T\bm{\beta}_0) - \psi(V^T \widehat{\bm{\beta}})\right\} \bigg| V^T\widehat{\bm{\beta}} \right]\bigg| \{(X_i,Y_i)\}_{i=1}^n\right]\\
&=& E\left[\left\{ \psi(V^T \widehat{\bm{\beta}}) - \widehat F(V^T \widehat{\bm{\beta}})\right\}\left\{\psi(V^T \widehat{\bm{\beta}}) - \psi(V^T \widehat{\bm{\beta}})\right\}  \bigg| \{(X_i,Y_i)\}_{i=1}^n\right] = 0.
\end{eqnarray*}
}
Furthermore, let
\begin{eqnarray*}
R(V, \widehat{\bm{\beta}}) = F_0(V^T\bm{\beta}_0) - F_0(V^T\widehat{\bm{\beta}}) - F_0'(V^T\widehat{\bm{\beta}})V^T(\bm{\beta}_0 - \widehat{\bm{\beta}}).
\end{eqnarray*}
Then
\begin{eqnarray*}
|R(V, \widehat{\bm{\beta}})| \le 0.5\sup_{u\in \bbR}|F_0''(u)|\sup_{\bm{v} \in \mathX-\mathX}\|\bm{v}\|_2^2\|\widehat{\bm{\beta}} - \bm{\beta}_0\|^2\lesssim \|\widehat{\bm{\beta}} - \bm{\beta}_0\|^2.
\end{eqnarray*}
Hence, there exists a constant $C_R>0$ such that
\begin{eqnarray}
\label{https://urldefense.proofpoint.com/v2/url?u=http-3A__upper.cr&d=DwIGAg&c=bKRySV-ouEg_AT-w2QWsTdd9X__KYh9Eq2fdmQDVZgw&r=i_cgqzHIka1v417jAcDgz0IaIOFsFESIEvu690tlsDs&m=Fp5s2fQN5ZzgsskonkXk7ibavWeWsly3wAMupYGOTNU&s=BCTd9BGKbY65REYJpVe66otdmiWgEqyVAfP8mcrufes&e=}
|R(V, \widehat{\bm{\beta}})| \le C_R \|\widehat{\bm{\beta}} - \bm{\beta}_0\|^2.
\end{eqnarray}
Noting (\ref{https://urldefense.proofpoint.com/v2/url?u=http-3A__upper.cr&d=DwIGAg&c=bKRySV-ouEg_AT-w2QWsTdd9X__KYh9Eq2fdmQDVZgw&r=i_cgqzHIka1v417jAcDgz0IaIOFsFESIEvu690tlsDs&m=Fp5s2fQN5ZzgsskonkXk7ibavWeWsly3wAMupYGOTNU&s=BCTd9BGKbY65REYJpVe66otdmiWgEqyVAfP8mcrufes&e=}) and Condition 5, we have
\begin{eqnarray}
\label{thm-1-3} &&E\left[\left\{F_0(V^T\bm{\beta}_0) - \psi(V^T \widehat{\bm{\beta}})\right\}^2\Big| \widehat{\bm{\beta}}\right]\\
&=& E\bigg[\Big\{F_0'(V^T\widehat{\bm{\beta}})(\bm{\beta}_0 - \widehat{\bm{\beta}})^T \left( V- E(V|V^T\widehat{\bm{\beta}}, \widehat{\bm{\beta}})\right)  \nonumber \\
&&\hspace{1in} + R(V, \widehat{\bm{\beta}}) - E(R(V,\widehat{\bm{\beta}})|V^T\widehat{\bm{\beta}}, \widehat{\bm{\beta}}) \Big\}^2                \Big| \widehat{\bm{\beta}}\bigg]\nonumber\\
&\ge& 0.5 E\left[\left\{F_0'(V^T\widehat{\bm{\beta}})(\bm{\beta}_0 - \widehat{\bm{\beta}})^T \left( V- E(V|V^T\widehat{\bm{\beta}}, \widehat{\bm{\beta}})\right)\right\}^2\Big| \widehat{\bm{\beta}}\right] \nonumber\\ && -  E\left[ \left\{R(V, \widehat{\bm{\beta}}) - E(R(V,\widehat{\bm{\beta}})|V^T\widehat{\bm{\beta}}, \widehat{\bm{\beta}}) \right\}^2                \Big| \widehat{\bm{\beta}}\right] \nonumber\\
&\gtrsim& (\bm{\beta}_0 - \widehat{\bm{\beta}})^T E\left[\left( V- E(V|V^T\widehat{\bm{\beta}}, \widehat{\bm{\beta}})\right)\left( V- E(V|V^T\widehat{\bm{\beta}}, \widehat{\bm{\beta}})\right)^T \Big|\widehat{\bm{\beta}}\right] \nonumber \\ &&
\times (\bm{\beta}_0 - \widehat{\bm{\beta}}) - C_R \|\widehat{\bm{\beta}} - \bm{\beta}_0\|^4 \nonumber\\
&=& (\bm{\beta}_0 - \widehat{\bm{\beta}})^T \mbox{var}\left[\left( V- E(V|V^T\widehat{\bm{\beta}})\right)\Big|\widehat{\bm{\beta}}\right] (\bm{\beta}_0 - \widehat{\bm{\beta}}) -  C_R \|\widehat{\bm{\beta}} - \bm{\beta}_0\|^4. \nonumber
\end{eqnarray}
{\color{black} Therefore, it is left to verify that
with probability arbitrarily large,
\begin{eqnarray*}
(\bm{\beta}_0 - \widehat{\bm{\beta}})^T \mbox{var}\left[\left( V- E(V|V^T\widehat{\bm{\beta}})\right)\Big|\widehat{\bm{\beta}}\right] (\bm{\beta}_0 - \widehat{\bm{\beta}}) \gtrsim \|\widehat{\bm{\beta}} - \bm{\beta}_0\|^2.
\end{eqnarray*}
To this end, note that based on Condition 6, there exists an orthonormal matrix $P$ (may depend on $\widehat{\bm{\beta}}$), and eigenvalues $0< \lambda_2(\widehat{\bm{\beta}}) \leq \ldots \leq \lambda_p(\widehat{\bm{\beta}})$ such that
\begin{eqnarray}
&&\mbox{var}\left[\left( V- E(V|V^T\widehat{\bm{\beta}})\right)\Big|\widehat{\bm{\beta}}\right] \nonumber \\&=& P^T \mbox{diag}(0, \lambda_2(\widehat{\bm{\beta}}), \ldots, \lambda_p(\widehat{\bm{\beta}})) P \nonumber\\
&=& \left(\widehat{\bm{\beta}}, P_{2:p}^T\right)  \mbox{diag}\left(0, \lambda_2(\widehat{\bm{\beta}}), \ldots, \lambda_p(\widehat{\bm{\beta}})\right)  \left(\widehat{\bm{\beta}}, P_{2:p}^T\right)^T \nonumber\\
&=& \bar P^T  \mbox{diag}\left(0, \lambda_2(\widehat{\bm{\beta}}), \ldots, \lambda_p(\widehat{\bm{\beta}})\right) \bar P, \label{eq-condition-6-added-1}
\end{eqnarray}
where $P_{2:p}$ denotes the matrix formed by the 2nd to pth row of $P$, $\bar P = \left(\widehat{\bm{\beta}}, P_{2:p}^T\right)^T$. We can verify that $\bar P$ is still orthonormal; this is because that the first row of $\bar P$ is $\widehat{\bm{\beta}}^T$, which satisfies $\|\widehat{\bm{\beta}}\|_2 = 1$, and
\begin{eqnarray*}
0&=& \widehat{\bm{\beta}}^T \mbox{var}\left[\left( V- E(V|V^T\widehat{\bm{\beta}})\right)\Big|\widehat{\bm{\beta}}\right] \widehat{\bm{\beta}}  \\ &=& \widehat{\bm{\beta}}^T\bar P^T  \mbox{diag}\left(0, \lambda_2(\widehat{\bm{\beta}}), \ldots, \lambda_p(\widehat{\bm{\beta}})\right) \bar P \widehat{\bm{\beta}}
\geq \lambda_2(\widehat{\bm{\beta}})   \left\| \bar P_{2:p} \widehat{\bm{\beta}} \right\|_2^2, \label{eq-condition-6-added-2}
\end{eqnarray*}
which concludes $\bar P_{2:p} \widehat{\bm{\beta}} = 0$. Denote by $a_i$, $b_i$, $i=1,\ldots,p$, the entries of $\bar P \bm{\beta}_0$ and $\bar P \widehat{\bm{\beta}}$ respectively; then we have $a_1 = \widehat{\bm{\beta}}^T \bm{\beta}_0 > 0.5$ with probability arbitrarily large for sufficiently large $n$, and $b_1 = 1 > 0.5$; and $\sum_{i} a_i^2 = 1$, $\sum_{i} b_i^2 = 1$. Then, denoting by $\bar P_1$ the first row of $\bar P$,  we have
\begin{eqnarray}
&&\left\{\bar P_1 \left( \bm{\beta}_0 -  \widehat{\bm{\beta}}\right)\right\}^2  \nonumber\\&=&  (a_1 - b_1)^2 = \left(\frac{a_1^2 - b_1^2}{a_1 + b_1}\right)^2 \nonumber \\
&\lesssim& \left\{ (a_2^2 - b_2^2) + \ldots + (a_p^2 - b_p^2)   \right\}^2 \nonumber \\
&=& \left\{(a_2 - b_2, \ldots, a_p-b_p)\cdot (a_2 + b_2, \ldots, a_p+b_p)^T \right\}^2 \nonumber \\
&\leq& \left\{ (a_2 - b_2)^2 + \ldots + (a_p-b_p)^2  \right\} \cdot \left\{(a_2 + b_2)^2 + (a_p+b_p)^2\right\} \nonumber \\
&\lesssim& (a_2 - b_2)^2 + \ldots + (a_p-b_p)^2  \nonumber \\
&=&  \left\| \bar P_{2:p} \left( \bm{\beta}_0 -  \widehat{\bm{\beta}}\right) \right\|_2^2. \label{eq-condition-6-added-3}
\end{eqnarray}
Combining (\ref{eq-condition-6-added-1}) with (\ref{eq-condition-6-added-3}), and noting that $\bar P$ is orthonormal, we have
\begin{eqnarray*}
&&(\bm{\beta}_0 - \widehat{\bm{\beta}})^T \mbox{var}\left[\left( V- E(V|V^T\widehat{\bm{\beta}})\right)\Big|\widehat{\bm{\beta}}\right] (\bm{\beta}_0 - \widehat{\bm{\beta}}) \\ &\geq& \lambda_2(\widehat{\bm{\beta}}) \left\| \bar P_{2:p}  (\bm{\beta}_0 - \widehat{\bm{\beta}}) \right\|_2^2
\gtrsim \left\| \bar P   (\bm{\beta}_0 - \widehat{\bm{\beta}}) \right\|_2^2 = \left\| \bm{\beta}_0 - \widehat{\bm{\beta}} \right\|_2^2,
\end{eqnarray*}
where we have used Condition 6. This together with (\ref{thm-1-3}) and (\ref{thm-1-2}) completes the proof of Part (b). We complete the proof of Theorem \ref{theorem-1}.
}
\epf

\section{Technical Details for Lemma \ref{lemma-psi-0}, Theorem \ref{thm-normality}, and Corollary  \ref{corollary-normality}} \label{section-proof-normality}

{

This section is composed of five subsections. Section \ref{proof-lemma-psi-0} gives the proof of Lemma \ref{lemma-psi-0}; Sections \ref{sec-pf-part-1}--\ref{sec-pf-part-3} presents proof of Theorem \ref{thm-normality} Parts (1)--(3); Section \ref{section-proof-corrollary} proves  Corollary \ref{corollary-normality}.

\subsection{Proof of Lemma \ref{lemma-psi-0}} \label{proof-lemma-psi-0}

\proof The proof for Part (1), i.e., $\psi(\bm{\beta}_0) = 0$, is straightforward.

We proceed to show Parts (2) and (3). For Part (2), let $V = X_1 -X_2$ and $U= Y_1 -Y_2$. We then have
\begin{eqnarray*}
\psi_0(\bm{\beta}) &=& E\left[V\left\{I(U>0) - F_{\bm{\beta}}(V^T \bm{\beta})\right\}\right],
\end{eqnarray*}
and therefore based on Conditions A1 and A2, $\psi_0'(\bm{\beta})$ exists and
\begin{eqnarray}
\frac{\partial \psi_0(\bm{\beta})}{\partial\bm{\beta}}\Big|_{\bm{\beta} = \bm{\beta}_0} &=& -E\left\{V \frac{\partial F_{\bm{\beta}}(V^T \bm{\beta})}{\partial\bm{\beta}} \right\} \bigg|_{\bm{\beta} = \bm{\beta}_0} \nonumber \\
&=& -E\left\{ V \int \frac{\partial F_0(V^T \bm{\beta} + \bm{v}^T (\bm{\beta}_0 - \bm{\beta}))}{\partial \bm{\beta}} f_{V|V^T\bm{\beta}} (\bm{v}) d\bm{v} \right\}\bigg|_{\bm{\beta} = \bm{\beta}_0} \nonumber \\
&& -E\left\{ V \int  F_0(V^T \bm{\beta} + \bm{v}^T (\bm{\beta}_0 - \bm{\beta})) \frac{\partial f_{V|V^T\bm{\beta}} (\bm{v})}{\partial \bm{\beta}} d\bm{v} \right\}\bigg|_{\bm{\beta} = \bm{\beta}_0} \nonumber \\
&=& -E\left\{ V \int (V^T-\bm{v}^T) f_0(V^T \bm{\beta}_0) f_{V|V^T\bm{\beta}_0 } (\bm{v}) d\bm{v}  \right\} + 0 \nonumber \\
&=& -E[V\{V^T - E(V^T|V^T\bm{\beta}_0)\} f_0(V^T \bm{\beta}_0)] \nonumber \\
&=& -E[\{V-E(V|V^T\bm{\beta}_0)\}\{V^T - E(V^T|V^T\bm{\beta}_0)\} f_0(V^T \bm{\beta}_0)]. \label{eq-psi-0-derivative-1}
\end{eqnarray}
As a consequence, based on Conditions 6 and A1, {\color{black} we immediately conclude that $\psi_0'(\bm{\beta}_0)$ has rank $p-1$}. This completes our proof of Part (2).

{\color{black}  Last, we show Part (3). Without loss of generality, assume that $\bm{\beta}_{0,1} \neq 0$, i.e, the first element of $\bm{\beta}_0$ is nonzero.
Let
\begin{eqnarray*}
\check \psi(\bm{\beta}) = \psi_0(\bm{\beta}) + \left( \begin{matrix} 0.5\bm{\beta}^T\bm{\beta}\\ \bm{0} \end{matrix}\right).
\end{eqnarray*}
Then by the fact that $\|\bm{\beta}\|_2^2 = 1$, we have $ \psi_0(\bm{\beta}) = \check \psi(\bm{\beta}) - \check \psi(\bm{\beta}_0) $; by Lemma \ref{lemma-rank-psi-0-prime} given below and the fact that $\psi_0(\bm{\beta}_0) \cdot \bm{\beta}_0 = \bm{0}$, we immediately conclude that
\begin{eqnarray*}
\check \psi'(\bm{\beta}_0) = \psi'(\bm{\beta}_0) +  \left(\begin{matrix} \bm{\beta}_0^T \\ \bm{0} \end{matrix} \right)
\end{eqnarray*}
is of full rank. Define $A(\bm{\beta}_0) = \check \psi'(\bm{\beta}_0)$; by Taylor expansion, we conclude
\begin{eqnarray*}
\psi_0(\bm{\beta}) = \check \psi(\bm{\beta}) - \check \psi(\bm{\beta}_0) = A(\bm{\beta}_0) (\bm{\beta}-\bm{\beta}_0) + o(\bm{\beta} - \bm{\beta}_0).
\end{eqnarray*}
This completes the proof of this lemma. \epf

}

{\color{black} 
\begin{lemma} \label{lemma-rank-psi-0-prime}

Assume Condition 6 and $\bm{\beta}_{0,1} \neq 0$, i.e., the first component of $\bm{\beta}_0$ is non-zero, then
$\psi_0'(\bm{\beta}_0)_{2:p}$ is of full rank.
\end{lemma}

 \pf Based on Conditions 6 and A1, and (\ref{eq-psi-0-derivative-1}), there exist eigenvalues of $\psi_0'(\bm{\beta}_0)$: $0< \lambda_2 \leq\ldots\leq \lambda_p$ and orthonormal matrix $P$, such that
 \begin{eqnarray*}
 \Sigma \equiv \psi_0'(\bm{\beta}_0) = P \mbox{diag}(0, \lambda_2, \ldots,  \lambda_p) P^T.
 \end{eqnarray*}
To show that $\psi_0'(\bm{\beta}_0)_{2:p}$ is of full rank, it suffices to show that $\psi_0'(\bm{\beta}_0)_{-1, -1} = \Sigma_{-1,-1}$ is of full rank.

 Denote $\widetilde \Lambda = \mbox{diag}(\lambda_2, \lambda_2, \ldots, \lambda_p)$ and $\widetilde I_p = \mbox{diag}(0, 1, \ldots, 1)$.
Then, we have
\begin{eqnarray}
\widetilde\Lambda^{-1/2} P^T \Sigma P \widetilde\Lambda^{-1/2} = \widetilde I_p. \label{eq-cond-6-1}
\end{eqnarray}

On the other hand, let $U$ be an orthonormal matrix whose first row is given by
\begin{eqnarray}
U_1 = \frac{\bm{\beta}_0^T P \widetilde\Lambda^{1/2}}{\sqrt{\bm{\beta}_0^T \widetilde \Sigma \bm{\beta}_0}}, \label{eq-cond-6-1-1}
\end{eqnarray}
where $\widetilde\Sigma = P \widetilde\Lambda P^T$. Let $Q = U \widetilde\Lambda^{-1/2} P^T$; then its first row is $Q_1 = \frac{\bm{\beta}_0^T}{\sqrt{\bm{\beta}_0^T \widetilde \Sigma \bm{\beta}_0}}$ and $Q^{-1} = P \widetilde\Lambda^{1/2} U^T$. We consider
\begin{eqnarray*}
\Sigma &=& \mbox{var}\left(V - E(V|V^T \bm{\beta}_0)\right) = Q^{-1} \mbox{var}\left( QV - E(QV|V^T \bm{\beta}_0) \right) (Q^{-1})^T\\
&=& Q^{-1} \mbox{var}\left(  \left(\begin{matrix} \frac{\bm{\beta}_0^T V}{\sqrt{\bm{\beta}_0^T \widetilde \Sigma \bm{\beta}_0}} \\ Q_{2:p}V \end{matrix} \right) - E\left(\left(\begin{matrix} \frac{\bm{\beta}_0^TV}{\sqrt{\bm{\beta}_0^T \widetilde \Sigma \bm{\beta}_0}} \\ Q_{2:p}V \end{matrix} \right)  \Bigg| V^T \bm{\beta}_0 \right)     \right)     (Q^{-1})^T \\
&=& Q^{-1} \mbox{var}\left(  \begin{matrix} 0 \\ Q_{2:p}\left( V - E(V|V^T \bm{\beta}_0) \right) \end{matrix}          \right)     (Q^{-1})^T \\
&=& Q^{-1} \left( \begin{matrix} 0 & \bm{0} \\ \bm{0} & Q_{2:p} \Sigma Q_{2:p}^T     \end{matrix} \right)  (Q^{-1})^T\\
&=& Q^{-1} \left( \begin{matrix} 0 & \bm{0} \\\bm{0} & U_{2:p} \widetilde I_p U_{2:p}^T     \end{matrix} \right)  (Q^{-1})^T,
\end{eqnarray*}
and therefore
\begin{eqnarray*}
\Sigma_{-1,-1} &=&  (Q^{-1})_{-1,-1} U_{2:p} \widetilde I_p U_{2:p}^T (Q^{-1})_{-1,-1}^T \\
&=& P_{2:p} \widetilde\Lambda^{1/2} U_{2:p}^T  U_{2:p} \widetilde I_p U_{2:p}^T  U_{2:p} \widetilde\Lambda^{1/2} P_{2:p}^T.
\end{eqnarray*}
Our goal is to show that $\Sigma_{-1,-1}$ is of full rank, and we prove this by contradiction. Suppose otherwise, then there exists $\bm{0} \neq \bm{x} \in \bbR^{p-1}$ such that $\bm{x}^T \Sigma_{-1, -1} \bm{x} = 0$, which is equivalent to
\begin{eqnarray*}
0 &=& \bm{x}^T P_{2:p} \widetilde\Lambda^{1/2} U_{2:p}^T  U_{2:p} \widetilde I_p U_{2:p}^T  U_{2:p} \widetilde\Lambda^{1/2} P_{2:p}^T \bm{x}\\
&=& \| \widetilde I_p U_{2:p}^T  U_{2:p} \widetilde\Lambda^{1/2} P_{2:p}^T \bm{x} \|_2^2,
\end{eqnarray*}
which is further equivalent to
\begin{eqnarray*}
\bm{0} &=& \widetilde I_p U_{2:p}^T  U_{2:p} \widetilde\Lambda^{1/2} P_{2:p}^T \bm{x}\\
&=& \left(\begin{matrix} 0 \\ U_{-1,-1}^T U_{2:p}\widetilde\Lambda^{1/2} P_{2:p}^T \bm{x} \end{matrix}\right)
\end{eqnarray*}
and therefore
\begin{eqnarray}
U_{-1,-1}^T U_{2:p}\widetilde\Lambda^{1/2} P_{2:p}^T \bm{x} = \bm{0}. \label{eq-cond-6-1-2}
\end{eqnarray}
Note that $U$ is orthonormal, its determinant $\mbox{det}(U) = 1$ or $-1$, and since $U^T = U^{-1} = \frac{1}{\mbox{det}(U)} U^*$, where $U^*$ denotes the adjoint matrix of $U$, its $(1,1)$th entry is $\mbox{det}(U_{-1,-1})$. As a consequence,
\begin{equation}
\mbox{det}(U_{-1,-1}) = U_{1,1} \cdot \mbox{det}(U)  = \frac{\bm{\beta}_0^T P (\sqrt{\lambda_2}, 0, \ldots, 0)^T}{\sqrt{\bm{\beta}_0^T \widetilde \Sigma \bm{\beta}_0}} \cdot \mbox{det}(U) \neq 0 \label{eq-cond-6-1-3}
\end{equation}
if
\begin{eqnarray}
\bm{\beta}_0^T P (\sqrt{\lambda_2}, 0, \ldots, 0)^T \neq 0. \label{eq-cond-6-1-4}
\end{eqnarray}
For presentational continuity, we postpone the proof of (\ref{eq-cond-6-1-4}) to the end of the proof for this lemma. (\ref{eq-cond-6-1-3}) indicates that $U_{-1,-1}$ is invertible, and this together with (\ref{eq-cond-6-1-2}) gives
\begin{eqnarray*}
U_{2:p}\widetilde\Lambda^{1/2} P_{2:p}^T \bm{x} =\bm{0}.
\end{eqnarray*}
This together with the fact that $U$ is orthonormal and the definition of $U_1$ in (\ref{eq-cond-6-1-1}) gives
\begin{eqnarray*}
0 &=& \bm{x}^T P_{2:p}\widetilde\Lambda^{1/2} U_{2:p}^T U_{2:p} \widetilde\Lambda^{1/2} P_{2:p}^T \bm{x} \\
 &=& \bm{x}^T P_{2:p}\widetilde\Lambda^{1/2} (I_p - U_1^T U_1) \widetilde\Lambda^{1/2} P_{2:p}^T \bm{x}\\
 &=& \bm{x}^T \widetilde \Sigma_{-1,-1} \bm{x} - \bm{x}^T P_{2:p}\widetilde\Lambda^{1/2}    \frac{\widetilde\Lambda^{1/2} P^T \bm{\beta}_0 }{\sqrt{\bm{\beta}_0^T \widetilde \Sigma \bm{\beta}_0}} \frac{\bm{\beta}_0^T P \widetilde\Lambda^{1/2}}{\sqrt{\bm{\beta}_0^T \widetilde \Sigma \bm{\beta}_0}}     \widetilde\Lambda^{1/2} P_{2:p}^T \bm{x},
\end{eqnarray*}
which leads to
\begin{eqnarray}
\bm{x}^T \widetilde \Sigma_{-1,-1} \bm{x} &=& \bm{x}^T P_{2:p}\widetilde\Lambda^{1/2}    \frac{\widetilde\Lambda^{1/2} P^T \bm{\beta}_0 }{\sqrt{\bm{\beta}_0^T \widetilde \Sigma \bm{\beta}_0}} \frac{\bm{\beta}_0^T P \widetilde\Lambda^{1/2}}{\sqrt{\bm{\beta}_0^T \widetilde \Sigma \bm{\beta}_0}}     \widetilde\Lambda^{1/2} P_{2:p}^T \bm{x} \nonumber \\
&=& \frac{\left(\bm{x}^T P_{2:p}\widetilde\Lambda^{1/2} \widetilde\Lambda^{1/2} P^T \bm{\beta}_0\right)^2}{\bm{\beta}_0^T \widetilde \Sigma \bm{\beta}_0} \nonumber \\
&\leq& \frac{\|\widetilde\Lambda^{1/2}  P_{2:p}^T \bm{x} \|_2^2 \cdot \| \widetilde\Lambda^{1/2} P^T \bm{\beta}_0\|_2^2}{\bm{\beta}_0^T \widetilde \Sigma \bm{\beta}_0} = \bm{x}^T \widetilde \Sigma_{-1,-1} \bm{x},  \label{eq-cond-6-1-5}
\end{eqnarray}
where ``$\leq$" is from Cauchy's inequality, and ``$=$" holds if and only if
\begin{eqnarray}
\widetilde\Lambda^{1/2}  P_{2:p}^T \bm{x} = a\cdot\widetilde\Lambda^{1/2} P^T \bm{\beta}_0 \label{eq-cond-6-1-6}
\end{eqnarray}
for some $a\in \bbR$.
Therefore, (\ref{eq-cond-6-1-5}) introduces a contradiction, if we can show that (\ref{eq-cond-6-1-6}) cannot hold.  In fact, if
\begin{eqnarray*}
\widetilde\Lambda^{1/2}  P_{2:p}^T \bm{x} = a\cdot \widetilde\Lambda^{1/2} P^T \bm{\beta}_0
\end{eqnarray*}
then
\begin{eqnarray*}
\widetilde\Lambda^{1/2} P^T \left\{ (0, \bm{x}^T)^T - a\cdot \bm{\beta}_0    \right\}  = 0,
\end{eqnarray*}
which leads to $(0, \bm{x}^T)^T = a \cdot \bm{\beta}_0$, since $\widetilde\Lambda^{1/2} P^T$ is invertible. Recall that the first element of $\bm{\beta}_0$ is nonzero by assumption, so we must have $a = 0$, which further implies that $\bm{x} = \bm{0}$; and this contradicts $\bm{x} \neq 0$.

Finally, we verify (\ref{eq-cond-6-1-3}). Consider
\begin{eqnarray}
\bm{\beta}_0^T P \Lambda P^T \bm{\beta}_0 = \bm{\beta}_0^T \Sigma \bm{\beta}_0 = 0. \label{eq-cond-6-1-3-1}
\end{eqnarray}
On the other hand, since $\widetilde \Sigma$ is of full rank,
\begin{eqnarray}
\bm{\beta}_0^T P \widetilde \Lambda P^T \bm{\beta}_0 = \bm{\beta}_0^T \widetilde \Sigma \bm{\beta}_0 > 0. \label{eq-cond-6-1-3-2}
\end{eqnarray}
Combining (\ref{eq-cond-6-1-3-1}) with (\ref{eq-cond-6-1-3-2}) leads to
\begin{eqnarray*}
(\bm{\beta}_0^T P (\sqrt{\lambda_2}, 0, \ldots, 0)^T)^2 &=& \bm{\beta}_0^T P \cdot \mbox{diag}(\lambda_2, 0, \ldots, 0)\cdot P^T \bm{\beta}_0  \\
&=& \bm{\beta}_0^T P (\widetilde \Lambda - \Lambda) P^T \bm{\beta}_0 > 0,
\end{eqnarray*}
giving (\ref{eq-cond-6-1-3}). We have completed the proof of this lemma. \epf

}

\subsection{Proof of Theorem \ref{thm-normality} Part (1)} \label{sec-pf-part-1}

In this section, we show the existence of the zero-crossing for $\psi_n(\bm{\beta})$.
We need the following lemma, which establishes the $L_2$ consistency of $\widehat F_{\bm{\beta}}$.
Let
\begin{eqnarray*}
D_{\bm{\beta}}(F_2, F_1) = \left[\int\int\left\{ F_1((\bm{x}_1 - \bm{x}_2)^T\bm{\beta}) - F_2((\bm{x}_1 - \bm{x}_2)^T\bm{\beta}) \right\}^2 d F_X(\bm{x}_1) d F_X(\bm{x}_2) \right]^{1/2}.
\end{eqnarray*}
\begin{lemma} \label{lemma-D-beta}
Assume Conditions 0--2 and Condition A1. We have
\begin{eqnarray*}
\sup_{\bm{\beta}\in \mathB} D_{\bm{\beta}}(\widehat F_{\bm{\beta}}, F_{\bm{\beta}}) = O_p(n^{-1/3}),
\end{eqnarray*}
where we recall that $\widehat F_{\bm{\beta}}$ is defined by {(3.4) in the main article}.
\end{lemma}
\proof The proof follows the same lines as that of Theorem \ref{theorem-1} Part (a); the details are omitted. \epf

We return to the proof of Theorem \ref{thm-normality} Part (1).
Consider
\begin{eqnarray}
\psi_n(\bm{\beta}) &=& \int \int (\bm{x}_1 - \bm{x}_2)\left\{I(y_1>y_2) - \widehat F_{\bm{\beta}}((\bm{x}_1 - \bm{x}_2)^T \bm{\beta})\right\} \nonumber\\ && \hspace{2in} \times d \bbF_{X,Y}(\bm{x}_1, y_1) d \bbF_{X,Y}(\bm{x}_2, y_2) \nonumber\\
&=& \int \int (\bm{x}_1 - \bm{x}_2)\left\{I(y_1>y_2) -  F_{\bm{\beta}}((\bm{x}_1 - \bm{x}_2)^T \bm{\beta})\right\} \nonumber \\ && \hspace{2in} \times d \bbF_{X,Y}(\bm{x}_1, y_1) d \bbF_{X,Y}(\bm{x}_2, y_2) \nonumber\\
&&+\int \int (\bm{x}_1 - \bm{x}_2)\left\{F_{\bm{\beta}}((\bm{x}_1 - \bm{x}_2)^T \bm{\beta}) -  \widehat F_{\bm{\beta}}((\bm{x}_1 - \bm{x}_2)^T \bm{\beta})\right\} \nonumber \\ && \hspace{2in} \times d \bbF_{X,Y}(\bm{x}_1, y_1) d \bbF_{X,Y}(\bm{x}_2, y_2) \nonumber \\
&=& I_1 + I_2. \label{eq-pf-normality-1}
\end{eqnarray}
We consider $I_1$ first. Note that we can further decompose $I_1 = I_{1,1} + I_{1,2} + I_{1,3}$, where
\begin{eqnarray*}
I_{1,1} &=& \int \int (\bm{x}_1 - \bm{x}_2)\left\{I(y_1>y_2) -  F_{\bm{\beta}}((\bm{x}_1 - \bm{x}_2)^T \bm{\beta})\right\}  \\
&&\hspace{1in} \times d \left\{\bbF_{X,Y}(\bm{x}_1, y_1) -  F_{X,Y}(\bm{x}_1, y_1) \right\} d \bbF_{X,Y}(\bm{x}_2, y_2)\\
I_{1,2} &=& \int \int (\bm{x}_1 - \bm{x}_2)\left\{I(y_1>y_2) -  F_{\bm{\beta}}((\bm{x}_1 - \bm{x}_2)^T \bm{\beta})\right\}  \\
&&\hspace{1in} \times d F_{X,Y}(\bm{x}_1, y_1) d \left\{\bbF_{X,Y}(\bm{x}_2, y_2) - F_{X,Y}(\bm{x}_2, y_2) \right\}\\
I_{1,3} &=& \int \int (\bm{x}_1 - \bm{x}_2)\left\{I(y_1>y_2) -  F_{\bm{\beta}}((\bm{x}_1 - \bm{x}_2)^T \bm{\beta})\right\} \nonumber\\ && \hspace{1in} \times d F_{X,Y}(\bm{x}_1, y_1) d F_{X,Y}(\bm{x}_2, y_2).
\end{eqnarray*}

Without loss of generality, we assume that $I_{1,1}$, $I_{1,2}$, and $I_{1,3}$ are one-dimensional. If not, the following developments can be applied entry-wise. Furthermore, this argument is applicable to the similar developments in Sections \ref{sec-pf-part-2} and \ref{sec-pf-part-3}.

For $I_{1,1}$,
we consider the function class
\begin{eqnarray*}
\mathJ_1 = \left\{(\bm{x}_1 - \bm{x}_2)\left\{I(y_1>y_2) -  F_{\bm{\beta}}((\bm{x}_1 - \bm{x}_2)^T \bm{\beta})\right\}: \bm{x}_2 \in \mathX, y_2 \in \mathY, \bm{\beta} \in \mathB \right\},
\end{eqnarray*}
and with Lemma \ref{lemma-1} and Condition A1 that $F_{\bm{\beta}}(t)$ is monotone in $t$, it is straightforward to verify that
\begin{eqnarray*}
\sup_{g\in \mathJ_1}|g| &\lesssim& 1,\\
H_{q, B}(\delta, \mathJ_1, F_{X,Y}) &\lesssim& 1/\delta.
\end{eqnarray*}
Applying Lemma \ref{lemma-p-1} leads to
\begin{eqnarray*}
&& \sup_{\bm{x}_2 \in \mathX, y_2 \in \mathY, \bm{\beta} \in \mathB}\Big| \int (\bm{x}_1 - \bm{x}_2)\left\{I(y_1>y_2) -  F_{\bm{\beta}}((\bm{x}_1 - \bm{x}_2)^T \bm{\beta})\right\} \\  && \hspace{2in} \times d \left\{\bbF_{X,Y}(\bm{x}_1, y_1) -  F_{X,Y}(\bm{x}_1, y_1) \right\} \Big| \\&=& O_p(n^{-1/2}),
\end{eqnarray*}
and therefore
\begin{eqnarray*}
\sup_{\bm{\beta} \in \mathB} |I_{1,1}| = O_p(n^{-1/2}).
\end{eqnarray*}
For $I_{1,2}$, we consider the function class
\begin{eqnarray*}
\mathJ_2 = \left\{\int (\bm{x}_1 - \bm{x}_2)\left\{I(y_1>y_2) -  F_{\bm{\beta}}((\bm{x}_1 - \bm{x}_2)^T \bm{\beta})\right\} d F_{X,Y}(\bm{x}_1, y_1): \bm{\beta \in \mathB}  \right\}.
\end{eqnarray*}
Clearly, $\sup_{g\in \mathJ_2}|g| \lesssim 1$.
With the same strategy used to compute the bracketing number for $\mathG_I$ in Step 2 of the proof of Lemma \ref{lemma-u-stat}, we have
\begin{eqnarray*}
H_{q,B}(\delta, \mathJ_2, F_{X,Y}) \lesssim 1/\delta.
\end{eqnarray*}
Applying Lemma \ref{lemma-p-1}, we have
\begin{eqnarray*}
\sup_{\bm{\beta} \in \mathB} I_{1,2} = O_p(n^{-1/2}).
\end{eqnarray*}
Furthermore, we have $I_{1,3} = \psi_0(\bm{\beta})$.
As a consequence, we have established
\begin{eqnarray}
\sup_{\bm{\beta}\in \mathB} |I_1 - \psi_0(\bm{\beta})| = O_p(n^{-1/2}). \label{eq-pf-normality-2}
\end{eqnarray}

For $I_2$, we can use the same strategy to decompose $I_2 = I_{2,1} + I_{2,2} + I_{2,3}$ and verify $\sup_{\bm{\beta}\in \mathB} I_{2,1} = O_p(n^{-1/2})$, $\sup_{\bm{\beta}\in \mathB}I_{2,2} = O_p(n^{-1/2})$.  Applying the Cauchy--Schwarz inequality and Lemma \ref{lemma-D-beta}, we have
\begin{eqnarray*}
&&\sup_{\bm{\beta}\in \mathB} |I_{2,3}| \\
&=& \sup_{\bm{\beta}\in \mathB}\Big|\int \int (\bm{x}_1 - \bm{x}_2)\left\{F_{\bm{\beta}}((\bm{x}_1 - \bm{x}_2)^T \bm{\beta}) -  \widehat F_{\bm{\beta}}((\bm{x}_1 - \bm{x}_2)^T \bm{\beta})\right\} \\ && \hspace{2in} \times d F_{X,Y}(\bm{x}_1, y_1) d F_{X,Y}(\bm{x}_2, y_2)\Big|\\
&\lesssim& \sup_{\bm{\beta}\in \mathB} D_{\bm{\beta}}(\widehat F_{\bm{\beta}}, F_{\bm{\beta}}) = O_p(n^{-1/3}).
\end{eqnarray*}
As a consequence,
\begin{eqnarray}
\sup_{\bm{\beta}\in \mathB} |I_2| = O_p(n^{-1/3}). \label{eq-pf-normality-3}
\end{eqnarray}
Combining (\ref{eq-pf-normality-1})--(\ref{eq-pf-normality-3}) leads to
\begin{eqnarray}
\psi_n(\bm{\beta}) = \psi_0(\bm{\beta}) + O_p(n^{-1/3}), \label{eq-pf-normality-4}
\end{eqnarray}
uniformly in $\bm{\beta} \in \mathB$.

With Lemma \ref{lemma-psi-0}, the rest of the proof for Part (1) of Theorem \ref{thm-normality} follows the same lines as those of Part 1 of Theorem 4.1 in Groeneboom~and~Hendrickx~(2018); the details are omitted. \epf

\subsection{Proof of Theorem \ref{thm-normality} Part (2)} \label{sec-pf-part-2}

Similarly to the proof of Lemma \ref{lemma-7}, we establish Part (2) by showing the following:
\begin{itemize}
\item[(i)] $\psi_0(\widetilde{\bm{\beta}}) = o_p(1)$;
\item[(ii)] $\psi_0(\bm{\beta}) = 0$ implies that $\bm{\beta} = \bm{\beta}_0$;
\item[(iii)] $\psi_0(\bm{\beta})$ is continuous in $\mathB$.
\end{itemize}

(iii) holds because of Condition A1. We proceed to show (i).
Since $\widetilde{\bm{\beta}}$ is the zero-crossing of $\psi_n(\bm{\beta})$, there exist $\widetilde{\bm{\beta}}^{(U)}$ and $\widetilde{\bm{\beta}}^{(L)}$ such that
\begin{eqnarray}
\widetilde{\bm{\beta}}^{(U)}= \widetilde{\bm{\beta}} +  O(1/n), \quad && \widetilde{\bm{\beta}}^{(L)}= \widetilde{\bm{\beta}} +  O(1/n),\nonumber\\ \mbox{and} \hspace{1in} \psi_n(\widetilde{\bm{\beta}}^{(U)}) \geq 0, \quad && \psi_n(\widetilde{\bm{\beta}}^{(L)}) \leq 0. \label{eq-pf-normality-5}
\end{eqnarray}
As a consequence, there exists $\bm{\alpha}_n = (\alpha_{n,1}, \ldots, \alpha_{n,p})^T$, $0\leq \alpha_{n,j}\leq 1$ for $j=1,\ldots,p$, such that
\begin{eqnarray}
\bm{\alpha}_n^T \psi_n(\widetilde{\bm{\beta}}^{(U)}) + (1-\bm{\alpha}_n)^T \psi_n(\widetilde{\bm{\beta}}^{(L)}) = 0. \label{eq-pf-normality-6}
\end{eqnarray}
On the other hand, applying (\ref{eq-pf-normality-4}) and based on the continuity of $\psi_0(\cdot)$, we have
\begin{eqnarray*}
&&\bm{\alpha}_n^T \psi_n(\widetilde{\bm{\beta}}^{(U)}) + (1-\bm{\alpha}_n)^T \psi_n(\widetilde{\bm{\beta}}^{(L)})\\
&=& \bm{\alpha}_n^T\psi_0(\widetilde{\bm{\beta}}^{(U)}) + (1-\bm{\alpha}_n)^T \psi_0(\widetilde{\bm{\beta}}^{(L)}) + O_p(n^{-1/3})\\
&=& \psi_0(\widetilde{\bm{\beta}}) + O(n^{-1}) + O_p(n^{-1/3}),
\end{eqnarray*}
which together with (\ref{eq-pf-normality-6}) proves (i). We now show (ii). Based on the definition of $\psi_0(\bm{\beta})$,
\begin{eqnarray*}
\bm{\beta}_0^T\psi_0(\bm{\beta}) &=& E\left[ \bm{\beta}_0^TV\Big\{I(U>0) - E(I(U>0)|V^T\bm{\beta})\Big\}  \right]\\
&=& E\left[ \bm{\beta}_0^TV\Big\{F_0(\bm{\beta}_0^T V) - E(F_0(\bm{\beta}_0^T V)|V^T\bm{\beta})\Big\}  \right]\\
&=& E\left[\mbox{cov}\left\{\bm{\beta}_0^TV, F_0(\bm{\beta}_0^T V)\Big|V^T\bm{\beta} \right\}\right]\\
&=& E\left[\mbox{cov}\left\{Z_1, F_0(Z_1)\Big|Z_2 \right\} \right],
\end{eqnarray*}
where we denote $Z_1 = \bm{\beta}_0^T V$ and $Z_2 = V^T\bm{\beta}$, with $Z_1\geq C$, for some $C>-\infty$. Using the same arguments as in the proof of Part (ii) of Lemma 4.1 in Groeneboom and Hendrickx (2018), we conclude that
\begin{eqnarray*}
\mbox{cov}\left\{Z_1, F_0(Z_1)\Big|Z_2 \right\} \geq 0,
\end{eqnarray*}
almost surely. As a consequence, if there exists some $\bm{\beta}\neq \bm{\beta}_0$ such that $\psi_0(\bm{\beta}) = 0$, we must have
\begin{eqnarray*}
\mbox{cov}\left\{\bm{\beta}_0^TV, F_0(\bm{\beta}_0^T V)\Big|V^T\bm{\beta} \right\} = 0
\end{eqnarray*}
almost surely, which contradicts Condition A3. This completes the proof of (ii). \epf

\subsection{Proof of Theorem \ref{thm-normality} Part (3)} \label{sec-pf-part-3}

Recall the definition of $\varphi_{\bm{\beta}}$ in (\ref{def-varphi}).
For any piecewise constant distribution function $F$ with finitely many jumps $\tau_1< \tau_2 <\ldots$, we define
\begin{eqnarray*}
\overline{\varphi}_{\bm{\beta}, F}(t) = \left\{ \begin{array}{ll} \varphi_{\bm{\beta}}(\tau_i) & \mbox{if } F_{\bm{\beta}}(t) > F(\tau_i),\ \ t \in [\tau_i, \tau_{i+1}) \\
\varphi_{\bm{\beta}}(s) & \mbox{if } F_{\bm{\beta}}(t) = F(s), \ \mbox{ for some } s\in [\tau_i, \tau_{i+1}) \\
\varphi_{\bm{\beta}}(\tau_{i+1}) & \mbox{if } F_{\bm{\beta}}(t) < F(\tau_i),\ \ t \in [\tau_i, \tau_{i+1}) \end{array} \right. .
\end{eqnarray*}

Then, we have the following lemma.
\begin{lemma} \label{lemma-bar-varphi}
Assume Conditions 1, A1, and A2. For $\bm{\beta}\in\mathB$ and any piecewise constant distribution function $F$ with finitely many jumps $\tau_1< \tau_2 <\ldots$, there exists a constant $C>0$ not depending on $F$ and $\bm{\beta}$ such that
\begin{eqnarray*}
\left|\varphi_{\bm{\beta}}(t) -  \overline{\varphi}_{\bm{\beta}, F}(t) \right| \leq C|F(t) - F_{\bm{\beta}}(t)|,
\end{eqnarray*}
where if $\varphi_{\bm{\beta}}(\cdot)$ and $\overline{\varphi}_{\bm{\beta}, F}(\cdot)$ are multi-dimensional, this inequality holds for $\varphi_{\bm{\beta}}(t) -  \overline{\varphi}_{\bm{\beta}, F}(t)$ entry-wise.

\end{lemma}

\proof Without loss of generality, assume that $\varphi_{\bm{\beta}}(\cdot)$ and $\overline{\varphi}_{\bm{\beta}, F}(\cdot)$ are one-dimensional. For any $t\in [\tau_i, \tau_{i+1})$, we have
\begin{eqnarray*}
\varphi_{\bm{\beta}}(t) -  \overline{\varphi}_{\bm{\beta}, F}(t) = \left\{ \begin{array}{ll} \varphi_{\bm{\beta}}(t)-\varphi_{\bm{\beta}}(\tau_i) & \mbox{if } F_{\bm{\beta}}(t) > F(\tau_i),\ \ t \in [\tau_i, \tau_{i+1}) \\
\varphi_{\bm{\beta}}(t) - \varphi_{\bm{\beta}}(s) & \mbox{if } F_{\bm{\beta}}(t) = F(s), \ \mbox{ for some } s\in [\tau_i, \tau_{i+1}) \\
\varphi_{\bm{\beta}}(t) - \varphi_{\bm{\beta}}(\tau_{i+1}) & \mbox{if } F_{\bm{\beta}}(t) < F(\tau_i),\ \ t \in [\tau_i, \tau_{i+1}) \end{array} \right..
\end{eqnarray*}
We need to show only that if $F_{\bm{\beta}}(t) > F(\tau_i)$ for $t \in [\tau_i, \tau_{i+1})$,
\begin{eqnarray*}
\left|\varphi_{\bm{\beta}}(t) -  \overline{\varphi}_{\bm{\beta}, F}(t) \right|\lesssim |F_{\bm{\beta}}(t) - F(t)|
\end{eqnarray*}
up to a constant not depending on $F$ and $\bm{\beta}$. The other two cases are similar. In fact,
based on Conditions 1, A1, and A2,
\begin{eqnarray*}
\left|\varphi_{\bm{\beta}}(t) -  \overline{\varphi}_{\bm{\beta}, F}(t)\right| &=& \left|\varphi_{\bm{\beta}}(t)-\varphi_{\bm{\beta}}(\tau_i)\right| \lesssim |t-\tau_i|\\
&\le& \frac{1}{\inf_{\bm{\beta} \in \mathB, t\in [\tau_i, \tau_{i+1})} f_{\bm{\beta}}(t) } |F_{\bm{\beta}}(t) - F_{\bm{\beta}}(\tau_i)| \lesssim |F_{\bm{\beta}}(t) - F_{\bm{\beta}}(\tau_i)|\\
&\leq& |F_{\bm{\beta}}(t) - F(\tau_i)| = |F_{\bm{\beta}}(t) - F(t)|,
\end{eqnarray*}
where both ``$\lesssim$" are up to constants not depending on $\bm{\beta}$.
\epf

The next lemma establishes the bracketing numbers of the function classes
\begin{eqnarray*}
\mathJ_{3,1} &=& \{\overline{\varphi}_{\bm{\beta}, F}((\bm{x}_1- \bm{x}_2)^T \bm{\beta}): \bm{\beta}\in \mathB, F\in \widetilde{\mathF}\}, \\
\mathJ_{3,2} &=& \{\overline{\varphi}_{\bm{\beta}, F}((\bm{x}_1- \bm{x}_2)^T \bm{\beta}): \bm{x}_2 \in \mathX, \bm{\beta}\in \mathB, F\in \widetilde{\mathF}\},
\end{eqnarray*}
where $\widetilde{\mathF}$ is the class of piecewise constant distribution functions with finitely many jumps, and all jumps are uniformly bounded, i.e., there exists a constant $M>0$ such that if $\tau$ is a jump for some $F\in \widetilde{F}$, then $|\tau|\leq M$.

\begin{lemma} \label{lemma-J-3}
Assume Conditions 1, 2, and A2. For any arbitrary $1\leq q\leq \infty$, we have
\begin{eqnarray*}
H_{q,B}(\delta, \mathJ_{3,1}, F_{X_1, X_2}) &\lesssim& 1/\delta,\\
H_{q,B}(\delta, \mathJ_{3,2}, F_{X_1, X_2}) &\lesssim& 1/\delta.
\end{eqnarray*}
\end{lemma}
\proof Based on the definition of $\overline{\varphi}_{\bm{\beta}, F}(\cdot)$, for any $\bm{\beta}\in \mathB$ and $F \in \widetilde{\mathF}$ with jumps $\tau_0 \leq \ldots \leq \tau_m$, there exist $\xi_j \in [\tau_{j-1}, \tau_j]$, $j=1,\ldots,m$, which may depend on $\bm{\beta}$ and $F$, such that
\begin{eqnarray*}
\overline{\varphi}_{\bm{\beta}, F}(t) &=& \sum_{j=1}^m \varphi_{\bm{\beta}}(\xi_j) I\left(t\in [\tau_{j-1}, \tau_j)\right)\\
&=& \varphi_{\bm{\beta}}(\xi_1) + \sum_{\{j: j\geq 2, t \geq \tau_j\}} \left\{\varphi_{\bm{\beta}}(\xi_j) - \varphi_{\bm{\beta}}(\xi_{j-1}) \right\} \\
&=& \varphi_{\bm{\beta}}(\xi_1) + \sum_{\{j: j\geq 2, t \geq \tau_j\}} \left\{\varphi_{\bm{\beta}}(\xi_j) - \varphi_{\bm{\beta}}(\xi_{j-1}) \right\}_{+} - \sum_{\{j: j\geq 2, t \geq \tau_j\}} \left\{\varphi_{\bm{\beta}}(\xi_j) - \varphi_{\bm{\beta}}(\xi_{j-1}) \right\}_{-}\\
&=& \overline{\varphi}_{\bm{\beta}, F, +}(t) - \overline{\varphi}_{\bm{\beta}, F, -}(t),
\end{eqnarray*}
where
\begin{eqnarray*}
\overline{\varphi}_{\bm{\beta}, F, +}(t)&=&\varphi_{\bm{\beta}}(\xi_1) + \sum_{\{j: j\geq 2, t \geq \tau_j\}} \left\{\varphi_{\bm{\beta}}(\xi_j) - \varphi_{\bm{\beta}}(\xi_{j-1}) \right\}_{+} \\
\overline{\varphi}_{\bm{\beta}, F, -}(t)&=&  \sum_{\{j: j\geq 2, t \geq \tau_j\}} \left\{\varphi_{\bm{\beta}}(\xi_j) - \varphi_{\bm{\beta}}(\xi_{j-1}) \right\}_{-}
\end{eqnarray*}
are both monotone in $t$. Here for any $a\in \bbR$, $a_{+} = \max\{a, 0\}$, $a_{-} = \min\{a, 0\}$.  Based on Conditions 1 and A2, and the fact that by definition all functions in $\widetilde{\mathF}$ have uniformly bounded jumps, we have
\begin{eqnarray*}
\left|\overline{\varphi}_{\bm{\beta}, F, +}(t)\right|&=& \left|\varphi_{\bm{\beta}}(\xi_1) + \sum_{\{j: j\geq 2, t \geq \tau_j\}} \left\{\varphi_{\bm{\beta}}(\xi_j) - \varphi_{\bm{\beta}}(\xi_{j-1}) \right\}_{+} \right| \\
&\leq& \left|\varphi_{\bm{\beta}}(\xi_1)\right| + \sum_{j=1}^m \left|\varphi_{\bm{\beta}}(\xi_j) - \varphi_{\bm{\beta}}(\xi_{j-1}) \right|\\
&\lesssim &
\left| \varphi_{\bm{\beta}}(\xi_1)\right| + \sum_{j=1}^m |\xi_j-\xi_{j-1}| = \left|\varphi_{\bm{\beta}}(\xi_1)\right|+ \xi_m - \xi_1 \lesssim 1
\end{eqnarray*}
uniformly in $\bm{\beta}$ and $F$, and likewise
\begin{eqnarray*}
\overline{\varphi}_{\bm{\beta}, F, -}(t)\lesssim 1
\end{eqnarray*}
uniformly in $\bm{\beta}$ and $F$. With Lemma \ref{lemma-1}, we can immediately conclude that
\begin{eqnarray*}
\mathJ_{3,1,+} &=& \{\overline{\varphi}_{\bm{\beta}, F, +}((\bm{x}_1- \bm{x}_2)^T \bm{\beta}): \bm{\beta} \in \mathB, F\in \widetilde{\mathF}\}\\
\mathJ_{3,1,-} &=& \{\overline{\varphi}_{\bm{\beta}, F, -}((\bm{x}_1- \bm{x}_2)^T \bm{\beta}): \bm{\beta} \in \mathB, F\in \widetilde{\mathF}\}
\end{eqnarray*}
are uniformly bounded function classes and satisfy
\begin{eqnarray*}
H_{2,B}(\delta, \mathJ_{3,1, +}, F_{X_1, X_2}) \lesssim 1/\delta \quad \mbox{and} \quad H_{2,B}(\delta, \mathJ_{3,1, -}, F_{X_1, X_2}) \lesssim 1/\delta.
\end{eqnarray*}
Now it is straightforward to verify that $\mathJ_{3,1}  \subset \mathJ_{3,1, +} - \mathJ_{3,1, -}$ satisfies
\begin{eqnarray*}
H_{q,B}(\delta, \mathJ_{3,1}, F_{X_1, X_2}) \lesssim 1/\delta.
\end{eqnarray*}
With the same strategy, we can verify that
\begin{eqnarray*}
H_{q,B}(\delta, \mathJ_{3,2}, F_{X_1, X_2}) \lesssim 1/\delta.  \endpf
\end{eqnarray*}

Consider an estimator $\widecheck{\bm{\beta}}$ of $\bm{\beta}$ that satisfies
\begin{eqnarray}
\widecheck{\bm{\beta}} \to \bm{\beta}_0 \label{eq-con-beta-check}
\end{eqnarray}
in probability as $n\to \infty$. We have
\begin{eqnarray}
\psi_n(\widecheck{\bm{\beta}}) &=& \int \int (\bm{x}_1 - \bm{x}_2)\left\{I(y_1>y_2) - \widehat F_{\widecheck{\bm{\beta}}}((\bm{x}_1 - \bm{x}_2)^T \widecheck{\bm{\beta}})\right\}  \nonumber \\ && \hspace{1.2in}\times d \bbF_{X,Y}(\bm{x}_1, y_1) d \bbF_{X,Y}(\bm{x}_2, y_2) \nonumber \\
&=& \int \int \left\{\bm{x}_1 - \bm{x}_2 - \varphi_{\widecheck{\bm{\beta}}}\left((\bm{x}_1 - \bm{x}_2)^T \widecheck{\bm{\beta}}\right)\right\} \nonumber\\ &&   \times \left\{I(y_1>y_2) - \widehat F_{\widecheck{\bm{\beta}}}\left((\bm{x}_1 - \bm{x}_2)^T \widecheck{\bm{\beta}}\right)\right\} d \bbF_{X,Y}(\bm{x}_1, y_1) d \bbF_{X,Y}(\bm{x}_2, y_2) \nonumber  \\
&& + \int \int \left\{\varphi_{\widecheck{\bm{\beta}}}\left((\bm{x}_1 - \bm{x}_2)^T \widecheck{\bm{\beta}}\right) - \bar\varphi_{\widecheck{\bm{\beta}}, \widehat F_{\widecheck{\bm{\beta}}}}\left((\bm{x}_1 - \bm{x}_2)^T \widecheck{\bm{\beta}}\right)\right\} \nonumber \\
&&\times \left\{I(y_1>y_2) - \widehat F_{\widecheck{\bm{\beta}}}\left((\bm{x}_1 - \bm{x}_2)^T \widecheck{\bm{\beta}}\right)\right\} d \bbF_{X,Y}(\bm{x}_1, y_1) d \bbF_{X,Y}(\bm{x}_2, y_2) \nonumber \\
&=& I_3 + I_4,
\label{eq-con-I-3-I-4}
\end{eqnarray}
where we have used the fact that
\begin{eqnarray*}
\int \int  \bar\varphi_{\widecheck{\bm{\beta}}, \widehat F_{\widecheck{\bm{\beta}}}}\left((\bm{x}_1 - \bm{x}_2)^T \widecheck{\bm{\beta}}\right) \left\{I(y_1>y_2) - \widehat F_{\widecheck{\bm{\beta}}}\left((\bm{x}_1 - \bm{x}_2)^T \widecheck{\bm{\beta}}\right)\right\} \\ \times d \bbF_{X,Y}(\bm{x}_1, y_1) d \bbF_{X,Y}(\bm{x}_2, y_2) = 0.
\end{eqnarray*}
This is because $\bar\varphi_{\widecheck{\bm{\beta}}, \widehat F_{\widecheck{\bm{\beta}}}}\left(t\right)$ is a piecewise constant function with the same jumps as $\widehat F_{\widecheck{\bm{\beta}}}\left(t\right)$; on the other hand, recall the definition of $\widehat F_{\widecheck{\bm{\beta}}}\left(t\right)$: it is the slope of the greatest convex minorant of the corresponding cusum diagram, based on the values of $I(Y_i>Y_j)$ in the order of $(X_i - X_j)^T \widecheck{\bm{\beta}}$. Therefore, for every constant piece $[\tau_j, \tau_{j+1})$ of $\widehat F_{\widecheck{\bm{\beta}}}\left(t\right)$,
\begin{eqnarray*}
\int_{(\bm{x}_1 - \bm{x}_2)^T \widecheck{\bm{\beta}}\in [\tau_j, \tau_{j+1})}  \left\{I(y_1>y_2) - \widehat F_{\widecheck{\bm{\beta}}}\left((\bm{x}_1 - \bm{x}_2)^T \widecheck{\bm{\beta}}\right)\right\} \\ \times d \bbF_{X,Y}(\bm{x}_1, y_1) d \bbF_{X,Y}(\bm{x}_2, y_2) = 0.
\end{eqnarray*}

Without loss of generality, we assume that $\psi_n(\widecheck{\bm{\beta}})$ is one-dimensional; if not, the rest of the proof in this subsection can be applied entry-wise.

We derive the asymptotic properties of $I_3$ and $I_4$ given in (\ref{eq-con-I-3-I-4}) separately. We consider $I_4$ first. We will show that
\begin{eqnarray}
I_4 = O_p(n^{-2/3}), \label{eq-norm-part-3-4}
\end{eqnarray}
using a strategy similar to that in the proof of Lemma \ref{lemma-u-stat}. Denote
\begin{eqnarray}
&&g_{4, \bm{\beta}, F}(\bm{x}_1 -\bm{x}_2, y_1 - y_2) \nonumber\\ &=& \left\{\varphi_{\bm{\beta}}\left((\bm{x}_1 - \bm{x}_2)^T \bm{\beta}\right) - \overline\varphi_{\bm{\beta}, F}\left((\bm{x}_1 - \bm{x}_2)^T \bm{\beta}\right)\right\} \nonumber \\ && \times
\left\{I(y_1>y_2) - F\left((\bm{x}_1 - \bm{x}_2)^T \bm{\beta}\right)\right\}. \label{def-g-4}
\end{eqnarray}
Then
\begin{equation}
I_4 = \int \int g_{4, \widecheck{\bm{\beta}}, \widehat F_{\widecheck{\bm{\beta}}}}(\bm{x}_1 -\bm{x}_2, y_1 - y_2) d \bbF_{X,Y}(\bm{x}_1, y_1) d \bbF_{X,Y}(\bm{x}_2, y_2). \label{eq-norm-part-3-4-1}
\end{equation}

Consider the function class
\begin{eqnarray*}
\mathJ_4 = \bigg\{ g_{4, \bm{\beta}, F}(\bm{x}_1 -\bm{x}_2, y_1 - y_2): \bm{x}_2 \in \mathX, y_2 \in \mathY,  \bm{\beta}\in \mathB, F \in \widetilde{\mathF}\bigg\}.
\end{eqnarray*}
With Conditions 1 and A2, we can verify that for any $\bm{\beta}_1, \bm{\beta}_2\in \mathB$, we have
\begin{eqnarray*}
\left|\varphi_{\bm{\beta}_1}((\bm{x}_1-\bm{x}_2)\bm{\beta}_1)- \varphi_{\bm{\beta}_2}((\bm{x}_1-\bm{x}_2)\bm{\beta}_2)\right| \lesssim \|\bm{\beta}_1 - \bm{\beta}_2\|_2.
\end{eqnarray*}
Based on Theorem 9.23 in Kosorok (2008), we immediately conclude that the funciton class $$\mathJ_{4,1} = \left\{\varphi_{\bm{\beta}}((\bm{x}_1-\bm{x}_2)^T \bm{\beta}): \bm{\beta}\in \bm{\beta}\right\}$$
satisfies for any $1\leq q \leq \infty$,
\begin{eqnarray*}
H_{q,B}(\delta, \mathJ_{4,1}, F_{X_1, X_2}) \lesssim -\log\delta,
\end{eqnarray*}
up to a constant depending only on $q$. This, together with Lemma \ref{lemma-1} and Lemma \ref{lemma-J-3}, leads to
\begin{eqnarray*}
H_{q,B}(\delta, \mathJ_4, F_{X_1, X_2}) \lesssim 1/\delta.
\end{eqnarray*}
With the same strategy as in the derivation of (\ref{lem-3-17}), we can show that
\begin{eqnarray}
\label{eq-norm-part-3-5} &&\int \left| \int g_{4, \widecheck{\bm{\beta}}, \widehat F_{\widecheck{\bm{\beta}}}}(\bm{x}_1 - \bm{x}_2, y_1 - y_2) d\left\{ \bbF_{X,Y}(\bm{x}_1,y_1)-F_{X,Y}(\bm{x}_1,y_1)\right\}  \right| \\ && \hspace{3in} \times d\bbF_{X,Y}(\bm{x}_2, y_2) \nonumber\\
&=& O_p(n^{-2/3})\vee O_p(n^{-1/2}) \frac{1}{n}\sum_{i=1}^n \left\|g_{4, \widecheck{\bm{\beta}}, \widehat F_{\widecheck{\bm{\beta}}}}(\bm{x}_1 - X_i, y_1 - Y_i)\right\|_{2, F_{X,Y}}^{1/2} \nonumber
\end{eqnarray}
and
\begin{eqnarray}
&&\frac{1}{n}\sum_{i=1}^n \left\|g_{4, \widecheck{\bm{\beta}}, \widehat F_{\widecheck{\bm{\beta}}}}(\bm{x}_1 - X_i, y_1 - Y_i)\right\|_{2, F_{X,Y}}^{1/2}\nonumber\\
&\leq & \left\{ \frac{1}{n}\sum_{i=1}^n \left\|g_{4, \widecheck{\bm{\beta}}, \widehat F_{\widecheck{\bm{\beta}}}}(\bm{x}_1 - X_i, y_1 - Y_i)\right\|_{2, F_{X,Y}} \right\}^{1/2} \nonumber\\
&\lesssim & \left\{ \frac{1}{n}\sum_{i=1}^n \left\| \widehat F_{\widecheck{\bm{\beta}}}((\bm{x}_1 - X_i)^T\widecheck{\bm{\beta}}) - F_{\widecheck{\bm{\beta}}}((\bm{x}_1 - X_i)^T\widecheck{\bm{\beta}}) \right\|_{2, F_{X,Y}} \right\}^{1/2} \nonumber\\
&=& \left\{ \frac{1}{n}\sum_{i=1}^n g_{5, \widecheck{\bm{\beta}}, \widehat F_{\widecheck{\bm{\beta}}}}(X_i) \right\}^{1/2}, \label{eq-norm-part-3-6}
\end{eqnarray}
where
\begin{eqnarray*}
g_{5, \bm{\beta}, F}(\bm{x}_2) &=& \left\|F((\bm{x}_1 - \bm{x}_2)^T\bm{\beta}) - F_{\bm{\beta}}((\bm{x}_1 - \bm{x}_2)^T\bm{\beta}) \right\|_{2, F_{X,Y}} \\
&=& \left[ \int \left\{F((\bm{x}_1 - \bm{x}_2)^T\bm{\beta}) - F_{\bm{\beta}}((\bm{x}_1 - \bm{x}_2)^T\bm{\beta})\right\}^2d F_X(\bm{x}_1)  \right]^{1/2};
\end{eqnarray*}
and to derive ``$\lesssim$" we have applied Lemma \ref{lemma-bar-varphi}.
With the same development as Part (3) of Lemma \ref{lemma-1}, we can show that the function class
\begin{eqnarray*}
\mathJ_5 = \left\{g_{5, \bm{\beta}, F}(\bm{x}_2): \bm{\beta} \in \mathB, F \in \widetilde F \right\}
\end{eqnarray*}
satisfies for any arbitrary $1\leq q\leq  \infty$,
\begin{eqnarray*}
H_{q,B}(\delta, \mathJ_5, F_X) \lesssim 1/\delta,
\end{eqnarray*}
with ``$\lesssim$" being up to a universal constant depending only on $p$ and $q$. Therefore, applying Lemma \ref{lemma-p-2} given at the beginning of the proof of Lemma \ref{lemma-u-stat}, we have
\begin{eqnarray*}
&&\frac{1}{n}\sum_{i=1}^n g_{5, \widecheck{\bm{\beta}}, \widehat F_{\widecheck{\bm{\beta}}}}(X_i) = \int g_{5, \widecheck{\bm{\beta}}, \widehat F_{\widecheck{\bm{\beta}}}}(\bm{x}_2) dF_X(\bm{x}_2)   + O_p(n^{-1/2})\\
&=& \int \left[ \int \left\{\widehat F_{\widecheck{\bm{\beta}}}((\bm{x}_1 - \bm{x}_2)^T\widecheck{\bm{\beta}}) - F_{\widecheck{\bm{\beta}}}((\bm{x}_1 - \bm{x}_2)^T\widecheck{\bm{\beta}})\right\}^2d F_X(\bm{x}_1)  \right]^{1/2} dF_X(\bm{x}_2) \\ && \hspace{3.5in}  + O_p(n^{-1/2})\\
&\leq& \left[ \int  \int \left\{\widehat F_{\widecheck{\bm{\beta}}}((\bm{x}_1 - \bm{x}_2)^T\widecheck{\bm{\beta}}) - F_{\widecheck{\bm{\beta}}}((\bm{x}_1 - \bm{x}_2)^T\widecheck{\bm{\beta}})\right\}^2d F_X(\bm{x}_1)   dF_X(\bm{x}_2)\right]^{1/2} \\
&& \hspace{3.5in} + O_p(n^{-1/2})\\
&=& D_{\widecheck{\bm{\beta}}}(\widehat F_{\widecheck{\bm{\beta}}}, F_{\widecheck{\bm{\beta}}}) + O_p(n^{-1/2})\\
&=& O_p(n^{-1/3}),
\end{eqnarray*}
by Lemma \ref{lemma-D-beta}. This, together with (\ref{eq-norm-part-3-5}) and (\ref{eq-norm-part-3-6}), leads to
\begin{eqnarray}
\int \left| \int g_{4, \widecheck{\bm{\beta}}, \widehat F_{\widecheck{\bm{\beta}}}}(\bm{x}_1 - \bm{x}_2, y_1 - y_2) d\left\{ \bbF_{X,Y}(\bm{x}_1,y_1)-F_{X,Y}(\bm{x}_1,y_1)\right\}  \right| \nonumber \\   \times d\bbF_{X,Y}(\bm{x}_2, y_2) = O_p(n^{-2/3}).
\label{eq-norm-part-3-7}
\end{eqnarray}

Furthermore, using a development similar to the proof of (\ref{lem-3-3-4}), we can derive
\begin{eqnarray*}
\int  \int g_{4, \widecheck{\bm{\beta}}, \widehat F_{\widecheck{\bm{\beta}}}}(\bm{x}_1 - \bm{x}_2, y_1 - y_2)   dF_{X,Y}(\bm{x}_1, y_1) d\left\{ \bbF_{X,Y}(\bm{x}_2,y_2)-F_{X,Y}(\bm{x}_2,y_2)\right\} \\ = O_p(n^{-2/3}),
\end{eqnarray*}
which together with (\ref{eq-norm-part-3-4-1}), (\ref{eq-norm-part-3-7}), and the definition of $g_{4, \bm{\beta}, F}(\cdot, \cdot)$ given in (\ref{def-g-4}) leads to
\begin{eqnarray*}
|I_4| &=&  \left|\int  \int g_{4, \widecheck{\bm{\beta}}, \widehat F_{\widecheck{\bm{\beta}}}}(\bm{x}_1 - \bm{x}_2, y_1 - y_2)   dF_{X,Y}(\bm{x}_1, y_1) dF_{X,Y}(\bm{x}_2,y_2) + O_p(n^{-2/3})\right|\\
&\leq&\left|\int \int \left\{\varphi_{\widecheck{\bm{\beta}}}\left((\bm{x}_1 - \bm{x}_2)^T \widecheck{\bm{\beta}}\right) - \bar\varphi_{\widecheck{\bm{\beta}}, \widehat F_{\widecheck{\bm{\beta}}}}\left((\bm{x}_1 - \bm{x}_2)^T \widecheck{\bm{\beta}}\right)\right\}\right. \\ &&\times
\left.\left\{I(y_1>y_2) - \widehat F_{\widecheck{\bm{\beta}}}\left((\bm{x}_1 - \bm{x}_2)^T \widecheck{\bm{\beta}}\right)\right\} d F_{X,Y}(\bm{x}_1, y_1) d F_{X,Y}(\bm{x}_2, y_2)\right| \\ &&\hspace{3.5in} + O_p(n^{-2/3})\\
&=& \left|\int \int \left\{\varphi_{\widecheck{\bm{\beta}}}\left((\bm{x}_1 - \bm{x}_2)^T \widecheck{\bm{\beta}}\right) - \bar\varphi_{\widecheck{\bm{\beta}}, \widehat F_{\widecheck{\bm{\beta}}}}\left((\bm{x}_1 - \bm{x}_2)^T \widecheck{\bm{\beta}}\right)\right\}\right. \\ &&\left. \times
\left\{F_{\widecheck{\bm{\beta}}}\left((\bm{x}_1 - \bm{x}_2)^T \widecheck{\bm{\beta}}\right) - \widehat F_{\widecheck{\bm{\beta}}}\left((\bm{x}_1 - \bm{x}_2)^T \widecheck{\bm{\beta}}\right)\right\} d F_X(\bm{x}_1) d F_X(\bm{x}_2)\right| \\ && \hspace{3.5in} + O_p(n^{-2/3})\\
&\leq& \left\|\varphi_{\widecheck{\bm{\beta}}}\left((\bm{x}_1 - \bm{x}_2)^T \widecheck{\bm{\beta}}\right) - \bar\varphi_{\widecheck{\bm{\beta}}, \widehat F_{\widecheck{\bm{\beta}}}}\left((\bm{x}_1 - \bm{x}_2)^T \widecheck{\bm{\beta}}\right) \right\|_{2, F_{X_1,X_2}} \\
&&\times \left\| F_{\widecheck{\bm{\beta}}}\left((\bm{x}_1 - \bm{x}_2)^T \widecheck{\bm{\beta}}\right) - \widehat F_{\widecheck{\bm{\beta}}}\left((\bm{x}_1 - \bm{x}_2)^T \widecheck{\bm{\beta}}\right) \right\|_{2, F_{X_1,X_2}} + O_p(n^{-2/3})\\
&\lesssim& \left\| F_{\widecheck{\bm{\beta}}}\left((\bm{x}_1 - \bm{x}_2)^T \widecheck{\bm{\beta}}\right) - \widehat F_{\widecheck{\bm{\beta}}}\left((\bm{x}_1 - \bm{x}_2)^T \widecheck{\bm{\beta}}\right) \right\|_{2, F_{X_1,X_2}}^2 + O_p(n^{-2/3}) \\
&=& O_p(n^{-2/3}),
\end{eqnarray*}
where ``$\lesssim$" is because of Lemma \ref{lemma-bar-varphi}, and the last ``$=$" is because of Lemma \ref{lemma-D-beta}. Therefore, we have completed the proof of (\ref{eq-norm-part-3-4}).

Next, we consider $I_3$. Denote
\begin{eqnarray}
\widetilde g_{6, \bm{\beta}, F} (\bm{x}_1 - \bm{x}_2, y_1 - y_2) &=& \left\{\bm{x}_1 - \bm{x}_2 - \varphi_{\bm{\beta}}\left((\bm{x}_1 - \bm{x}_2)^T \bm{\beta}\right)\right\} \nonumber \\ &&  \times \left\{I(y_1>y_2) - F\left((\bm{x}_1 - \bm{x}_2)^T \bm{\beta}\right)\right\} \label{def-g-6}
\end{eqnarray}
and
\begin{eqnarray*}
g_{6, \bm{\beta}, F} (\bm{x}_1 - \bm{x}_2, y_1 - y_2) &=& \widetilde g_{6, \bm{\beta}, F} (\bm{x}_1 - \bm{x}_2, y_1 - y_2)  \\&& -  \widetilde g_{6, \bm{\beta}_0, F_0}(\bm{x}_1 - \bm{x}_2, y_1 - y_2).
\end{eqnarray*}
Then
\begin{eqnarray*}
I_3 = \int \int \widetilde g_{6, \widecheck{\bm{\beta}}, \widehat F_{\widecheck{\bm{\beta}}}} (\bm{x}_1 - \bm{x}_2, y_1 - y_2) d \bbF_{X,Y}(\bm{x}_1, y_1) d \bbF_{X,Y}(\bm{x}_2, y_2).
\end{eqnarray*}
Based on Conditions 1 and A2, we have
\begin{eqnarray*}
\left|g_{6, \widecheck{\bm{\beta}}, \widehat F_{\widecheck{\bm{\beta}}}}(\bm{v}, \bm{u})\right| \lesssim \left|\widecheck{\bm{\beta}} - \bm{\beta}_0 \right| + \left| \widehat F_{\widecheck{\bm{\beta}}}\left(\bm{v}^T \widecheck{\bm{\beta}}\right) - F_{\widecheck{\bm{\beta}}}\left(\bm{v}^T \widecheck{\bm{\beta}}\right)\right| \\ + \left|F_{\widecheck{\bm{\beta}}}\left(\bm{v}^T \widecheck{\bm{\beta}}\right) - F_0(\bm{v}^T \bm{\beta}_0) \right|,
\end{eqnarray*}
up to a constant not depending on $\bm{v}\in \mathX-\mathX$ and $\bm{u}\in \mathY-\mathY$.
With this inequality, working on function classes formed by $g_{6, \bm{\beta}, F} (\bm{x}_1 - \bm{x}_2, y_1 - y_2)$ and using the same arguments as in the proof of Lemma \ref{lemma-u-stat}, we can derive
\begin{eqnarray*}
&&\int \int g_{6, \widecheck{\bm{\beta}}, \widehat F_{\widecheck{\bm{\beta}}}} (\bm{x}_1 - \bm{x}_2, y_1 - y_2) d \bbF_{X,Y}(\bm{x}_1, y_1) d \bbF_{X,Y}(\bm{x}_2, y_2) \\
&=& \int \int g_{6, \widecheck{\bm{\beta}}, \widehat F_{\widecheck{\bm{\beta}}}} (\bm{x}_1 - \bm{x}_2, y_1 - y_2) d F_{X,Y}(\bm{x}_1, y_1) d F_{X,Y}(\bm{x}_2, y_2) + o_p(n^{-1/2}).
\end{eqnarray*}
As a consequence,
\begin{eqnarray}
I_3 &=& \int \int \widetilde g_{6, \bm{\beta}_0, F_0}(\bm{x}_1 - \bm{x}_2, y_1 - y_2) d \bbF_{X,Y}(\bm{x}_1, y_1) d \bbF_{X,Y}(\bm{x}_2, y_2) \nonumber \\
&& + \int \int g_{6, \widecheck{\bm{\beta}}, \widehat F_{\widecheck{\bm{\beta}}}} (\bm{x}_1 - \bm{x}_2, y_1 - y_2) d F_{X,Y}(\bm{x}_1, y_1) d F_{X,Y}(\bm{x}_2, y_2)  \nonumber\\ && + o_p(n^{-1/2}).
\label{eq-norm-part-3-8}
\end{eqnarray}
We consider the second term on the right-hand side of the equality above. Note that it is straightforward to verify that
\begin{eqnarray*}
\int \int \widetilde g_{6, \bm{\beta}_0, F_0} (\bm{x}_1 - \bm{x}_2, y_1 - y_2) d F_{X,Y}(\bm{x}_1, y_1) d F_{X,Y}(\bm{x}_2, y_2) = 0.
\end{eqnarray*}
Therefore, using the notation $\bm{u} = y_1 - y_2$ and $\bm{v} = \bm{x}_1 -\bm{x}_2$, we have
\begin{eqnarray}
&&\int \int g_{6, \widecheck{\bm{\beta}}, \widehat F_{\widecheck{\bm{\beta}}}} (\bm{x}_1 - \bm{x}_2, y_1 - y_2) d F_{X,Y}(\bm{x}_1, y_1) d F_{X,Y}(\bm{x}_2, y_2) \nonumber\\
&=& \int \int \widetilde g_{6, \widecheck{\bm{\beta}}, \widehat F_{\widecheck{\bm{\beta}}}} (\bm{x}_1 - \bm{x}_2, y_1 - y_2) d F_{X,Y}(\bm{x}_1, y_1) d F_{X,Y}(\bm{x}_2, y_2) \nonumber\\
&=& \int \int \left\{\bm{v}- \varphi_{\widecheck{\bm{\beta}}}\left(\bm{v}^T \widecheck{\bm{\beta}}\right)\right\}\left\{I(u>0) - \widehat F_{\widecheck{\bm{\beta}}}\left(\bm{v}^T \widecheck{\bm{\beta}}\right)\right\} d F_{V,U}(\bm{v}, u) \nonumber\\
&=& \int \int \left\{\bm{v}- \varphi_{\widecheck{\bm{\beta}}}\left(\bm{v}^T \widecheck{\bm{\beta}}\right)\right\}\left\{I(u>0) - F_{\widecheck{\bm{\beta}}}\left(\bm{v}^T \widecheck{\bm{\beta}}\right)\right\} d F_{V,U}(\bm{v}, u)\nonumber\\
&&+\int \int \left\{\bm{v}- \varphi_{\widecheck{\bm{\beta}}}\left(\bm{v}^T \widecheck{\bm{\beta}}\right)\right\}\left\{F_{\widecheck{\bm{\beta}}}\left(\bm{v}^T \widecheck{\bm{\beta}}\right) - \widehat F_{\widecheck{\bm{\beta}}}\left(\bm{v}^T \widecheck{\bm{\beta}}\right)\right\} d F_{V}(\bm{v}) \label{eq-norm-part-3-9}\\
&=& \int \int \left\{\bm{v}- \varphi_{\widecheck{\bm{\beta}}}\left(\bm{v}^T \widecheck{\bm{\beta}}\right)\right\}\left\{I(u>0) - F_{\widecheck{\bm{\beta}}}\left(\bm{v}^T \widecheck{\bm{\beta}}\right)\right\} d F_{V,U}(\bm{v}, u) \label{eq-norm-part-3-10}\\
&=& \int \int \bm{v}\left\{I(u>0) - F_{\widecheck{\bm{\beta}}}\left(\bm{v}^T \widecheck{\bm{\beta}}\right)\right\} d F_{V,U}(\bm{v}, u) \label{eq-norm-part-3-11}\\
&=& \psi_0(\widecheck{\bm{\beta}}). \label{eq-norm-part-3-12}
\end{eqnarray}
Some of the above steps need explanation. From (\ref{eq-norm-part-3-9}) to (\ref{eq-norm-part-3-10}), we have used the fact that $\varphi_{\widecheck{\bm{\beta}}}\left(V^T \widecheck{\bm{\beta}}\right)$, $F_{\widecheck{\bm{\beta}}}\left(V^T \widecheck{\bm{\beta}}\right)$, and $\widehat F_{\widecheck{\bm{\beta}}}\left(V^T \widecheck{\bm{\beta}}\right)$ are all measurable functions of $V^T\widecheck{\bm{\beta}}, \widecheck{\bm{\beta}}, (X_1,Y_1) \ldots, (X_n,Y_n)$, so by considering $V$ as a random vector independent of $\widecheck{\bm{\beta}}, (X_1,Y_1) \ldots, (X_n,Y_n)$, we have
{\small
\begin{eqnarray*}
&&\int \int \left\{\bm{v}- \varphi_{\widecheck{\bm{\beta}}}\left(\bm{v}^T \widecheck{\bm{\beta}}\right)\right\}\left\{F_{\widecheck{\bm{\beta}}}\left(\bm{v}^T \widecheck{\bm{\beta}}\right) - \widehat F_{\widecheck{\bm{\beta}}}\left(\bm{v}^T \widecheck{\bm{\beta}}\right)\right\} d F_{V}(\bm{v})\\
&=& E \left[\left\{V- \varphi_{\widecheck{\bm{\beta}}}\left(V^T \widecheck{\bm{\beta}}\right)\right\}\left\{F_{\widecheck{\bm{\beta}}}\left(V^T \widecheck{\bm{\beta}}\right) - \widehat F_{\widecheck{\bm{\beta}}}\left(V^T \widecheck{\bm{\beta}}\right)\right\} \bigg| \{(X_i,Y_i)\}_{i=1}^n, \widecheck{\bm{\beta}} \right]\\
&=&E\left[\left\{E\left(V\Big| V^T \widecheck{\bm{\beta}}\right)- \varphi_{\widecheck{\bm{\beta}}}\left(V^T \widecheck{\bm{\beta}}\right)\right\}\left\{F_{\widecheck{\bm{\beta}}}\left(V^T \widecheck{\bm{\beta}}\right) - \widehat F_{\widecheck{\bm{\beta}}}\left(V^T \widecheck{\bm{\beta}}\right)\right\}\bigg| \{(X_i,Y_i)\}_{i=1}^n, \widecheck{\bm{\beta}}\right]\\
&=& E\left[\left\{\varphi_{\widecheck{\bm{\beta}}}\left(V^T \widecheck{\bm{\beta}}\right) - \varphi_{\widecheck{\bm{\beta}}}\left(V^T \widecheck{\bm{\beta}}\right)\right\}\left\{F_{\widecheck{\bm{\beta}}}\left(V^T \widecheck{\bm{\beta}}\right) - \widehat F_{\widecheck{\bm{\beta}}}\left(V^T \widecheck{\bm{\beta}}\right)\right\}\bigg|\{(X_i,Y_i)\}_{i=1}^n, \widecheck{\bm{\beta}} \right] \\ &=& 0.
\end{eqnarray*}
}
From (\ref{eq-norm-part-3-10}) to (\ref{eq-norm-part-3-11}), we have used the fact that $\varphi_{\widecheck{\bm{\beta}}}\left(V^T \widecheck{\bm{\beta}}\right)$ and $F_{\widecheck{\bm{\beta}}}\left(V^T \widecheck{\bm{\beta}}\right)$ are measurable functions of $V^T\widecheck{\bm{\beta}} , \widecheck{\bm{\beta}}, (X_1,Y_1), \ldots, (X_n,Y_n)$, so by considering $(U,V)$ as independent of $ \widecheck{\bm{\beta}}$, $(X_1,Y_1) \ldots, (X_n,Y_n)$, we have
\begin{eqnarray*}
&&\int \int \varphi_{\widecheck{\bm{\beta}}}\left(\bm{v}^T \widecheck{\bm{\beta}}\right)\left\{I(u>0) - F_{\widecheck{\bm{\beta}}}\left(\bm{v}^T \widecheck{\bm{\beta}}\right)\right\} d F_{V,U}(\bm{v}, u)\\
&=& E\left[\varphi_{\widecheck{\bm{\beta}}}\left(V^T \widecheck{\bm{\beta}}\right)\left\{I(U>0) - F_{\widecheck{\bm{\beta}}}\left(V^T \widecheck{\bm{\beta}}\right)\right\} \bigg| \widecheck{\bm{\beta}}, \{(X_1,Y_1)\}_{i=1}^n \right]\\
&=& E\left[\varphi_{\widecheck{\bm{\beta}}}\left(V^T \widecheck{\bm{\beta}}\right)\left\{E\left\{I(U>0)\Big|V^T\widecheck{\bm{\beta}}\right\} - F_{\widecheck{\bm{\beta}}}\left(V^T \widecheck{\bm{\beta}}\right)\right\} \bigg|  \widecheck{\bm{\beta}}, \{(X_1,Y_1)\}_{i=1}^n \right]\\
&=&E\left[\varphi_{\widecheck{\bm{\beta}}}\left(V^T \widecheck{\bm{\beta}}\right)\left\{F_{\widecheck{\bm{\beta}}}\left(V^T \widecheck{\bm{\beta}}\right) - F_{\widecheck{\bm{\beta}}}\left(V^T \widecheck{\bm{\beta}}\right)\right\} \bigg|  \widecheck{\bm{\beta}}, \{(X_1,Y_1)\}_{i=1}^n \right] = 0.
\end{eqnarray*}
From (\ref{eq-norm-part-3-11}) to (\ref{eq-norm-part-3-12}), we have used the definition of $\psi_0(\cdot)$ given by (\ref{def-psi-0}).

{\color{black} Applying Lemma \ref{lemma-psi-0}}, and combining
(\ref{eq-con-beta-check}), (\ref{def-g-6}), (\ref{eq-norm-part-3-8}), and (\ref{eq-norm-part-3-12}), we have
\begin{eqnarray}
I_3 &=& \int\int\left\{\bm{x}_1 - \bm{x}_2 - \varphi_0\left((\bm{x}_1 - \bm{x}_2)^T \bm{\beta}_0\right)\right\} \nonumber\\&& \times \left\{I(y_1>y_2) - F_0\left((\bm{x}_1 - \bm{x}_2)^T \bm{\beta}_0\right)\right\} d \bbF_{X,Y}(\bm{x}_1, y_1) d \bbF_{X,Y}(\bm{x}_2, y_2) \nonumber\\
&& + {\color{black} A}(\widecheck{\bm{\beta}} - \bm{\beta}_0) + o_p(\widecheck{\bm{\beta}} - \bm{\beta}_0) + o_p(n^{-1/2}).
 \label{con-I-3}
\end{eqnarray}

Combining (\ref{eq-con-I-3-I-4}), (\ref{eq-norm-part-3-4}), and (\ref{con-I-3}), we conclude that  for any estimator $\widecheck{\bm{\beta}}$ of $\bm{\beta}_0$ such that $\widecheck{\bm{\beta}} - \bm{\beta}_0 = o_p(1)$, we have
\begin{eqnarray}
\psi_n(\widecheck{\bm{\beta}}) &=& \int\int\left\{\bm{x}_1 - \bm{x}_2 - \varphi_0\left((\bm{x}_1 - \bm{x}_2)^T \bm{\beta}_0\right)\right\}\nonumber\\ &&\times \left\{I(y_1>y_2) - F_0\left((\bm{x}_1 - \bm{x}_2)^T \bm{\beta}_0\right)\right\} d \bbF_{X,Y}(\bm{x}_1, y_1) d \bbF_{X,Y}(\bm{x}_2, y_2) \nonumber\\
&& + {\color{black} A}(\widecheck{\bm{\beta}} - \bm{\beta}_0) + o_p(\widecheck{\bm{\beta}} - \bm{\beta}_0) + o_p(n^{-1/2}).
 \label{eq-con-psi-n}
\end{eqnarray}

Furthermore, the conclusion in Part (b) of this theorem implies that $\widetilde{\bm{\beta}}^{(U)}$ and $\widetilde{\bm{\beta}}^{(L)}$ given by (\ref{eq-pf-normality-5}) are both consistent estimators for $\bm{\beta}_0$. Combining (\ref{eq-pf-normality-6}) with (\ref{eq-con-psi-n}) leads to
\begin{eqnarray*}
0 &=& \bm{\alpha}_n^T \psi_n(\widetilde{\bm{\beta}}^{(U)}) + (1-\bm{\alpha}_n)^T \psi_n(\widetilde{\bm{\beta}}^{(L)}) \\
&=&\int\int\left\{\bm{x}_1 - \bm{x}_2 - \varphi_0\left((\bm{x}_1 - \bm{x}_2)^T \bm{\beta}_0\right)\right\}\\&&\times\left\{I(y_1>y_2) - F_0\left((\bm{x}_1 - \bm{x}_2)^T \bm{\beta}_0\right)\right\}d \bbF_{X,Y}(\bm{x}_1, y_1) d \bbF_{X,Y}(\bm{x}_2, y_2) \nonumber\\
&& + {\color{black} A}(\widetilde{\bm{\beta}} - \bm{\beta}_0) + o_p(\widetilde{\bm{\beta}} - \bm{\beta}_0) + o_p(n^{-1/2}),
\end{eqnarray*}
which completes our proof of Theorem \ref{thm-normality}. \epf

\subsection{Proof of Corollary \ref{corollary-normality}} \label{section-proof-corrollary}

{\color{black} Based on Theorem \ref{thm-normality}, we need to work with the asymptotics of the statistic
\begin{equation}
\bbU_n = \sum_{i\neq j}
\left\{X_i - X_j - \varphi_0\left((X_i - X_j)^T \bm{\beta}_0\right)\right\}\left\{I(Y_i>Y_j) - F_0\left((X_i - X_j)^T \bm{\beta}_0\right)\right\}. \label{pf-coro-normality-added-1-0}
\end{equation}

Without loss of generality, assume $\bm{\beta}_{0,1} \neq 0$, i.e., the first element of $\bm{\beta}_0$ is nonzero. Then we have $\widetilde A = \psi_0'(\bm{\beta}_0)\Sigma_X^{-1} + \left( \begin{matrix} \bm{\beta}_0^T \\ \bm{0}\end{matrix} \right)$; and it is invertible. If otherwise, there exists $\bm{0}\neq \bm{\alpha} \in \mathbb{R}^p$ such that
\begin{eqnarray}
\bm{\alpha}^T \psi_0'(\bm{\beta}_0) \Sigma_X^{-1} + \bm{\alpha}_1 \bm{\beta}_0^T = 0. \label{pf-coro-normality-added-1}
\end{eqnarray}
\begin{itemize}

\item If $\bm{\alpha}_1 \neq 0$, (\ref{pf-coro-normality-added-1}) leads to
\begin{eqnarray*}
\bm{\beta}_0^T \Sigma_X \bm{\beta}_0 = -\frac{\bm{\alpha}^T}{\bm{\alpha}_1} \psi_0'(\bm{\beta}_0) \bm{\beta}_0 = 0,
\end{eqnarray*}
which contradicts that $\bm{\beta}_0 \neq \bm{0}$ and $\Sigma_X$ is of full rank.

\item If $\bm{\alpha}_1 = 0$, (\ref{pf-coro-normality-added-1}) leads to
\begin{eqnarray*}
\bm{\alpha}_{2:p}^T \psi_0'(\bm{\beta}_0)_{2:p} \Sigma_X^{-1}=0,
\end{eqnarray*}
which together with our assumption that $\Sigma_X$ is of full rank contradicts the conclusion given by Lemma \ref{lemma-rank-psi-0-prime}.

\end{itemize}

For any $i\neq j$, noting that $\left\{\psi_0'(\bm{\beta}_0)\Sigma_X^{-1}\right\} \Sigma_X \bm{\beta}_0 =\bm{0}$ and the assumption given in (\ref{condition-distribution-X}),  we have
\begin{eqnarray*}
 &&\psi_0'(\bm{\beta}_0)\Sigma_X^{-1} \Big[X_i - X_j - E\left\{X_i-X_j|(X_i-X_j)^T \bm{\beta}_0\right\} \Big] \\
&=& \psi_0'(\bm{\beta}_0)\Sigma_X^{-1} (X_i-X_j) - E\left\{ \psi_0'(\bm{\beta}_0)\Sigma_X^{-1} (X_i - X_j)| (X_i-X_j)^T \bm{\beta}_0\right\}\\
&=& \psi_0'(\bm{\beta}_0)\Sigma_X^{-1} (X_i-X_j),
\end{eqnarray*}
and
\begin{eqnarray*}
\bm{\beta}_0^T \Big[X_i - X_j - E\left\{X_i-X_j|(X_i-X_j)^T \bm{\beta}_0\right\} \Big] = 0.
\end{eqnarray*}
As a consequence, for any $i \neq j$, recalling that $$\varphi_0\left((X_i - X_j)^T \bm{\beta}_0\right)=E\left\{X_i-X_j|(X_i-X_j)^T \bm{\beta}_0\right\},$$ we have
\begin{eqnarray*}
\widetilde A^{-1} \widetilde A \left\{X_i - X_j - \varphi_0\left((X_i - X_j)^T \bm{\beta}_0\right) \right\} = \widetilde A^{-1} \psi_0'(\bm{\beta}_0) \Sigma_X^{-1}(X_i-X_j).
\end{eqnarray*}
Therefore
\begin{eqnarray*}
\bbU_n &=& \widetilde A^{-1} \psi_0'(\bm{\beta}_0) \Sigma_X^{-1}\sum_{i\neq j} (X_i-X_j) \left\{I(Y_i>Y_j) - F_0\left((X_i - X_j)^T \bm{\beta}_0\right)\right\}\\
&=& \widetilde A^{-1} \psi_0'(\bm{\beta}_0) \Sigma_X^{-1}\sum_{i\neq j} h(X_i, Y_i; X_j, Y_j),
\end{eqnarray*}
where
\begin{eqnarray*}
h(\bm{x}_1, y_1; \bm{x}_2, y_2) = (\bm{x}_1-\bm{x}_2) \left\{I(y_1>y_2) - F_0\left((\bm{x}_1 - \bm{x}_2)^T \bm{\beta}_0\right) \right\}.
\end{eqnarray*}


Because of Conditions 0 and A1, with probability equal to 1, $Y_i \neq Y_j$ for $i\neq j$. Using the fact that $F_0(\cdot)$ is the c.d.f. of a symmetric distribution about 0, we have
\begin{eqnarray*}
&& I(y_1>y_2) - F_0\left( (\bm{x}_1-\bm{x}_2)^T\bm{\beta}_0 \right) \\&=& 1- I(y_1<y_2) - \left\{ 1-F_0\left( -(\bm{x}_1-\bm{x}_2)^T\bm{\beta}_0 \right) \right\}\\
&=& - \left\{ I(y_2>y_1) - F_0 \left((\bm{x}_2-\bm{x}_1)^T\bm{\beta}_0 \right) \right\},
\end{eqnarray*}
and
\begin{eqnarray*}
\bm{x}_1 - \bm{x}_2
= - \left(\bm{x}_2 - \bm{x}_1\right).
\end{eqnarray*}
Therefore,
\begin{eqnarray*}
h(\bm{x}_1, y_1; \bm{x}_2, y_2) = h(\bm{x}_2, y_2; \bm{x}_1, y_1),
\end{eqnarray*}
i.e., $h$ is a symmetric kernel for the U-statistic contained in $\bbU_n$.
Furthermore, it is straightforward to verify that $E\left\{h(X_1, Y_1; X_2, Y_2)\right\} = 0$; we have
\begin{eqnarray*}
\overline{\bbU}_n &=&  \sum_{i=1}^n E(\bbU_n|X_i, Y_i)\\
&=& \frac{2(n-1)}{n^2} \widetilde A^{-1} \psi_0'(\bm{\beta}_0) \Sigma_X^{-1}  \sum_{i=1}^n \hbar(X_i,Y_i),
\end{eqnarray*}
where we recall that
\begin{eqnarray*}
\hbar(\bm{x}_1, y_1) = E\left\{ h(\bm{x}_1, y_1; X_2, Y_2) \right\}.
\end{eqnarray*}
With the condition that $\lambda_{\max}(\Sigma_{\hbar}) <\infty$, and applying Theorem 12.3 in van der Vaart (1998), we have that
\begin{eqnarray}
\bbU_n - \overline{\bbU}_n = o_p(n^{-1/2}) \label{pf-coro-normality-2}
\end{eqnarray}
holds entry-wise. By the condition $0<\lambda_{\min}(\Sigma_{\hbar}) \leq \lambda_{\max}(\Sigma_{\hbar})<\infty$ and the Central Limit Theorem, we have
\begin{eqnarray}
\sqrt{n}\overline{\bbU}_n \to 2 \widetilde A^{-1} \psi_0'(\bm{\beta}_0) \Sigma_X^{-1}  N(0, \Sigma_{\hbar}), \label{pf-coro-normality-3}
\end{eqnarray}
where we have used the fact that $E\left\{\hbar(X_1, Y_1) \right\} = 0$. Combining (\ref{pf-coro-normality-added-1-0}), (\ref{pf-coro-normality-2}), (\ref{pf-coro-normality-3}) with (\ref{beta-tilde-expansion}) leads to (\ref{eq-coro-normality}). This completes the proof of this corollary. \epf

}

{\color{black} 

\section{Proof of Theorem \ref{theorem-F-asymptotics}} \label{sec-proof-normality-F-betahat}

In this section,  we shall adopt the following notations. For any function $f(\bm{x}_1, y_1; \bm{x}_2, y_2)$, we denote
\begin{eqnarray*}
\mathbb{P}^2 f = \int f(\bm{x}_1, y_1; \bm{x}_2, y_2) dF_{X, Y}(\bm{x}_1, y_1) d F_{X,Y}(\bm{x}_2, y_2),
\end{eqnarray*}
and $\mathbb{U}_n^2 f$ to be the second order $U$-statistic with $f$ being the kernel.
For any function $f(\bm{x}, y)$, we denote
\begin{eqnarray*}
\mathbb{P} f =\mathbb{P} f(\cdot, \cdot) = \int f(\bm{x},y) dF_{X, Y}(\bm{x}, y)\\
\mathbb{P}_n f = \mathbb{P}_n f(\cdot, \cdot) =  \frac{1}{n}\sum_{i=1}^n f(X_i,Y_i)\\
\mathbb{G}_n f = \sqrt{n}\left(\mathbb{P}_n f - \mathbb{P} f \right).
\end{eqnarray*}
We generally use ``$\bar F$" to denote the envelope function of a function class.
Denote $T_{i,j} = (X_i - X_j)^T\bm{\beta}$, $\widetilde T_{i,j} = (X_i - X_j)^T\widetilde{\bm{\beta}}$, $T_{i,j,0} = (X_i - X_j)^T \bm{\beta}_0$, and $\Delta_{i,j} = I(Y_i>Y_j)$; then the expression for $\ell(\bm{\beta}, F)$ can be rewritten to be
\begin{equation}
\ell(\bm{\beta}, F) = \sum_{i,j}\left[\Delta_{i,j} \log\{ F(T_{i,j})\}+(1-\Delta_{i,j}) \log \{1-F(T_{i,j})\}\right]. \label{F-con-eq-3}
\end{equation}
For any $\bm{\beta}$, let $G_{\bm{\beta}}(\cdot)$ and $G_n(\cdot)$ be the c.d.f. and empirical c.d.f. of $T_{i,j}$ respectively. Denote $G_0 = G_{\bm{\beta}_0}$, $g_{\bm{\beta}}(t) = \frac{\partial G_{\bm{\beta}}(t)}{\partial t}$, and $g_0(t)  = \frac{\partial G_0(t)}{\partial t}$. Let
\begin{eqnarray*}
V_n(t) = \frac{1}{n^2} \sum_{i,j} \Delta_{i,j} I(T_{i,j}\leq t). \label{F-con-eq-4}
\end{eqnarray*}
In addition, we add tilde to a notation to denote the corresponding expression whose $\bm{\beta}$ is replaced by $\widetilde{\bm{\beta}}$; for example,
\begin{eqnarray*}
\widetilde V_n(t)  = \frac{1}{n^2} \sum_{i,j} \Delta_{i,j} I(\widetilde T_{i,j}\leq t).
\end{eqnarray*}

We sketch a blueprint of the proof for this theorem first; the technical details are then organised in the lemmas afterwards.
Note that for every $\bm{\beta}$, $\widehat F_{\bm{\beta}}$ is the maximiser of $\ell(\bm{\beta}, F)$ given by (\ref{F-con-eq-3}); it is the left derivative of the greatest convex minorant of the cumulative sum diagram defined by $(G_n(t), V_n(t)), t \in \mathbb{R}$.
Furthermore, let
\begin{eqnarray}
U_n(s) = \mbox{argmin}_{t\in \mathbb{R}}\{ V_n(t) - s G_n(t) \}. \label{F-con-eq-5}
\end{eqnarray}
Following the discussion given at the beginning of the proof for Theorem 3.7 of Groeneboom and Jongbloed (2014),
we have the switching relation:
\begin{eqnarray*}
\widehat F_{\bm{\beta}}(t) \geq s \Leftrightarrow G_n(t) \geq G_n(U_n(s)) \Leftrightarrow t\geq U_n(s). \label{F-con-eq-6}
\end{eqnarray*}
Set $s_0 \equiv F_0(t)$; we have
\begin{eqnarray}
P\left(n^{1/2}\left\{ \widehat F_{\widetilde{\bm{\beta}}}(t) - F_0(t)\right\} \geq x \right) &=& P\left( \widehat F_{\widetilde{\bm{\beta}}}(t) \geq s_0 + n^{-1/2} x \right) \nonumber \label{F-con-eq-7}\\
&=& P\left(t\geq \widetilde U_n(s_0+n^{-1/2} x)\right) \label{F-con-eq-8} \nonumber\\
&=& P\left(n^{1/2}\left\{ \widetilde U_n(s_0+n^{-1/2} x) - t \right\}\leq 0\right). \label{F-con-eq-9}
\end{eqnarray}
Therefore, to derive the asymptotic distribution for $n^{1/2}\left\{ \widehat F_{\widetilde{\bm{\beta}}}(t) - F_0(t)\right\}$, we can work on the asymptotic distribution of $n^{1/2}\left\{ \widetilde U_n(s_0+n^{-1/2} x) - t \right\}$ instead. This is achieved by three lemmas.

\begin{itemize}
\item
We first establish that for any $M>0$,  $n^{1/3}\sup_{x\in[-M,M]}\left| \widetilde U_n(s_0+n^{-1/2} x) - t \right| = O_p(1)$ in Lemma \ref{lemma-U-cubic-con}.

\item We then show that $n^{1/2}\sup_{x\in[-M,M]} \left| \widetilde U_n(s_0+n^{-1/2} x) - t \right|= O_p(1)$ in Lemma \ref{lemma-U-root-con}.

\item Last with the argmax theorem reviewed by Lemma \ref{argmax}, we establish the asymptotic distribution for $n^{1/2}\left\{ \widetilde U_n(s_0+n^{-1/2} x) - t \right\}$ in Lemma \ref{lemma-U-asymptotic}:
\begin{eqnarray}
n^{1/2}\left\{ \widetilde U_n(s_0+n^{-1/2} x) - t \right\}  \rightsquigarrow \frac{g_0(t)x - \sigma(t) Z}{f_0(t) g_0(t)}. \label{F-con-eq-9-1}
\end{eqnarray}

\end{itemize}

Combining (\ref{F-con-eq-9-1}) with (\ref{F-con-eq-9}) leads to
\begin{eqnarray*}
P\left(n^{1/2}\left\{ \widehat F_{\widetilde{\bm{\beta}}}(t) - F_0(t)\right\} \geq x \right) &\to& P\left( \frac{g_0(t) x - \sigma(t) Z}{g_0(t)f_0(t)} \leq 0 \right)\\
&=& P\left(\frac{\sigma(t)}{g_0(t)}Z\geq x\right),
\end{eqnarray*}
where $Z\sim N(0,1)$, which implies $n^{1/2}\left\{ \widehat F_{\widetilde{\bm{\beta}}}(t) - F_0(t)\right\} \rightsquigarrow N\left(0,  \sigma^2(t)/g_0^2(t)\right)$. This completes the proof of this theorem. \epf

We first establish the tightness of $n^{1/3}\sup_{x\in[-M,M]}\left| \widetilde U_n(s_0+n^{-1/2} x) - t \right| = O_p(1)$ in the following lemma.

\begin{lemma} \label{lemma-U-cubic-con}
Assume that all conditions for Theorem \ref{theorem-F-asymptotics} are effective. For every $M>0$, we have
\begin{eqnarray}
n^{1/3}\sup_{x\in [-M, M]}\left| \widetilde U_n(s_0+n^{-1/2} x) - t \right| = O_p(1). \label{cubic-root-consistency}
\end{eqnarray}

\end{lemma}

\pf Based on Corollary \ref{corollary-normality}, $\widetilde{\bm{\beta}} - \bm{\beta}_0 = O_p\left(n^{-1/2}\right)$; therefore for any $\epsilon>0$, there exists a universal constant $C>0$, such that
\begin{eqnarray*}
P\left(\widetilde{\bm{\beta}} \in \mathcal{B}_{n, C}\right) > 1- \epsilon,
\end{eqnarray*}
where $\mathcal{B}_{n,C} = \left\{\bm{\beta}: \|\bm{\beta}- \bm{\beta}_0\|_2\leq n^{-1/2}C \right\}$. Hereafter, if appropriate we treat  $\widetilde{\bm{\beta}} \in \mathcal{B}_{n,C}$ to avoid some presentational complication; for example, in some occasions, we are to show that the probability of some event, which involves $\widetilde{\bm{\beta}}$, can be arbitrarily small; we can treat $\widetilde{\bm{\beta}} \in \mathcal{B}_{n,C}$, since otherwise we can use $\left\{\widetilde{\bm{\beta}} \in \mathcal{B}_{n, C}\right\}$ to intersect the event of interest.

Note that $U_n(s)$ is nondecreasing in $s$; this is because for any $s_1<s_2$, let $t_1 = U_n(s_1)$ and $t_2 = U_n(s_2)$, then we must have $t_1\leq t_2$; since if otherwise, recalling the definition of $U_n(\cdot)$ given by (\ref{F-con-eq-5}), we have
\begin{eqnarray*}
&&V_n(t_2) - s_2G_n(t_2) \\&=& \frac{1}{n^2} \sum_{i,j}(\Delta_{i,j} - s_2)I(T_{i,j}\leq t_2) \\
&=& \frac{1}{n^2} \sum_{i,j}(\Delta_{i,j} - s_1)I(T_{i,j}\leq t_2) + \frac{1}{n^2} \sum_{i,j}(s_1 - s_2)I(T_{i,j}\leq t_2) \\
&\geq&  \frac{1}{n^2} \sum_{i,j}(\Delta_{i,j} - s_1)I(T_{i,j}\leq t_1) + \frac{1}{n^2} \sum_{i,j}(s_1 - s_2)I(T_{i,j}\leq t_1)\\
&=& \frac{1}{n^2} \sum_{i,j}(\Delta_{i,j} - s_2)I(T_{i,j}\leq t_1)\\
&=& V_n(t_1) - s_2G_n(t_1), \label{F-con-eq-10}
\end{eqnarray*}
which contradicts the definition of $t_2$. Therefore, to show (\ref{cubic-root-consistency}), we need to show that for any $\epsilon>0$, there exists an $M'>0$, such that
\begin{eqnarray}
P\left(n^{1/3} \left\{\widetilde U_n(s_0+n^{-1/2} M) - t\right\} > M' \right)<\epsilon   \label{F-con-eq-11} \\
P\left(n^{1/3} \left\{\widetilde U_n(s_0-n^{-1/2} M) - t\right\} < -M' \right)<\epsilon. \label{F-con-eq-12}
\end{eqnarray}
We need to show only (\ref{F-con-eq-11}); (\ref{F-con-eq-12}) can be shown using exactly the same arguments.
Note that
\begin{eqnarray*}
&&P\left(n^{1/3} \left\{\widetilde U_n(s_0+n^{-1/2} M) - t\right\} > M' \right) \label{F-con-eq-13}\\
&\leq& P\bigg( \exists u > M': \widetilde V_n(t+n^{-1/3}u) - (s_0+n^{-1/2} M) \widetilde G_n(t+n^{-1/3}u) \\ && \hspace{0.3in}\leq \widetilde V_n(t) - (s_0+n^{-1/2} M) \widetilde G_n(t) \bigg) \label{F-con-eq-14}\\
&=& P\left( \exists u > M': \widetilde W_{n,1}(u) + \widetilde W_{n,2}(u) - \widetilde W_{n,3}(u) + \widetilde W_{n,4}(u) \leq 0 \right), \label{F-con-eq-15}
\end{eqnarray*}
where we recall that $s_0 = F_0(t)$, and
\begin{eqnarray}
W_{n,1}(u) &=& n^{2/3} \left\{ V_n(t+n^{-1/3}u) - V_n(t) \right\} \nonumber \\&& - \frac{n^{2/3}}{n^2}\sum_{i,j}F_0(T_{i,j,0})I(T_{i,j}\in[t, t+n^{-1/3}u]) \label{F-con-eq-16}\nonumber\\
W_{n,2}(u) &=& n^{2/3} \int_{w\in[t, t+n^{-1/3}u]}\{ F_0(w) - F_0(t)\} d G_n(w)  \label{F-con-eq-17} \nonumber \\
W_{n,3}(u) &=& n^{1/6} M  \left\{G_n(t+n^{-1/3}u) - G_n(t) \right\} \nonumber\\
W_{n,4}(u) &=& \frac{n^{2/3}}{n^2}\sum_{i,j}\Big[\left\{F_0(T_{i,j,0})-F_0(T_{i,j})\right\} \nonumber \\ && \hspace{0.6in}\times I(T_{i,j}\in[t, t+n^{-1/3}u]) \Big].\label{F-con-eq-18}
\end{eqnarray}

We derive the asymptotic properties of $\widetilde W_{n, i}(u)$ for $i=1,2,3,4$ separately. In particular, we show that
\begin{itemize}
\item[(a)]
for any $\eta>0, \epsilon>0$, and $M'>0$, when $M_1$ is sufficiently large, we have
\begin{eqnarray}
P\left(\exists u>M': \left|\widetilde W_{n,1}(u) \right| \geq \eta u^2 + M_1 \right) < \epsilon; \label{F-con-eq-19}
\end{eqnarray}

\item[(b)] there exists a universal constant $c>0$, such that for any $\epsilon>0$, when $M'$ is sufficiently large, we have
    \begin{eqnarray}
    P\left(\exists u>M': \widetilde W_{n,2}(u)  \geq c u^2 \right) > 1- \epsilon; \label{F-con-eq-19-1}
    \end{eqnarray}

\item[(c)] for any $\epsilon>0$ and $M'>0$, when $M_1$ is sufficiently large, we have
    \begin{eqnarray*}
    P\left(\exists u>M': \left|\widetilde W_{n,3}(u) \right| \geq M_1 \right) < \epsilon; \label{F-con-eq-19-2}
    \end{eqnarray*}

\item[(d)] for any $\epsilon>0$ and $M'>0$, when $M_1$ is sufficiently large, we have
\begin{eqnarray*}
    P\left(\exists u>M': \left|\widetilde W_{n,4}(u) \right| \geq M_1 \right) < \epsilon. \label{F-con-eq-19-3}
    \end{eqnarray*}

\end{itemize}

We show (a) first. Note that
\begin{eqnarray*}
W_{n,1}(u) &=& \frac{n^{2/3}}{n^2} \sum_{i,j} \left\{\Delta_{i,j} - F_0(T_{i,j,0})\right\}I(T_{i,j} \in [t, t+n^{-1/3}u]) \label{F-con-eq-20}\\
&\equiv& \frac{n^{2/3}}{n^2}  \sum_{i,j} f_{n, u}(X_i, Y_i, X_j, Y_j), \label{F-con-eq-21}
\end{eqnarray*}
where
\begin{equation}
f_{n, u}(X_i, Y_i; X_j, Y_j) =  \left\{\Delta_{i,j} - F_0(T_{i,j,0})\right\}I(T_{i,j} \in [t, t+n^{-1/3}u]). \label{F-con-eq-22}
\end{equation}
We observe that the main part of $W_{n,1}(u)$ is a second order $U$ statistic, and that $\mathbb{P}^2 f_{n,u}(\cdot, \cdot; \cdot, \cdot) = 0$; we have the decomposition (see Serfling, 1980, pages 177--178):
\begin{eqnarray}
W_{n,1}(u) = n^{2/3} \left\{\mathbb{U}_n^2 \bar f_{n,u} + \mathbb{P}_n f_{n,u,1} + \mathbb{P}_n f_{n,u,2} \right\}, \label{F-con-eq-23}
\end{eqnarray}
where $\bar f_{n,u} = f_{n,u} - f_{n,u,1} - f_{n,u,2}$; $\mathbb{U}_n^2 \bar f_{n,u}$ is a degenerate U-statistic. For $\mathbb{U}_n^2 \bar f_{n,u}$,
based on the discussion in van der Vaart and Wellner (1996) (page 98), we observe that the concept ``covering number" in (\ref{covering-F-n}) can be replaced with ``packing number" given by Definition 1 in Sherman (1994). Then, we immediately have that the function class
\begin{eqnarray}
\mathcal{F}_n = \{f_{n,u}(\cdot, \cdot, \cdot, \cdot): u\in \mathbb{R}, \bm{\beta}\in \mathcal{B} \} \label{F-con-eq-24}
\end{eqnarray}
is Euclidean$(C, 8p+12)$ for envelope function $F=2$ with $C$ being a universal constant not depending on $n$ and $t$; see Definition 3 in Sherman (1994) for the definition of Euclidean. With an application of Lemma 6 and then Corollary 4 in Sherman (1994), we conclude that
\begin{eqnarray}
n\sup_{u\in \mathbb{R}, \bm{\beta}\in \mathcal{B}}\left|\mathbb{U}_n^2 \bar f_{n,u}\right| = O_p(1). \label{F-con-eq-25}
\end{eqnarray}

For $\mathbb{P}_n f_{n,u,1}$, we consider $[M', \infty) \subset \cup_{j=\lfloor M' \rfloor}^\infty A_j \equiv \cup_{j=\lfloor M' \rfloor}^\infty [j, (j+1)]$, where $\lfloor M' \rfloor$ denotes the largest integer that is less than or equal to $M'$.  Then we have
\begin{eqnarray}
&&P\left(\exists u>M': n^{2/3} \sup_{\bm{\beta}\in \mathcal{B}_{n,C}}\left|\mathbb{P}_n f_{n,u,1}\right| \geq \eta u^2 + M_1 \right) \label{F-con-eq-26} \nonumber\\
&\leq&\sum_{j=\lfloor M' \rfloor}^\infty P\left(\exists u\in A_j:  n^{2/3} \sup_{\bm{\beta}\in \mathcal{B}_{n,C}}\left|\mathbb{P}_n f_{n,u,1}\right| \geq \eta  u^2 + M_1\right) \label{F-con-eq-27} \nonumber\\
&\leq& \sum_{j=\lfloor M' \rfloor}^\infty P\left(\exists u\in A_j:  n^{2/3} \sup_{\bm{\beta}\in \mathcal{B}_{n,C}}\left|\mathbb{P}_n f_{n,u,1}\right| \geq \eta j^2 + M_1\right) \label{F-con-eq-28} \nonumber\\
&\leq& n^{4/3}\sum_{j=1}^\infty \frac{E\left\{\sup_{u\in A_{j}, \bm{\beta}\in \mathcal{B}_{n,C}}\left|\mathbb{P}_n f_{n,u,1}\right|^2\right\}}{(\eta j^2 + M_1)^2}. \label{F-con-eq-29}
\end{eqnarray}
Note that $\mathcal{F}_{n, 1, j} = \{f_{n,u,1}: u\in A_j, \bm{\beta} \in \mathcal{B}_{n,C}\}$ is a subset of the function class
\begin{eqnarray*}
\mathcal{F}_{n, 1} =\{f_{n,u,1}(\boldsymbol{x}, y) = \mathbb{P} f_{n,u}(\cdot, \cdot, \boldsymbol{x}, y): u\in \mathcal{R}, \bm{\beta}\in \mathcal{B}\}, \label{F-con-eq-30}
\end{eqnarray*}
considered in Lemma \ref{lemma-covering-F}.
Based on the definition of $f_{n,u,1}$, $\mathcal{B}_{n,C}$, we can find a universal constant $C_1>0$, such that
\begin{eqnarray*}
&&\sup_{u\in A_j, \bm{\beta} \in \mathcal{B}_{n,C}} \left|f_{n,u,1}(\bm{x}, y)\right| \\
&\leq&  C_1 \int_{\bm{w}\in \mathcal{X}} I\left\{ (\bm{w}-\bm{x})^T \bm{\beta}_0 \in[t- C_1 n^{-1/2}, t+n^{-1/3}(j+1) + C_1 n^{-1/2}] \right\} \\
&& \hspace{3.5in} \times d F_X(\bm{w})\\
&=& C_1\Big[ F_{X^T\bm{\beta}_0}\left(\bm{x}^T \bm{\beta}_0 + t+n^{-1/3}(j+1) + C_1 n^{-1/2}\right)\\ && \hspace{0.3in}- F_{X^T\bm{\beta}_0}\left(\bm{x}^T \bm{\beta}_0 + t - C_1 n^{-1/2}\right)\Big]\\
&\equiv& \bar F(\bm{x}),
\end{eqnarray*}
which serves as an envelope function of $\mathcal{F}_{n, 1, j}$. Based on Condition 2, we have $\|\bar F\|_{2, P} \lesssim (j+1)n^{-1/3}$. Then, applying Theorem 11.1 in Kosorok (2008), we have
\begin{eqnarray}
\label{F-con-eq-31} && n E\left\{\sup_{u\in A_j, \bm{\beta}\in \mathcal{B}_{n,C}}\left|\mathbb{P}_n f_{n,u,1}\right|^2\right\} \\  &\lesssim& \{J^*(1,\mathcal{F}_{n,1})\}^2 \|\bar F\|_{2,P}^2 \lesssim (j+1)^2 n^{-2/3}, \nonumber
\end{eqnarray}
where we have used the fact that
\begin{eqnarray*}
J^*(1,\mathcal{F}_{n,1}) \equiv \sup_{Q} \int_0^1\sqrt{1+H_2(\epsilon \|\bar F\|_{2, Q}, \mathcal{F}_{n,1,j}, Q)} d\epsilon \lesssim 1. \label{F-con-eq-32}
\end{eqnarray*}
Combining (\ref{F-con-eq-29}) with (\ref{F-con-eq-31}) leads to
\begin{eqnarray}
&&P\left(\exists u>M': n^{2/3} \sup_{\bm{\beta}\in \mathcal{B}_{n,C}}\left|\mathbb{P}_n f_{n,u,1}\right| \geq \eta u^2 + M_1 \right) \nonumber \\
&\lesssim& n^{-1/3} \sum_{j=\lfloor M' \rfloor}^\infty \frac{(j+1)^2}{(\eta j^2 + M_1)^2} < \epsilon/3, \label{F-con-eq-34}
\end{eqnarray}
with $M_1$ chosen to be sufficiently large. Using exactly the same strategy as the derivation for (\ref{F-con-eq-34}), we can obtain
\begin{eqnarray}
&& P\left(\exists u>M': n^{2/3} \sup_{\bm{\beta}\in \mathcal{B}_{n,C}} \left|\mathbb{P}_n f_{n,u,2}\right| \geq \eta u^2 + M_1 \right) \nonumber \\  &\lesssim& n^{-1/3} \sum_{j=\lfloor M' \rfloor}^\infty \frac{(j+1)^2}{(\eta j^2 + M_1)^2} < \epsilon/3. \label{F-con-eq-36}
\end{eqnarray}
Combining  (\ref{F-con-eq-25}), (\ref{F-con-eq-34}), and (\ref{F-con-eq-36}) leads to (\ref{F-con-eq-19}); we complete the proof for (a).

We proceed to show (b). We can write
\begin{eqnarray}
W_{n,2} &=& \frac{n^{2/3}}{n^2} \sum_{i,j} \left\{F_0(T_{i,j}) - F_0(t)\right\} I\left(T_{i,j} \in [t, t+n^{-1/3} u)\right) \label{F-con-eq-37}\nonumber\\
&\equiv& \frac{n^{2/3}}{n^2} \sum_{i,j} g_{n,u}(X_i, X_j) \label{F-con-eq-38} \nonumber\\
&=& \frac{n^{2/3}}{n^2} \sum_{i,j} \left\{ g_{n,u}(X_i, X_j) - \mathbb{P}^2 g_{n,u}(\cdot, \cdot) \right\} + n^{2/3}\mathbb{P}^2 g_{n,u}(\cdot, \cdot).  \label{F-con-eq-39}
\end{eqnarray}
Using exactly the same procedure as the proof for (a), we are able to show that
for any $\eta>0, \epsilon>0$, and $M'>0$, when $M_1$ is sufficiently large, we have
\begin{eqnarray}
P\left(\exists u>M': \left|\frac{n^{2/3}}{n^2} \sum_{i,j} \left\{ \widetilde g_{n,u}(X_i, X_j) - \mathbb{P}^2 \widetilde g_{n,u}(\cdot, \cdot) \right\} \right| \geq \eta u^2 + M_1 \right) \nonumber \\  < \epsilon.  \label{F-con-eq-40}
\end{eqnarray}
Furthermore,  recall that $G_{\bm{\beta}}(w)$ is the c.d.f. of $T_{i,j}$; based on Conditions F1 and F2 and the discussion in Remark \ref{remark-condition-F}, we can verify that there exists a universal constant $c>0$, such that
\begin{eqnarray}
&& n^{2/3} \inf_{\bm{\beta} \in \mathcal{B}_{n,C}} \mathbb{P}^2 g_{n,u}(\cdot, \cdot)  \nonumber\\ &=& n^{2/3} \inf_{\bm{\beta} \in \mathcal{B}_{n,C}} \int_{w\in [t, t+n^{-1/3}u)} \{F_0(w) - F_0(t) \} dG_{\bm{\beta}}(w) \geq c u^2. \label{F-con-eq-41}
\end{eqnarray}
Combining (\ref{F-con-eq-39})--(\ref{F-con-eq-41}) leads to (\ref{F-con-eq-19-1}); we complete the proof for (b).

Last, the proofs for (c) and (d) are straightforward, and are omitted.

Now, combining (a), (b), (c), and (d), we have that for any $\epsilon>0$, when $M'$ is sufficiently large,
\begin{eqnarray}
\label{F-con-eq-42} &&P\left( \exists u > M': \widetilde W_{n,1}(u) + \widetilde W_{n,2}(u) - \widetilde W_{n,3}(u) + \widetilde W_{n,4}(u) \leq 0 \right)\\
&\leq& P\left( \exists u > M': \widetilde W_{n,2}(u) < cu^2  \right) \nonumber\\ && + P\Big(\exists u > M':  cu^2 \leq \widetilde W_{n,2}(u) \nonumber \\&& \hspace{1.2in} \leq - \widetilde W_{n,1}(u) + \widetilde W_{n,3}(u) - \widetilde W_{n,4}(u) \Big)\nonumber\\
&\leq& \epsilon + P\Big(\exists u > M': cu^2 \leq \widetilde W_{n,2}(u) \nonumber \\ && \hspace{1.2in}  \leq |\widetilde W_{n,1}(u)| + |\widetilde W_{n,3}(u)| + |\widetilde W_{n,4}(u)|\Big)\nonumber\\
&\leq & \epsilon + P\left(\exists u> M': |\widetilde W_{n,1}(u)|\geq cu^2/3\right) \nonumber \\ && + P\left(\exists u> M': |\widetilde W_{n,3}(u)|\geq cu^2/3\right) \nonumber \\&&  + P\left(\exists u> M': |\widetilde W_{n,4}(u)|\geq cu^2/3\right) \nonumber\\
&\leq& \epsilon + \epsilon + \epsilon + \epsilon. \nonumber
\end{eqnarray}
 This together with (\ref{F-con-eq-18}) and that $\epsilon$ is arbitrary leads to (\ref{F-con-eq-11}). We complete the proof of this lemma.

\vspace{0.2in}

In the proof of Lemma \ref{lemma-U-cubic-con}, we have used the entropy conditions given in the following lemma. \epf

\begin{lemma} \label{lemma-covering-F}
Recall $f_{n,u}$ defined by (\ref{F-con-eq-22}). Consider the function classes
\begin{eqnarray*}
\mathcal{F}_{n} &=& \{f_{n,u}(\cdot, \cdot; \cdot, \cdot): u\in \mathbb{R}; \bm{\beta}\in \mathcal{B}\}\\
\mathcal{F}_{n, 1} &=& \{f_{n,u,1}(\boldsymbol{x}, y) = \mathbb{P} f_{n,u}(\cdot, \cdot, \boldsymbol{x}, y): u\in \mathbb{R}, \bm{\beta}\in \mathcal{B}\}\\
\mathcal{F}_{n, 1} &=& \{f_{n,u,2}(\boldsymbol{x}, y) = \mathbb{P} f_{n,u}(\boldsymbol{x}, y, \cdot, \cdot): u\in \mathbb{R},  \bm{\beta}\in \mathcal{B}\}.
\end{eqnarray*}
Then
\begin{itemize}
\item[(a)]
for any probability measure $Q$ defined on the $\sigma$-algebra $\sigma(X_1, Y_1)\times \sigma(X_2, Y_2)$, $r\geq 1$ and any envelope function $\bar F$ with $\|\bar F\|_{r,Q}>0$; and any $0<\epsilon <1$, we have
\begin{eqnarray}
N_r(\epsilon\|\bar F\|_{r,Q}, \mathcal{F}_n, Q) \lesssim (1/\epsilon)^{4pr+6r}, \label{covering-F-n}
\end{eqnarray}
where recall that $p$ is the dimension of $\bm{\beta}$;

\item[(b)] for any probability measure $Q$ defined on the $\sigma$-algebra $\sigma(X,Y)$, $r\geq 1$ and any envelope function $\bar F$ with $\|\bar F\|_{r,Q}>0$, we have
\begin{eqnarray}
N_r(\epsilon\|\bar F\|_{r,Q}, \mathcal{F}_{n,1}, Q) \lesssim (1/\epsilon)^{{4pr+6r}} \label{covering-F-n-1}\\
N_r(\epsilon\|\bar F\|_{r,Q}, \mathcal{F}_{n,2}, Q) \lesssim (1/\epsilon)^{{4pr+6r}}. \label{covering-F-n-2}
\end{eqnarray}

\end{itemize}

\end{lemma}

\pf Note that for any given $t$ and $n$, based on Lemma 9.6 in Kosorok (2008), we have that the function classes
\begin{eqnarray*}
\left\{\bm{x}^T \bm{\beta} - t-n^{-1/2}u: \bm{\beta}\in \mathcal{B}, u\in \mathbb{R} \right\}\\
\left\{\bm{x}^T \bm{\beta} - t: \bm{\beta}\in \mathcal{B}\right\}
\end{eqnarray*}
are both VC-subgraph classes with VC indexes $p+3$ and $p+2$ respectively. Then, based on Lemma 9.9 (iii) and (iv) in Kosorok (2008), the sets
\begin{eqnarray*}
\mathcal{C}_1 &=& \left\{ \left\{\bm{x}^T \bm{\beta} - t-n^{-1/2}u \leq 0 \right\}: \bm{\beta}\in \mathcal{B}, u\in \mathbb{R} \right\}\\
\mathcal{C}_2 &=& \left\{ \left\{\bm{x}^T \bm{\beta} - t \geq 0 \right\}: \bm{\beta}\in \mathcal{B}, u\in \mathbb{R} \right\}
\end{eqnarray*}
are both VC-classes of sets with $V_{\mathcal{C}_1} = p+3$ and $V_{\mathcal{C}_2} = p+2$. Applying Lemma 9.7 (ii) in Kosorok (2008), we conclude that
\begin{eqnarray*}
\left\{ \left\{\bm{x}^T \bm{\beta}  \in [t, t+n^{-1/2}u] \right\}: \bm{\beta}\in \mathcal{B}, u\in \mathbb{R} \right\} \subset \mathcal{C}_1 \sqcap \mathcal{C}_2
\end{eqnarray*}
is VC with index $\leq 2p+5-1$.  By Lemma 9.8, and Lemma 9.9 (vi) and (vii) in Kosorok (2008), we have that the function class $\mathcal{F}_n$ is VC with index $\leq 2(2p+4)-1 = 4p+7$.

Consequently, applying Theorem 9.3 in Kosorok (2008) leads to (\ref{covering-F-n}), based on which and applying Jensen inequality leads to (\ref{covering-F-n-1}) and (\ref{covering-F-n-2}). We complete the proof of this lemma. \epf

\begin{lemma}  \label{lemma-U-root-con}
Assume all conditions for Theorem \ref{theorem-F-asymptotics} are effective. For every $M>0$, we have
\begin{eqnarray}
n^{1/2}\sup_{x\in [-M, M]}\left| \widetilde U_n(s_0+n^{-1/2} x) - t \right| = O_p(1). \label{root-n-consistency}
\end{eqnarray}
\end{lemma}
\pf Similar in strategy to the proof of Lemma \ref{lemma-U-cubic-con}, because of the monotonicity of $U_n(\cdot)$, we need to show only that for any $\epsilon> 0$, there exists $M'>0$ such that
\begin{eqnarray}
P\left(n^{1/2} \left\{\widetilde U_n(s_0+n^{-1/2} M) - t\right\} > M' \right)<\epsilon   \label{F-con-eq-42-1} \\
P\left(n^{1/2} \left\{\widetilde U_n(s_0-n^{-1/2} M) - t\right\} < -M' \right)<\epsilon. \label{F-con-eq-43}
\end{eqnarray}
We need to show only (\ref{F-con-eq-42-1});  (\ref{F-con-eq-43}) can be shown using exactly the same arguments. Note that
\begin{eqnarray}
&&P\left(n^{1/2} \left\{\widetilde U_n(s_0+n^{-1/2} M) - t\right\} > M' \right) \label{F-con-eq-44} \nonumber\\
&\leq& P\left(t+ n^{-1/2}M'< \widetilde U_n(s_0+n^{-1/2} M) \leq t+M_2n^{-1/3} \right) \nonumber \\ && + P\left(\widetilde U_n(s_0+n^{-1/2} M) \geq t+M_2n^{-1/3} \right). \label{F-con-eq-45}
\end{eqnarray}
Consider the right hand side of the above expression; based on Lemma \ref{lemma-U-cubic-con}, the second term can be arbitrarily small for sufficiently large $M_2$; therefore, we only need to consider the first term. Recall that $s_0 = F_0(t)$, we have
{\small
\begin{eqnarray}
&&P\left(t+ n^{-1/2}M'< \widetilde U_n(s_0+n^{-1/2} M) \leq t+M_2n^{-1/3} \right) \label{F-con-eq-48} \\
&=& P\Big(\exists u\in[M',  M_2 n^{1/6}]:  \widetilde V_n(t+n^{-1/2}u) - (s_0+n^{-1/2} M) \widetilde G_n(t+n^{-1/2}u) \nonumber \\ && \hspace{1.5in} \leq \widetilde V_n(t) - (s_0+n^{-1/2} M) \widetilde G_n(t) \Big) \label{F-con-eq-47} \nonumber \\
&=& P\left(\exists u\in[M',  M_2 n^{1/6}]: \widetilde\mathW_{n,1}(u) + \widetilde\mathW_{n,2}(u) - \widetilde\mathW_{n,3}(u) + \widetilde\mathW_{n,4} \leq 0 \right),  \nonumber
\end{eqnarray}
}
where
\begin{eqnarray}
\mathW_{n,1}(u) &=& n \left\{ V_n(t+n^{-1/2}u) - V_n(t) \right\} \label{F-con-eq-49} \\ && - \frac{n}{n^2}\sum_{i,j}F_0(T_{i,j,0})I(T_{i,j}\in[t, t+n^{-1/2}u])  \nonumber\\
\mathW_{n,2}(u) &=& n \int_{w\in[t, t+n^{-1/2}u]}\{ F_0(w) - F_0(t)\} d G_n(w)  \label{F-con-eq-50} \\
\mathW_{n,3}(u) &=& n^{1/2} M  \left\{G_n(t+n^{-1/2}u) - G_n(t) \right\} \label{F-con-eq-51}\\
\mathW_{n,4}(u) &=& \frac{n}{n^2}\sum_{i,j}\bigg[\left\{F_0(T_{i,j,0})-F_0(T_{i,j})\right\}\label{F-con-eq-51-1} \\ && \hspace{0.5in} \times I(T_{i,j}\in[t, t+n^{-1/2}u])\bigg]. \nonumber
\end{eqnarray}
In analogous to ``(a), (b), (c), (d)" in Lemma \ref{lemma-U-cubic-con}, we proceed to verify
\begin{itemize}
\item[(a')] for any $\eta>0, \epsilon>0$, $M_2>0$, and $M'>0$, when $M_1$ is sufficiently large, we have
\begin{equation}
P\left(\exists u\in[M',  M_2 n^{1/6}]: \left|\widetilde\mathW_{n,1}(u) \right| \geq \eta u^2 + M_1 \right) < \epsilon; \label{F-con-eq-52}
\end{equation}

\item[(b')] there exists a universal constant $c>0$, such that for any $\epsilon>0$ and $M_2>0$, when $M'$ is sufficiently large, we have
    \begin{equation}
    P\left(\exists u\in[M',  M_2 n^{1/6}]: \widetilde\mathW_{n,2}(u)  \geq c u^2 \right) > 1- \epsilon; \label{F-con-eq-52-1}
    \end{equation}

\item[(c')] there exists a universal constant $C>0$, such that for any $\epsilon>0$, $M_2>0$, and $M>0$, when $M'$ and $M_1$ are sufficiently large, we have
\begin{eqnarray}
P\left(\exists u\in[M',  M_2 n^{1/6}]: \widetilde\mathW_{n,3}(u)  \geq C u + M_1 \right) < \epsilon; \label{F-con-eq-52-2}
\end{eqnarray}

\item[(d')] there exists a universal constant $C>0$, such that for any $\epsilon>0$, $M_2>0$, and $M>0$, when $M'$ and $M_1$ are sufficiently large, we have
\begin{eqnarray*}
    P\left(\exists u\in[M',  M_2 n^{1/6}]: \left|\widetilde \mathW_{n,4}(u) \right| \geq Cu +M_1 \right) < \epsilon.
    \end{eqnarray*}

\end{itemize}

We show (a') first; note that
\begin{eqnarray*}
\mathW_{n,1}(u) &=& \frac{1}{n} \sum_{i,j} \left\{\Delta_{i,j} - F_0(T_{i,j,0})\right\}I(T_{i,j} \in [t, t+n^{-1/2}u]) \label{F-con-eq-53}\\
&\equiv& \frac{1}{n}  \sum_{i,j} \pounds_{n, u}(X_i, Y_i, X_j, Y_j), \label{F-con-eq-54}
\end{eqnarray*}
where
\begin{eqnarray*}
\pounds_{n, u}(X_i, Y_i, X_j, Y_j) =  \left\{\Delta_{i,j} - F_0(T_{i,j,0})\right\}I(T_{i,j} \in [t, t+n^{-1/2}u]). \label{F-con-eq-55}
\end{eqnarray*}
Using similar developments as those from (\ref{F-con-eq-23}) to (\ref{F-con-eq-25}), we have the U-statistic decomposition
\begin{eqnarray}
\mathW_{n,1}(u) = n \left\{\mathbb{U}_n^2 \overline{\pounds}_{n,u} + \mathbb{P}_n \pounds_{n,u,1} + \mathbb{P}_n \pounds_{n,u,2} \right\},\label{F-con-eq-56}
\end{eqnarray}
where $\overline{\pounds}_{n,u} = \pounds_{n,u} - \pounds_{n,u,1} - \pounds_{n,u,2}$, such that $\mathbb{U}_n^2 \overline{\pounds}_{n,u}$ is a degenerate second order $U$-statistic, and that the function class
\begin{eqnarray*}
\Im_n = \left\{n^{1/6}\pounds_{n,u}: u \in[M',  M_2 n^{1/6}]; \bm{\beta}\in \mathcal{B}_{n,C}\right\} \label{F-con-eq-57}
\end{eqnarray*}
is Euclidean$(C, 8p+12)$ with an envelope function
\begin{eqnarray*}
\bar F(X_i, Y_i, X_j, Y_j) = 2 n^{1/6}I\left(T_{i,j} \in [t-C_1n^{-1/2}, t+n^{-1/2}M_2n^{1/6} + C_1 n^{-1/2}]\right)  \label{F-con-eq-58}
\end{eqnarray*}
for some universal constant $C_1>0$; based on Condition F2 and the discussion in Remark \ref{remark-condition-F}, it satisfies $\|\bar F\|_{2,P}\lesssim 1$. With an application of
 Lemma 6 and then Corollary 4 in Sherman (1994), we conclude that
\begin{eqnarray}
n^{7/6}\sup_{u \in[M',  M_2 n^{1/6}], \bm{\beta}\in \mathcal{B}_{n,C}}\left|\mathbb{U}_n^2 \overline{\pounds}_{n,u}\right| = O_p(1). \label{F-con-eq-59}
\end{eqnarray}
Furthermore, using exactly the same arguments as the proof for (\ref{F-con-eq-34}) and (\ref{F-con-eq-36}), we can establish 
\begin{eqnarray}
&&P\left(\exists u>M': n \sup_{\bm{\beta}\in \mathcal{B}_{n,C}} \left|\mathbb{P}_n \pounds_{n,u,1}\right| \geq \eta u^2 + M_1 \right) \nonumber\\
&\lesssim& \sum_{j=\lfloor M' \rfloor}^\infty \frac{(j+1)^2}{(\eta j^2 + M_1)^2} < \epsilon/3 \label{F-con-eq-60}\\
&& P\left(\exists u>M': n \sup_{\bm{\beta}\in \mathcal{B}_{n,C}}\left|\mathbb{P}_n \pounds_{n,u,2}\right| \geq \eta u^2 + M_1 \right) \nonumber\\
&\lesssim& \sum_{j=\lfloor M' \rfloor}^\infty \frac{(j+1)^2}{(\eta j^2 + M_1)^2} < \epsilon/3, \label{F-con-eq-61}
\end{eqnarray}
when $M_1$ is sufficiently large.

Combining (\ref{F-con-eq-59})--(\ref{F-con-eq-61}) leads to (\ref{F-con-eq-52}); we complete the proof for (a').

We consider (b') next. We can write
\begin{eqnarray}
\mathW_{n,2}(u) &=& n \int_{w\in[t, t+n^{-1/2}u]}\{ F_0(w) - F_0(t)\} d \left\{G_n(w)-G_{\bm{\beta}}(w)\right\} \nonumber \\
&& + n \int_{w\in[t, t+n^{-1/2}u]}\{ F_0(w) - F_0(t)\} dG_{\bm{\beta}}(w) \label{F-con-eq-62} \nonumber \\
&\equiv& \mathW_{n,2,1}(u) + \mathW_{n,2,2}(u). \label{F-con-eq-63}
\end{eqnarray}
Using exactly the same procedure as the proof for (a'), we have the decomposition
\begin{equation}
\mathW_{n,2,1} = n \left[\mathbb{U}_n^2 \overline{\varrho}_{n,u} + (\mathbb{P}_n \varrho_{n,u,1} - \mathbb{P}^2 \varrho_{n,u}) + \{\mathbb{P}_n \varrho_{n,u,2} - \mathbb{P}^2 \varrho_{n,u}\} \right], \label{F-con-eq-63-1}
\end{equation}
where $\overline{\varrho}_{n,u} = \varrho_{n,u} - \varrho_{n,u,1} - \varrho_{n,u,2} + \mathbb{P}^2 \varrho_{n,u}$, $\varrho_{n,u,1}(X_j, Y_j) = \mathbb{P} \varrho_{n,u}(\cdot, \cdot; X_j, Y_j) $,   $\varrho_{n,u,2}(X_i, Y_i) = \mathbb{P} \varrho_{n,u}(X_i, Y_i; \cdot, \cdot)$ with
\begin{eqnarray}
\varrho_{n,u}(X_i, Y_i; X_j, Y_j) &=& \left\{F_0(T_{i,j}) - F_0(t)\right\} I\left(T_{i,j} \in [t, t+n^{-1/2} u) \right)  \label{con-eq-63-2} \nonumber\\
 \mathbb{P}^2 \varrho_{n,u} &=& \mathW_{n,2,2}(u) =  \int_{w\in[t, t+n^{-1/2} u)} \left\{F_0(w) - F_0(t)\right\} dG_{\bm{\beta}}(w). \label{con-eq-63-2-1}
\end{eqnarray}
Note that $\mathbb{U}_n^2 \overline{\varrho}_{n,u}$ is a degenerate $U$-statistic. Using exactly the same developments as those for (\ref{F-con-eq-59}), we have
\begin{eqnarray}
n^{7/6}\sup_{u \in[M',  M_2 n^{1/6}], \bm{\beta}\in \mathcal{B}_{n,C}}\left|\mathbb{U}_n^2 \overline{\varrho}_{n,u}\right| = O_p(1) \label{F-con-eq-63-3}
\end{eqnarray}

Then, using exactly the same arguments as from (\ref{F-con-eq-56}) to (\ref{F-con-eq-61}), we
are able to verify that for any $\eta>0, \epsilon>0$, $M_2$, and $M'>0$, when $M_1$ is sufficiently large,
\begin{eqnarray}
P\left( \exists u \in[M',  M_2 n^{1/6}]: \widetilde \mathW_{n,2,1}(u)\geq \eta u^2 + M_1  \right)< \epsilon.  \label{F-con-eq-64}
\end{eqnarray}
Furthermore, based on Conditions F1 and F2 and the discussion given in Remark \ref{remark-condition-F}, we can find a universal constant $c>0$, such that
\begin{eqnarray}
\mathW_{n,2,2} &=& n \int_{w\in [t, t+n^{-1/2}u)} \{F_0(w) - F_0(t) \} dG_{\bm{\beta}}(w) \geq c u^2. \label{F-con-eq-65}
\end{eqnarray}
Combining (\ref{F-con-eq-63})--(\ref{F-con-eq-65}) leads to (\ref{F-con-eq-52-1}); we complete the proof for (b').

We proceed to consider (c'). Using a similar strategy as the proof for (a'), we can show that
\begin{eqnarray}
&& \sup_{u\in[M', M_2n^{1/6}], \bm{\beta}\in \mathcal{B}_{n,C}} n^{1/2} M \bigg|  \left\{G_n(t+n^{-1/2}u) - G_n(t) \right\} \nonumber \\ && \hspace{1.8in} - \left\{ G_{\bm{\beta}}(t+n^{-1/2}u) - G_{\bm{\beta}}(t) \right\} \bigg| = o_p(1).   \label{F-con-eq-66}
\end{eqnarray}
Furthermore, based on Condition F2 and the discussion given in Remark \ref{remark-condition-F}, there exists a universal constant $C_1>0$, such that
\begin{eqnarray}
n^{1/2}\sup_{\bm{\beta}\in \mathcal{B}_{n,C}}\left| G_{\bm{\beta}}(t+n^{-1/2}u) - G_{\bm{\beta}}(t) \right| \leq C_1 u.   \label{F-con-eq-67}
\end{eqnarray}
Combining (\ref{F-con-eq-66}) with (\ref{F-con-eq-67}) leads to (\ref{F-con-eq-52-2}), and we complete the proof for (c').

Last, we consider (d'). Note that $\widetilde{\bm{\beta}} - \bm{\beta}_0 = O_p\left(n^{-1/2}\right)$ and based on Condition F1 and the discussion given in Remark \ref{remark-condition-F}, with probability arbitrarily large,
\begin{eqnarray*}
|\mathW_{n,4}(u)| &\lesssim& \frac{n^{1/2}}{n^2} \sum_{i,j} I(T_{i,j}\in [t, t+n^{-1/2}u]) \\ &=& n^{1/2} \left\{G_n(t+n^{-1/2}u) - G_n(t) \right\}\\
&=& \mathW_{n,3}(u)/M.
\end{eqnarray*}
Therefore, the conclusion of (d') follows by using the result of (c').

Finally, combining (a')--(d'), and using a similar development to (\ref{F-con-eq-42}), and incorporating (\ref{F-con-eq-45}) and (\ref{F-con-eq-48}) leads to (\ref{F-con-eq-42-1}). We complete the proof of this lemma. \epf

\vspace{0.2in}

We need to apply the following Lemma \ref{argmax}, which is the argmax theorem adapted from Theorem 14.1 in Kosorok (2008) (see also Theorem 3.2.2 in van der Vaart and Wellner, 1996) to establish the asymptotic normality of $n^{1/2}\left\{ \widetilde U_n(s+n^{-1/2} x) - t \right\}$.

\begin{lemma} \label{argmax}

Let $W_n$, $W$ be stochastic processes indexed by a metric space $\mathcal{H}$, such that $W_n \rightsquigarrow W$ in $L^{\infty}(H)$ for every compact $H \subset \mathcal{H}$. Suppose also that almost all sample paths $h \mapsto W(h)$ are upper semicontinuous and possess a unique maximum at a (random) point $\widehat h$, which as a random map in $\mathcal{H}$ is tight. If the sequence $\widehat h_n$ is uniformly tight and satisfies $W_n(\widehat h_n) \geq \sup_{h\in H} W_n(h) - o_p(1)$, then $\widehat h_n \rightsquigarrow \widehat h$ in $\mathcal{H}$.

\end{lemma}

\begin{lemma} \label{lemma-U-asymptotic}
Assume all conditions for Theorem \ref{theorem-F-asymptotics} are effective. We have
\begin{eqnarray}
n^{1/2}\left\{ \widetilde U_n(s_0+n^{-1/2} x) - t \right\}  \rightsquigarrow \frac{g_0(t)x - \sigma(t) Z}{f_0(t) g_0(t)}. \label{eq-asymptotic-U-n}
\end{eqnarray}

\end{lemma}
\pf We shall apply Lemma \ref{argmax} to establish the asymptotic result (\ref{eq-asymptotic-U-n}). Note that in Lemma \ref{lemma-U-root-con}, we have verified that $n^{1/2}\left\{ \widetilde U_n(s_0+n^{-1/2} x) - t \right\}$ is uniformly tight. Therefore, to apply the argmax theorem, i.e., Lemma \ref{argmax}, we need to establish the asymptotic convergence of the corresponding stochastic process that it minimises.

Based on the  definition of $U_n$ given in (\ref{F-con-eq-5})  and note that $s_0 = F_0(t)$, we have
\begin{eqnarray}
&&n^{1/2}\left\{ \widetilde U_n(s_0+n^{-1/2} x) - t \right\} \nonumber \\ &=& \arg\min_{u} \left[n\left\{\widetilde V_n(t + n^{-1/2}u) + (s_0+n^{-1/2} x) \widetilde G_n(t+n^{-1/2}u)\right\}\right]\nonumber\\
&=& \arg\min_{u} \left\{\widetilde\mathW_{n,1}(u) +  \widetilde\mathW_{n,2}(u) - \widetilde\mathW_{n,5}(u) +\widetilde \mathW_{n,4}(u) \right\} \label{F-norm-eq-1}
\end{eqnarray}
where $\widetilde\mathW_{n,1}(u)$, $\widetilde\mathW_{n,2}(u)$, and $\widetilde\mathW_{n,4}(u)$ are respectively defined by (\ref{F-con-eq-49}), (\ref{F-con-eq-50}), and (\ref{F-con-eq-51-1}); and
\begin{eqnarray*}
\mathW_{n,5}(u) = n^{1/2} x  \left\{G_n(t+n^{-1/2}u) - G_n(t) \right\}. \label{F-norm-eq-2}
\end{eqnarray*}
Consider $\mathW_{n,4}(u)$. Recall
\begin{eqnarray}
\mathW_{n,4}(u) &=& \frac{n}{n^2}\sum_{i,j}\left\{F_0(T_{i,j,0})-F_0(T_{i,j})\right\}I(T_{i,j}\in[t, t+n^{-1/2}u]) \label{F-norm-eq-2-1}\nonumber\\
&=& \frac{n}{n^2} \sum_{i,j}h_{n,u}(X_i, X_j) \label{F-norm-eq-2-2} \nonumber\\
&=& \frac{n}{n^2} \sum_{i,j}\left\{ h_{n,u}(X_i, X_j) - \mathbb{P}^2 h_{n,u} \right\} + n \mathbb{P}^2 h_{n,u}. \label{F-norm-eq-2-3}
\end{eqnarray}
Then, applying exactly the same arguments as  (\ref{F-con-eq-56}) and (\ref{F-con-eq-59}), we have uniformly in $u\in [-M, M]$ and $\bm{\beta} \in \mathcal{B}_{n,C}$,
\begin{eqnarray}
&&\frac{n}{n^2} \sum_{i,j}\left\{ h_{n,u}(X_i, X_j) - \mathbb{P}^2 h_{n,u} \right\}\nonumber\\
&=& n \left\{\mathbb{P}_n h_{n,u, 1} - \mathbb{P}^2 h_{n,u}\right\} + n\left\{\mathbb{P}_n h_{n,u, 2} - \mathbb{P}^2 h_{n,u} \right\} +o_p(1), \label{F-norm-eq-2-4}
\end{eqnarray}
where $h_{n,u, 1}(\bm{x}) = \mathbb{P} h_{n,u}(\cdot, \bm{x})$ and $h_{n,u, 2}(\bm{x}) = \mathbb{P} h_{n,u}(\bm{x}, \cdot)$.
Furthermore,  recall the definition $F_{\bm{\beta}}(s) = E\left\{I(Y_1>Y_2)\Big| (X_1-X_2)^T\bm{\beta} = s\right\}$ and denote $\dot{F}_0(s) = \frac{\partial F_{\bm{\beta}}(s)}{\partial \bm{\beta}}\Big|_{\bm{\beta} = \bm{\beta}_0}$. Then based on Condition A1, $\dot{F}_0(s)$ is continuous in $s$; and based on Condition F2, $g_{\bm{\beta}}(s) = G_{\bm{\beta}}'(s)$ is continuous in the neighbourhood of $s = t$ and $\bm{\beta} = \bm{\beta}_0$. We have that uniformly in $u\in[-M, M]$ and $\bm{\beta}\in \mathcal{B}_{n,C}$,
\begin{eqnarray}
n \mathbb{P}^2 h_{n,u} &=& nE\left[\left\{\Delta_{i,j} - F_0(T_{i,j})\right\} I(T_{i,j} \in [t, t+n^{-1/2}u]) \right]\label{F-norm-eq-2-9}\nonumber\\
&=& n E\left(E\left[\left\{\Delta_{i,j} - F_0(T_{i,j})\right\} I(T_{i,j} \in [t, t+n^{-1/2}u]) \Big| T_{i,j} \right] \right) \label{F-norm-eq-2-10} \nonumber\\
&=& n E\left[ \left\{F_{\bm{\beta}}(T_{i,j}) - F_0(T_{i,j}) \right\} I(T_{i,j} \in [t, t+n^{-1/2}u]) \right]\label{F-norm-eq-2-11} \nonumber\\
&=& n(\bm{\beta}- \bm{\beta}_0)^T E\left\{ \dot{F}_0(T_{i,j}) I(T_{i,j} \in [t, t+n^{-1/2}u] )\right\} \left\{1+ o(1)\right\} \label{F-norm-eq-2-12} \nonumber\\
&=& n(\bm{\beta}- \bm{\beta}_0)^T \left\{1+o(1)\right\}\int_{t}^{t+n^{-1/2} u} \dot{F}_0(w) d G_{\bm{\beta}}(w) \label{F-norm-eq-2-13} \nonumber\\
&=& \sqrt{n}(\bm{\beta}-\bm{\beta}_0)^T \dot{F}_0(t) u g_0(t) \left\{1+o(1)\right\}. \label{F-norm-eq-2-14}
\end{eqnarray}

%


Combining (\ref{F-con-eq-56}), (\ref{F-con-eq-59}), (\ref{F-con-eq-63}), (\ref{F-con-eq-63-1}), (\ref{F-con-eq-63-3}), (\ref{F-con-eq-66}), (\ref{F-norm-eq-2-3}), and (\ref{F-norm-eq-2-4}), we have uniformly in $u\in[-M,M]$ and $\bm{\beta}\in \mathcal{B}_{n,C}$,
\begin{eqnarray}
&& \mathW_{n,1}(u) +  \mathW_{n,2}(u) - \mathW_{n,5}(u) +\mathW_{n,4}(u) \nonumber\\
&=& n \mathbb{P}_n (\pounds_{n,u,1}  + \pounds_{n,u,2} + \varrho_{n,u,1} + \varrho_{n,u,2} + h_{n,u,1} + h_{n,u,2}) \nonumber\\ && -  2 n \mathbb{P}^2\left(\varrho_{n,u} + h_{n,u}\right)
+ n \mathbb{P}^2 \varrho_{n,u} + n\mathbb{P}^2 h_{n,u} \nonumber\\ && -n^{1/2} x  \left\{G_{\bm{\beta}}(t+n^{-1/2}u) - G_{\bm{\beta}}(t) \right\} + o_p(1) \nonumber\\
&=& \mathbb{G}_n (p_{n,u,1} + p_{n,u,2}) + n\mathbb{P}^2 \varrho_{n,u} + n\mathbb{P}^2 h_{n,u}  \nonumber\\ && -n^{1/2} x  \left\{G_{\bm{\beta}}(t+n^{-1/2}u) - G_{\bm{\beta}}(t) \right\} + o_p(1), \label{F-norm-eq-2-16}
\end{eqnarray}
where
$p_{n,u,1}(X_j, Y_j) = \mathbb{P} p_{n,u}(\cdot, \cdot; X_j, Y_j)$, $p_{n,u,2}(X_i, Y_i) = \mathbb{P} p_{n,u}(X_i, Y_i; \cdot, \cdot)$ with
\begin{eqnarray*}
p_{n,u}(X_i, Y_i; X_j, Y_j) &=& \sqrt{n} \left\{\Delta_{i,j} - F_0(t)\right\}I(T_{i,j} \in [t, t+n^{-1/2}u]), \label{F-norm-eq-7}
\end{eqnarray*}
and therefore
\begin{eqnarray*}
&& p_{n,u,1}(\boldsymbol{x}, y) \\ &=&  \sqrt{n}\int_{\bm{w}^T\bm{\beta} \in [\boldsymbol{x}^T\boldsymbol{\beta}+t, \boldsymbol{x}^T\boldsymbol{\beta}+t + n^{-1/2}u]} \left\{1-F_\epsilon(H(y) - \bm{w}^T \bm{\beta}_0) - F_0(t)  \right\} d F_{X}(\bm{w}) \label{F-norm-eq-8}\\
&& p_{n,u,2}(\boldsymbol{x}, y)\\&=&\sqrt{n}\int_{\bm{w}^T \bm{\beta} \in [\boldsymbol{x}^T\boldsymbol{\beta}-t-n^{-1/2}u, \boldsymbol{x}^T\boldsymbol{\beta}-t]} \left\{F_\epsilon(H(y) - \bm{w}^T \bm{\beta}_0) - F_0(t)  \right\} d F_{X}(\bm{w}). \label{F-norm-eq-9}
\end{eqnarray*}

Denote
\begin{eqnarray*}
&& \breve p_{n,u,1}(\boldsymbol{x}, y)\\ &=&  \sqrt{n}\int_{\bm{w}^T\bm{\beta} \in [\boldsymbol{x}^T\boldsymbol{\beta}+t, \boldsymbol{x}^T\boldsymbol{\beta}+t + n^{-1/2}u]} \left\{1-F_\epsilon(H(y) - \bm{w}^T \bm{\beta}) - F_0(t)  \right\} d F_{X}(\bm{w})\\
&=& \sqrt{n}\int_{\boldsymbol{x}^T\boldsymbol{\beta}+t}^{\boldsymbol{x}^T\boldsymbol{\beta}+t + n^{-1/2}u} \left\{1-F_\epsilon(H(y) - w) - F_0(t)  \right\} d F_{X^T\bm{\beta}}(w)\\
&& \breve p_{n,u,2}(\boldsymbol{x}, y)\\ &=&\sqrt{n}\int_{\bm{w}^T \bm{\beta} \in [\boldsymbol{x}^T\boldsymbol{\beta}-t-n^{-1/2}u, \boldsymbol{x}^T\boldsymbol{\beta}-t]} \left\{F_\epsilon(H(y) - \bm{w}^T \bm{\beta}) - F_0(t)  \right\} d F_{X}(\bm{w})\\
&=& \sqrt{n}\int_{\boldsymbol{x}^T\boldsymbol{\beta}-t-n^{-1/2}u}^{\boldsymbol{x}^T\boldsymbol{\beta}-t} \left\{F_\epsilon(H(y) - w) - F_0(t)  \right\} d F_{X^T \bm{\beta}}(w).
\end{eqnarray*}
Then, based on Condition F1, considering the function class
\begin{eqnarray*}
\left\{p_{n,u,1} - \breve p_{n,u,1}: u\in [-M, M], \bm{\beta}\in \mathcal{B}_{n,C} \right\},
\end{eqnarray*}
and applying Theorem 11.1 in Kosorok (2008), we can establish
\begin{eqnarray}
\sup_{u \in [-M, M], \bm{\beta}\in \mathcal{B}_{n,C}}\left|\mathbb{G}_n (p_{n,u,1} - \breve p_{n,u,1}) \right| \lesssim n^{-1/2}. \label{F-norm-eq-2-17}
\end{eqnarray}
Similarly,
\begin{eqnarray}
\sup_{u \in [-M, M], \bm{\beta}\in \mathcal{B}_{n,C}}\left|\mathbb{G}_n (p_{n,u,2} - \breve p_{n,u,2}) \right| \lesssim n^{-1/2}. \label{F-norm-eq-2-18}
\end{eqnarray}
We further consider
\begin{eqnarray*}
\bar p_{n,u,1, \bm{\beta}}(\boldsymbol{x}, y)
&=&  u\left\{1-F_\epsilon(H(y) - \bm{x}^T \bm{\beta}-t) - F_0(t)  \right\} f_{X^T\bm{\beta}}(\bm{x}^T \bm{\beta}+t)\\
\bar p_{n,u,2, \bm{\beta}}(\boldsymbol{x}, y)
&=&  u\left\{F_\epsilon(H(y) - \boldsymbol{x}^T\boldsymbol{\beta}+t) - F_0(t)  \right\}f_{X^T\bm{\beta}}(\bm{x}^T \bm{\beta}-t).
\end{eqnarray*}
Based on Conditions F1 and F2, considering the function classes
\begin{eqnarray*}
\left\{\breve p_{n,u,1} - \bar p_{n,u,1,\bm{\beta}}: u \in [-M, M], \bm{\beta}\in \mathcal{B}_{n,C} \right\}\\
\left\{\bar p_{n,u,1,\bm{\beta}}: u\in[-M, M], \bm{\beta}\in \mathcal{B}_{n,C} \right\},
\end{eqnarray*}
and applying Theorem 11.1 in Kosorok (2008), we can obtain
\begin{eqnarray}
\sup_{u\in[-M,M], \bm{\beta}\in \mathcal{B}_{n,C}}\left|\mathbb{G}_n(\breve p_{n,u,1} - \bar p_{n,u,1,\bm{\beta}})\right| &=& O_p(n^{-1/2}) \label{F-norm-eq-2-19}\\
\sup_{u\in[-M,M], \bm{\beta}\in \mathcal{B}_{n,C}}\left|\mathbb{G}_n(\bar p_{n,u,1,\bm{\beta}}-\bar p_{n,u,1,\bm{\beta}_0})\right| &=& O_p(n^{-1/2}).  \label{F-norm-eq-2-20}
\end{eqnarray}
Likewise, we are able to derive
\begin{eqnarray}
\sup_{u\in[-M,M], \bm{\beta}\in \mathcal{B}_{n,C}}\left|\mathbb{G}_n(\breve p_{n,u,2} - \bar p_{n,u,2,\bm{\beta}})\right| &=& O_p(n^{-1/2}) \label{F-norm-eq-2-21}\\
\sup_{u\in[-M,M], \bm{\beta}\in \mathcal{B}_{n,C}}\left|\mathbb{G}_n(\bar p_{n,u,2,\bm{\beta}}-\bar p_{n,u,2,\bm{\beta}_0})\right| &=& O_p(n^{-1/2}). \label{F-norm-eq-2-22}
\end{eqnarray}
Combining (\ref{F-norm-eq-2-16})--(\ref{F-norm-eq-2-22}) and noting (\ref{con-eq-63-2-1}) and (\ref{F-norm-eq-2-14}), we have uniformly in $u\in[-M, M]$ and $\bm{\beta}\in \mathcal{B}_{n,C}$,
\begin{eqnarray*}
&&\mathW_{n,1}(u) +  \mathW_{n,2}(u) - \mathW_{n,5}(u) +\mathW_{n,4}(u) \\
&=& \mathbb{G}_n(\bar p_{n,u,1,\bm{\beta}_0} + \bar p_{n,u,2,\bm{\beta}_0})
+ n \int_{w\in[t, t+n^{-1/2} u]} \left\{F_0(w) - F_0(t)\right\} dG_{\bm{\beta}}(w) \\
&&+ \sqrt{n}(\bm{\beta}-\bm{\beta}_0)^T  \dot{F}_0(t) u g_0(t) \left\{1+o(1)\right\} -n^{1/2} x  \left\{G_{\bm{\beta}}(t+n^{-1/2}u) - G_{\bm{\beta}}(t) \right\} +o_p(1)\\
&=&  \mathbb{G}_n(\bar p_{n,u,1,\bm{\beta}_0} + \bar p_{n,u,2,\bm{\beta}_0})
+ n \int_{w\in[t, t+n^{-1/2} u]} \left\{F_0(w) - F_0(t)\right\} dG_0(w) \\
&&+ \sqrt{n}(\bm{\beta}-\bm{\beta}_0)^T \dot{F}_0(t) u g_0(t) \left\{1+o(1)\right\} -x  u g_0(t) +o_p(1) \\
&=&  \mathbb{G}_n(\bar p_{n,u,1,\bm{\beta}_0} + \bar p_{n,u,2,\bm{\beta}_0}) + \sqrt{n}(\bm{\beta}-\bm{\beta}_0)^T  \dot{F}_0(t) u g_0(t) + \frac12 g_0(t)f_0(t) u^2 - g_0(t) x u + o_p(1).
\end{eqnarray*}
Replacing $\bm{\beta}$ with $\widetilde{\bm{\beta}}$ in the expression above and based on Theorem \ref{thm-normality} and the proof of Corollary \ref{corollary-normality}  that
\begin{eqnarray*}
\sqrt{n} \left(\widetilde{\bm{\beta}} - \bm{\beta}_0\right) = 2 A^{-1} \widetilde A^{-1} \psi_0'(\bm{\beta}_0) \Sigma_X^{-1} \mathbb{G}_n \hbar +o_p(1)
\end{eqnarray*}
leads to
\begin{eqnarray*}
&&\widetilde \mathW_{n,1}(u) +  \widetilde \mathW_{n,2}(u) - \widetilde \mathW_{n,5}(u) +\widetilde \mathW_{n,4}(u) \\
&=& u \mathbb{G}_n \zeta_t + \frac12 g_0(t)f_0(t) u^2 - g_0(t) x u + o_p(1),
\end{eqnarray*}
where
\begin{eqnarray*}
\zeta_t(\bm{x}, y) &=& \left\{1-F_\epsilon(H(y) - \bm{x}^T \bm{\beta}_0-t) - F_0(t)  \right\} f_{X^T\bm{\beta}_0}(\bm{x}^T \bm{\beta}_0+t)\\
&& +\left\{F_\epsilon(H(y) - \boldsymbol{x}^T\boldsymbol{\beta}_0+t) - F_0(t)  \right\}f_{X^T\bm{\beta}_0}(\bm{x}^T \bm{\beta}_0-t)\\
&&+2 g_0(t) \dot{F}_0^T(t) A^{-1} \widetilde A^{-1} \psi_0'(\bm{\beta}_0) \Sigma_X^{-1}  \hbar(\bm{x}, y).
\end{eqnarray*}
Denoting $\sigma(t) = \mbox{var} (\zeta_t(X_1,Y_1)$,  and applying central limit theorem, we have
\begin{eqnarray*}
\widetilde \mathW_{n,1}(u) +  \widetilde \mathW_{n,2}(u) - \widetilde \mathW_{n,5}(u) +\widetilde \mathW_{n,4}(u) \\ \rightsquigarrow u \sigma(t) Z +  \frac12 g_0(t)f_0(t) u^2 - g_0(t) x u,
\end{eqnarray*}
where $Z$ is a standard normal random variable.
Applying Lemma \ref{argmax}, we have
\begin{eqnarray*}
n^{1/2}\left\{ \widetilde U_n(s+n^{-1/2} x) - t \right\}  \rightsquigarrow \frac{g_0(t)x - \sigma(t) Z}{f_0(t) g_0(t)},
\end{eqnarray*}
which completes our proof of this lemma. \epf

\section{Proof of Corollary \ref{coro-F-beta-asymptotic}} \label{sec-proof-of-normality-F-beta}

We give a sketched proof of Corollary \ref{coro-F-beta-asymptotic}; we omit details that are similar to those in the proof of Theorem \ref{theorem-F-asymptotics} to avoid lengthy repetition.
With the switching relation and using similar developments to (\ref{F-con-eq-9}), we can obtain
\begin{eqnarray}
&&P\left(n^{1/2}\left\{ \widehat F_{\bm{\beta}}(t) - F_{\bm{\beta}}(t)\right\} \geq x \right) \label{coro-F-beta-eq-1} \\ &=& P\left(n^{1/2}\left\{U_n\left(s_{\bm{\beta}}+n^{-1/2} x\right) - t \right\} \leq 0\right),  \nonumber
\end{eqnarray}
where $s_{\bm{\beta}} = F_{\bm{\beta}}(t)$.
Therefore, we only need to consider the asymptotics of the process $n^{1/2}\left\{U_n\left(s_{\bm{\beta}}+n^{-1/2} x\right) - t \right\}$.

For any $t\in \mathbb{R}$ and $\bm{\beta}\in \mathcal{\bm{\beta}}$, using similar arguments as Lemmas \ref{lemma-U-cubic-con} and \ref{lemma-U-root-con}, assuming that all the technical conditions are effective, we are able to verify that for any $M>0$,
\begin{eqnarray*}
n^{1/2}\sup_{x\in[-M,M]}\left|U_n\left(s_{\bm{\beta}}+n^{-1/2} x\right) - t \right| = O_p(1).
\end{eqnarray*}
Then, similar to the developments in (\ref{F-norm-eq-1}) and (\ref{F-norm-eq-2-16}), we can verify
\begin{eqnarray*}
&&n^{1/2}\left[U_n\left(s_{\bm{\beta}}+n^{-1/2} x\right) - t \right\}\\
&=&\arg\min_{u}\bigg[ \mathbb{G}_n (q_{n,u,1} + q_{n,u,2}) +n \int_{w\in[t, t+n^{-1/2} u]} \left\{F_{\bm{\beta}}(w) - F_{\bm{\beta}}(t)\right\} dG_{\bm{\beta}}(w) \\
&&\hspace{0.6in}+ n^{1/2} x  \left\{G_{\bm{\beta}}(t+n^{-1/2}u) - G_{\bm{\beta}}(t) \right\} + o_p(1) \bigg],
\end{eqnarray*}
where $o_p(1)$ is uniform in $u\in [-M, M]$ and $\bm{\beta}\in \mathcal{B}$;
$q_{n,u,1}(X_j, Y_j) = \mathbb{P} q_{n,u}(\cdot, \cdot; X_j, Y_j)$, $q_{n,u,2}(X_i, Y_i) = \mathbb{P} q_{n,u}(X_i, Y_i; \cdot, \cdot)$ with
\begin{eqnarray*}
q_{n,u} &=& \sqrt{n} \left\{\Delta_{i,j} - F_{\bm{\beta}}(t)\right\}I(T_{i,j} \in [t, t+n^{-1/2}u]). \\
q_{n,u,1}(\boldsymbol{x}, y)&=&  \sqrt{n}\int_{\boldsymbol{x}^T\boldsymbol{\beta}+t}^{\boldsymbol{x}^T\boldsymbol{\beta}+t + n^{-1/2}u} \left\{1-F_{\epsilon, \bm{\beta}}(y, w) - F_{\bm{\beta}}(t)  \right\} d F_{X^T\bm{\beta}}(w) \\
q_{n,u,2}(\boldsymbol{x}, y)&=&\sqrt{n}\int_{\boldsymbol{x}^T\boldsymbol{\beta}-t-n^{-1/2}u}^{\boldsymbol{x}^T\boldsymbol{\beta}-t} \left\{F_{\epsilon, \bm{\beta}}(y, w) - F_{\bm{\beta}}(t)  \right\} d F_{X^T\bm{\beta}}(w)\\
F_{\epsilon, \bm{\beta}}(y, w) &=& E\left\{F_\epsilon(H(y) - X^T\bm{\beta_0})\Big| X^T\bm{\beta} = w\right\}.
\end{eqnarray*}
Consider
\begin{eqnarray*}
\bar q_{n,u,1}(\boldsymbol{x}, y)
&=&  u\left\{1-F_{\epsilon, \bm{\beta}}(y, \bm{x}^T\bm{\beta} + t) - F_{\bm{\beta}}(t)  \right\} f_{X^T\bm{\beta}}(\bm{x}^T \bm{\beta}+t)\\
\bar q_{n,u,2}(\boldsymbol{x}, y)
&=&  u\left\{F_{\epsilon, \bm{\beta}}(y,  \boldsymbol{x}^T\boldsymbol{\beta}-t) - F_{\bm{\beta}}(t)  \right\}f_{X^T\bm{\beta}}(\bm{x}^T \bm{\beta}-t).
\end{eqnarray*}
Then based on Conditions F1' and F2',  considering the function classes
\begin{eqnarray*}
\Big\{q_{n,u,1} - \bar q_{n,u,1}: u \in [-M, M], \bm{\beta}\in \mathcal{B} \Big\}
\end{eqnarray*}
and applying Theorem 11.1 in Kosorok (2008), we can obtain
\begin{eqnarray}
\sup_{u\in[-M,M], \bm{\beta}\in \mathcal{B} }\left|\mathbb{G}_n(q_{n,u,1} - \bar q_{n,u,1})\right| &=& O_p(n^{-1/2}),
\end{eqnarray}
for any $M>0$.
Similarly, we have
\begin{eqnarray}
\sup_{u\in[-M,M], \bm{\beta}\in \mathcal{B} }\left|\mathbb{G}_n(q_{n,u,2} - \bar q_{n,u,2})\right| &=& O_p(n^{-1/2}).
\end{eqnarray}
Therefore, we have
\begin{eqnarray*}
&&\mathbb{G}_n (q_{n,u,1} + q_{n,u,2}) +n \int_{w\in[t, t+n^{-1/2} u]} \left\{F_{\bm{\beta}}(w) - F_{\bm{\beta}}(t)\right\} dG_{\bm{\beta}}(w) \\ && + n^{1/2} x  \left\{G_{\bm{\beta}}(t+n^{-1/2}u) - G_{\bm{\beta}}(t) \right\}\\
&=& \mathbb{G}_n (\bar q_{n,u,1} + \bar q_{n,u,2}) + \frac{u^2}{2} f_{\bm{\beta}}(t)g_{\bm{\beta}}(t) + u x g_{\bm{\beta}}(t) + o_p(1) \\
&\rightsquigarrow& u \sigma_{\bm{\beta}}(t) Z + \frac{u^2}{2} f_{\bm{\beta}}(t)g_{\bm{\beta}}(t) + u x g_{\bm{\beta}}(t),
\end{eqnarray*}
where $o_p(1)$ is uniform in $u\in[-M, M]$ and $\bm{\beta}\in  \mathcal{B}$; the weak convergence is in $L^\infty[-M, M]$ for any given $\bm{\beta}\in \mathcal{B}$; $Z$ is a standard normal random variable; and
\begin{eqnarray*}
\sigma_{\bm{\beta}}(t) &=& \mbox{var}\left\{\bar q_{n,u,1}(X_1, Y_1) + \bar q_{n,u,2}(X_1, Y_1)\right\}\\
&=& E\bigg[\mbox{var}\Big\{f_{X^T\boldsymbol{\beta}}(X_1^T\boldsymbol{\beta}+t)F_{\epsilon, \bm{\beta}}(Y_1, X_1^T\bm{\beta} + t) \\ && \hspace{0.5in}  - f_{X^T\boldsymbol{\beta}}(X_1^T\boldsymbol{\beta}-t) F_{\epsilon, \bm{\beta}}(Y_1, X_1^T\bm{\beta} - t)  \Big| X_1 \Big\} \bigg],
\end{eqnarray*}
by noting that $E\left\{ \bar q_{n,u,1}(X_1, Y_1)\right\} = E\left\{\bar q_{n,u,2}(X_1, Y_1) \right\} = 0$.
Applying Lemma \ref{argmax}, we have
\begin{eqnarray*}
n^{1/2}\left\{ U_n\left(s_{\bm{\beta}}+n^{-1/2} x\right) - t \right\}  \rightsquigarrow \frac{g_{\bm{\beta}}(t)x - \sigma_{\bm{\beta}}(t) Z}{f_{\bm{\beta}}(t) g_{\bm{\beta}}(t)},
\end{eqnarray*}
which together with (\ref{coro-F-beta-eq-1}) completes the proof of this corollary. \epf

}

{\color{black} 
\section{Some Details for the Numerical Algorithms of our Estimates} \label{details-numerical-imp}

In this section, we provide some details of the numerical algorithms for computing our $\widehat{\bm{\beta}}$ and $\widetilde{\bm{\beta}}$ estimates.

\subsection{Some details of the numeral algorithm for  $\widehat{\bm \beta}$} \label{section-numerical-details-1}
Recall that our $\widehat{\bm{\beta}}$ estimate is obtained by maximizing  the profile pairwise rank log-likelihood:
\begin{eqnarray*}
\ell(\bm{\beta},\widehat{F}_{\bm{\beta}})&=&\sum_{i\neq j}\Big[I(Y_i>Y_j)\log\{ \widehat{F}_{\bm{\beta}}((X_i-X_j)^T\bm{\beta})\} \nonumber \\ && \hspace{0.3in} +I(Y_i\leq Y_j)\log \{1-\widehat{F}_{\bm{\beta}}((X_i-X_j)^T\bm{\beta})\}\Big],
\end{eqnarray*}
subject to the constraint $\|\bm{\beta}\|_2 = 1$.

To incorporate the constraint $\|\bm{\beta}\|_2 = 1$, we
consider the following polar transformation for $\bm{\beta}=(\beta_1,\ldots,\beta_p)^T$:
\begin{eqnarray*}
\beta_1&=&\cos \theta_1,\\
\beta_2&=&\sin \theta_1 \cos\theta_2,\\
\cdots&\cdots&\cdots\\
\beta_{p-1}&=& \sin \theta_1\cdots\sin\theta_{p-2} \cos\theta_{p-1},\\
\beta_p&=&\sin \theta_1\cdots\sin\theta_{p-2} \sin\theta_{p-1},
\end{eqnarray*}
where $\theta_1,\ldots,\theta_{p-1}\in[-\pi,\pi]$.
Denote $\bm{\theta}=(\theta_1,\ldots,\theta_{p-1})^T$, and let
$
pl(\bm{\theta})=\ell(\bm{\beta},\widehat{F}_{\bm{\beta}})
$
with $\bm{\beta}$ being replaced by the polar transformation. We have
\begin{eqnarray}
\widehat{\bm\theta}=\arg\max_{{\bm\theta}}pl(\bm{\theta}).  \label{eq-likelihood-theta}
\end{eqnarray}

In our implementation,
we have used  the R function optim() with the default method ``Nelder--Mead". We observe that the likelihood function \eqref{eq-likelihood-theta} is neither concave nor smooth, so the algorithm can be trapped in a local maximum. To tackle this problem, we used multiple initial values in our numerical studies. Specifically, we used 25 randomly generated initial values and chose the one that led to the greatest value of $pl(\bm{\theta})$.
This strategy worked well.
We tested the method using different sets of initial values in a number of simulation examples. In all the examples, the different sets of initial values gave the same $\widehat{\bm{\beta}}$ estimates to the third decimal.

\subsection{Some details of the numerical algorithm for $\widetilde{\bm \beta}$}
Recall that $\widetilde{\bm\beta}$ is the zero-crossing of $\psi_n(\bm{\beta})$, where
\begin{equation*}
\psi_n(\bm{\beta}) =\frac{1}{n^2} \sum_{i\neq j}(X_i-X_j)\left\{I(Y_i>Y_j)- \widehat F_{\bm{\beta}}\left((X_i-X_j)^T \bm{\beta} \right)\right\}.
\end{equation*}
Let $S_n(\bm\theta)=\psi_n(\bm{\beta}) $ with  $\bm{\beta}$ being replaced by the polar transformation given in Section 9.1.
We define $\widetilde{\bm\theta}$ to be the zero-crossing of $S_n(\bm\theta)$. Clearly, once $\widetilde{\bm{\theta}}$ is obtained, $\widetilde{\bm{\beta}}$ can be obtained accordingly based on the polar transformation.

Therefore, we focus on how to calculate $\widetilde{\bm\theta}$ numerically.
Note that the dimension of $S_n(\bm\theta)$ is $p$, while the dimension of $\bm\theta$ is $p-1$. Therefore, the zero-crossing of $S_n(\bm\theta)$ may not exist when $n$ is small, although we have theoretically established its asymptotic existence in Theorem \ref{thm-normality}. To bypass this difficulty, we define $S_n^{(-k)}(\bm\theta)$ to be the vector obtained by removing the $k$th element of $S_n(\bm\theta)$, so that the dimensions of $S_n^{(-k)}(\bm\theta)$ and $\bm{\theta} $ are both $p-1$.
If we can obtain $\widetilde{\bm\theta} ^{(-k)}$   the zero-crossing of $S_n^{(-k)}(\bm\theta)$, we can then combine $\widetilde{ \bm\theta }^{(-1)}, \cdots, \widetilde{ \bm\theta }^{(-p)}$ to get
$$
\widetilde{\bm\theta}=\frac{1}{p}\sum_{k=1}^p\widetilde{ \bm\theta }^{(-k)}.
$$

Next, we discuss how to obtain $\widetilde{\bm\theta} ^{(-k)}$ from  $S_n^{(-k)}(\bm\theta)$. 
When $p=2$, the dimension of $S_n^{(-k)}(\bm\theta) $ is 1. 
We can apply the R function {\it uniroot.all()} in the rootSolve package to find  the zero-crossings of $S_n^{(-k)}(\bm\theta)$; if the solution is not unique, 
we select the one that leads to the largest value of $pl(\bm\theta)$.
When $p>2$, the dimension of $S_n^{(-k)}(\bm\theta) $ is greater than 1. The existing R functions for finding the roots of multivariate equations may not lead to stable results, since it is necessary to 
evaluate the Jacobian, which may not exist
for $S_n^{(-k)}(\bm\theta) $. 
Following a similar strategy  used by Balabdaoui et al. (2019), 
we suggest minimizing $\|S_n^{(-k)}(\bm\theta) \|_2 $ to obtain $\widetilde{ \bm\theta }^{(-k)}$. 
This can be done by using R function optim() with the default method  ``Nelder--Mead" 
and multiple initial values, say 25.
If multiple  zero-crossings of $S_n^{(-k)}(\bm\theta)$ exist, we select the one that leads to the largest value of $pl(\bm\theta)$.

}

%
%
%
%
%

\section{Extension to Data with Ties in Responses} \label{section-ties}

In the main article, we have assumed that the responses $Y_i$ have continuous distributions, and there are no ties. However, our pairwise rank likelihood method can readily be extended to accommodate the case where the observed responses $Y_i$ contain ties. In particular, noting the derivation of (3.1) in the main article, we have
\begin{eqnarray*}
P(Y_i\leq Y_j|X_i,X_j)&=&P\left(\epsilon_i-\epsilon_j\leq (X_j-X_i)^T\bm{\beta}|X_i,X_j\right)
=F((X_j-X_i)^T\bm{\beta}),\label{probij1}
\end{eqnarray*}
where we have used only the fact that $F(\cdot)$ is the c.d.f. of $\epsilon_i - \epsilon_j$, and therefore
\begin{eqnarray*}
P(Y_i>Y_j|X_i,X_j) = 1 - P(Y_i\leq Y_j|X_i,X_j) = 1- F((X_j-X_i)^T\bm{\beta}).
\end{eqnarray*}
In the spirit of (3.2) in the main article, the pairwise rank log-likelihood can be defined to be
\begin{eqnarray*}
\widetilde \ell(\bm{\beta},F)&=& \sum_{i\neq j}\Big[I(Y_i>Y_j)\log\{ P(Y_i>Y_j|X_i,X_j) \} \\ && \hspace{0.3in}+I(Y_i\leq Y_j)\log \{P(Y_i\leq Y_j|X_i,X_j)\}\Big]\\
&=& \sum_{i\neq j}\Big[I(Y_i>Y_j)\log\{ 1- F((X_j-X_i)^T\bm{\beta})\}\\&&\hspace{0.3in}+I(Y_i\leq Y_j)\log \{F((X_j-X_i)^T\bm{\beta})\}\Big].
\end{eqnarray*}
As a consequence, the algorithm for estimating $\bm{\beta}$ and $F(\cdot)$ is based on the procedures given in Section 3.1 of the main article.
{We observe that $\widetilde \ell(\bm{\beta}, F)$ accounts for the potential effects of ties in the responses;} the reason is as follows. When there are ties, the effects of the event $Y_i = Y_j$ and its probability $P(Y_i = Y_j|X_i, X_j)$ have been respectively accounted for in $I(Y_i \leq Y_j)$ and the corresponding probability $P(Y_i \leq Y_j|X_i,X_j)$. For a more intuitive view of this, note that for every given $i\neq j$, the only terms in $\widetilde \ell(\bm{\beta},F)$ that contain data from both the $i$th and $j$th subjects are
$$
I(Y_i>Y_j)\log\{ 1- F((X_j-X_i)^T\bm{\beta})\} + I(Y_i\leq Y_j)\log\{ F((X_j-X_i)^T\bm{\beta})\}
$$
and
$$
I(Y_i<Y_j)\log\{ 1- F((X_i-X_j)^T\bm{\beta})\} + I(Y_i\geq Y_j)\log\{F((X_i-X_j)^T\bm{\beta})\}.
$$
As a consequence, from the table below, we observe that for different relationships of $Y_i$ and $Y_j$ (i.e., $Y_i = Y_j$ or $Y_i>Y_j$ or $Y_i<Y_j$), the contribution to $\widetilde \ell(\bm{\beta},F)$ is different.

\begin{center}

\begin{tabular}{c|c}
\hline
&Contribution of observations $i$ and $j$ to $\widetilde \ell(\bm{\beta},F)$\\ \hline
$Y_i = Y_j$ & $\log\{ F((X_j-X_i)^T\bm{\beta})\} +  \log\{F((X_i-X_j)^T\bm{\beta})\}$\\
$Y_i > Y_j$ &  $\log\{ 1- F((X_j-X_i)^T\bm{\beta})\} + \log\{F((X_i-X_j)^T\bm{\beta})\}$  \\
$Y_i < Y_j$ & $\log\{ F((X_j-X_i)^T\bm{\beta})\} + \log\{ 1- F((X_i-X_j)^T\bm{\beta})\}$  \\ \hline
\end{tabular}

\end{center}

{\color{black} 
\section{Extension to Right-Censored Data} \label{section-right-censor}

In this section, we extend our methods to accommodate the right-censored data.
Let $T_i$ be the survival time of a subject $i$, and let $C_i$ be the corresponding censoring time, $i=1,\ldots, n$. Denote by $\{Y_i, \Delta_i, X_i\}_{i=1}^n$ the observed i.i.d. data, where $Y_i = \min\{T_i, C_i\}$ is the observed survival time of the $i$th subject with censoring and $\Delta_i$ is the censoring indicator; namely $\Delta_i = 1$ if $Y_i = T_i$, and $\Delta_i = 0$ otherwise.

\subsection{The pairwise rank likelihood for the right-censored data} \label{section-likelihood-censor}

Assuming that $T_i, i=1,\ldots, n$, are observed,  the pairwise rank likelihood, based on $(T_i, X_i),i=1,\ldots,n$, is given by
\begin{eqnarray}
\ell(\bm{\beta},F)
&=&\sum_{i<j}\Big[I(T_i>T_j)\log\{ F((X_i-X_j)^T\bm{\beta})\} \nonumber \\&&\hspace{0.3in}+I(T_i\leq T_j)\log \{1-F((X_i-X_j)^T\bm{\beta})\Big]. \nonumber
\end{eqnarray}
When the data are subject to censoring, we can incorporate the idea of Cheng et~al. (1995), i.e., the inverse probability estimation method, in this objective function. In particular, we replace $I(T_i>T_j)$ by $\Delta_jI(Y_i>Y_j)\{G^{-2}(Y_j)\}$, where $G(y) = P(C_i>y)$ is the unknown survival function for the censoring time. As a consequence, the pairwise rank likelihood based on $(Y_i, \Delta_i, X_i),i=1,\ldots,n$, is given by
\begin{eqnarray}\nonumber
\ell(\bm{\beta},F, G)&=&\sum_{i<j}\left[\frac{\Delta_j I(Y_i>Y_j)}{G^2(Y_j)}\log\{ F((X_i-X_j)^T\bm{\beta})\}\right.\\
&&\hspace{0.5in}+\left.\frac{\Delta_i I(Y_i\leq Y_j)}{G^2(Y_i)}\log \{1-F((X_i-X_j)^T\bm{\beta})\}\right].\label{wtloglike}
\end{eqnarray}
We observe that $\ell(\bm{\beta},F, G)$ is an unbiased estimator of $\ell(\bm{\beta}, F)$ if we assume that $T_i$ and $C_i$ are independent and $G(y)$ is continuous, since
{
\begin{eqnarray}
\label{eq-unbias-survival} &&E\left\{\frac{\Delta_j I(Y_i>Y_j)}{G^2(Y_j)}\bigg|X_i,X_j\right\} \\
&=&E\left\{\frac{I(T_j\leq C_j) I(\min(T_i,C_i)>\min(T_j,C_j))}{G^2(Y_j)}\bigg|X_i,X_j\right\} \nonumber\\
&=&E\left\{\frac{I(T_j\leq C_j) I(\min(T_i,C_i)>T_j)}{G^2(T_j)}\bigg|X_i,X_j\right\} \nonumber\\
&=&E\left[ E\left\{ \frac{I(T_j\leq C_j) I(T_i>T_j) I(C_i>T_j)}{G^2(T_j)}\bigg|T_j,X_i,X_j\right\} \bigg| X_i,X_j \right] \nonumber\\
&=&E\left\{I(T_i>T_j)|X_i,X_j\right\}. \nonumber
\end{eqnarray}
}
Note that the survival function $G(\cdot)$ can be estimated by the Kaplan--Meier estimator, denoted by $\widehat G(\cdot)$. As a consequence, $\bm{\beta}$ and $F$ are estimated by
\begin{eqnarray*}
(\widehat{\bm{\beta}}, \widehat F) = \mbox{argmax}_{\bm{\beta}\in \mathB, F \in \mathF} \ell(\bm{\beta}, F, \widehat G),
\end{eqnarray*}
where $\mathB$ and $\mathF$ share the same definitions as those without censoring in this article. This optimization problem can be solved using similar techniques to Section 3.1 in the main article and Section \ref{section-numerical-details-1}.

\subsection{Score-function-based method for right-censored data}

Similar in strategy to the developments of Section 3.2 in the main article, we can also derive a score-function-based method to estimate $\bm{\beta}$ for right-censored data. The procedure is sketched as follows.

Based on \eqref{wtloglike}, we can define
\begin{eqnarray*}
\widehat F_{\bm{\beta}, G}(\cdot) = \mbox{argmax}_{F \in \mathF}\ell(\bm{\beta}, F, G)
\end{eqnarray*}
for each $G(\cdot)$ and $\bm{\beta}\in \mathB$. This leads to the profile likelihood for $\bm{\beta}$: $\ell\left(\bm{\beta}, \widehat F_{\bm{\beta}}, G\right)$, whose score function (if it exists) has the form:
\begin{eqnarray*}
\overline{\psi}_n(\bm{\beta}, G) &=& \sum_{i\neq j} \xi(X_i,X_j;  \widehat F_{\bm{\beta}, G}, \bm{\beta})\bigg[\frac{\Delta_j I(Y_i>Y_j)}{G^2(Y_j)} \\ &&- \left\{ \frac{\Delta_j I(Y_i>Y_j)}{G^2(Y_j)} + \frac{\Delta_i I(Y_i\leq Y_j)}{G^2(Y_i)}  \right\} \widehat F_{\bm{\beta}, G}\left((X_i-X_j)^T \bm{\beta} \right)\bigg].
\end{eqnarray*}
By replacing $\xi(X_i,X_j;  \widehat F_{\bm{\beta}, G}, \bm{\beta})$ with $X_i-X_j$, this score function is reduced to
\begin{eqnarray*}
\psi_n(\bm{\beta}, G) &=& \sum_{i\neq j} (X_i-X_j)\bigg[\frac{\Delta_j I(Y_i>Y_j)}{G^2(Y_j)} \\ &&- \left\{ \frac{\Delta_j I(Y_i>Y_j)}{G^2(Y_j)} + \frac{\Delta_i I(Y_i\leq Y_j)}{G^2(Y_i)}  \right\} \widehat F_{\bm{\beta}, G}\left((X_i-X_j)^T \bm{\beta} \right)\bigg].
\end{eqnarray*}
Consequently, we can define $\widetilde{\bm{\beta}}$, the score-function-based estimate for $\bm{\beta}$, to be the zero crossing of $\psi_n(\bm{\beta}, \widehat G)$, where similar to Section \ref{section-likelihood-censor}, $\widehat G(\cdot)$ is the Kaplan--Meier estimator for the survival function $G(\cdot)$.
}

\end{document}